\documentclass[a4paper,aps,prl,twocolumn,showpacs,amsmath,amssymb,superscriptaddress,preprintnumbers]{revtex4-2}
\usepackage[english]{babel}

\usepackage[letterpaper,top=2cm,bottom=2cm,left=3cm,right=3cm,marginparwidth=1.75cm]{geometry}
\usepackage{amsmath}
\usepackage{amsfonts}
\usepackage{bm}
\usepackage{graphicx}
\usepackage[colorlinks=true, allcolors=blue]{hyperref}
\usepackage{xcolor}
\usepackage{booktabs}
\usepackage[capitalise,noabbrev]{cleveref}
\usepackage{siunitx}
\DeclareSIUnit\hartree{\text {Ha}}

\newcommand{\Lgap}{L_\text{bandgap}}

\begin{document}
\title{
    Algorithmic differentiation for plane-wave DFT:\\
    materials design, error control and learning model parameters
}

\author{Niklas Frederik Schmitz}
\email{niklas.schmitz@epfl.ch}
\thanks{(corresponding author)}
\affiliation{Mathematics for Materials Modelling (MatMat)\char`,{} Institute of Mathematics \& Institute of Materials,
École Polytechnique Fédérale de Lausanne, 1015 Lausanne, Switzerland}
\affiliation{National Centre for Computational Design and Discovery of Novel Materials (MARVEL),
\'Ecole Polytechnique F\'ed\'erale de Lausanne, 1015 Lausanne, Switzerland}

\author{Bruno Ploumhans}
\affiliation{Mathematics for Materials Modelling (MatMat)\char`,{} Institute of Mathematics \& Institute of Materials,
École Polytechnique Fédérale de Lausanne, 1015 Lausanne, Switzerland}
\affiliation{National Centre for Computational Design and Discovery of Novel Materials (MARVEL),
\'Ecole Polytechnique F\'ed\'erale de Lausanne, 1015 Lausanne, Switzerland}

\author{Michael F. Herbst}
\email{michael.herbst@epfl.ch}
\thanks{(corresponding author)}
\affiliation{Mathematics for Materials Modelling (MatMat)\char`,{} Institute of Mathematics \& Institute of Materials,
École Polytechnique Fédérale de Lausanne, 1015 Lausanne, Switzerland}
\affiliation{National Centre for Computational Design and Discovery of Novel Materials (MARVEL),
\'Ecole Polytechnique F\'ed\'erale de Lausanne, 1015 Lausanne, Switzerland}

\begin{abstract}
\noindent
We present a differentiation framework for plane-wave
density-functional theory (DFT) that combines the strengths of forward-mode algorithmic
differentiation~(AD) and density-functional perturbation theory~(DFPT).
In the resulting AD-DFPT framework
derivatives of any DFT output quantity
with respect to any input parameter~(e.g.~geometry, density functional or pseudopotential)
can be computed accurately without deriving gradient expressions by hand.
We implement AD-DFPT into the Density-Functional ToolKit (DFTK)
and show its broad applicability. Amongst others we consider
the inverse design of a semiconductor band gap,
the learning of exchange-correlation functional parameters, or the propagation
of DFT parameter uncertainties to relaxed structures.
These examples demonstrate a number of promising research avenues
opened by gradient-driven workflows in first-principles materials modeling.
\end{abstract}
\maketitle

\section{Introduction}

\noindent
The central goal of first-principles modeling is to provide access to
accurate predictions of atomistic properties.
Considering the most widely used approach, density-functional theory (DFT)~\cite{HohenbergKohn1964,KohnSham1965},
properties are commonly
obtained from the response of the electronic structure to an external perturbation,
i.e.~as derivatives of DFT simulation outcomes.
For example, phonon dispersion relations can be obtained from
the second-order derivative of the DFT energy wrt.~atomic positions~\cite{Baroni2001},
or dielectric susceptibility as the derivative of polarization wrt.~electric field strength~\cite{Gonze1997dynamicalmatrices}.

Density-functional perturbation theory (DFPT)~\cite{Baroni2001}
provides the rigorous framework to compute DFT derivatives.
It has taken decades of joint community effort
to iron out many subtleties of DFT gradient computation%
~\cite{Baroni1987elastic,Gonze1997conjugategradientdfpt,Hamann2005metrictensor,Wu2005systematicdfpt,gajdos2006vaspdfpt,LaflammeJanssen2016precise,Cances2023,gonze_variational_2024},
resulting in the efficient implementations available nowadays in widespread DFT codes~\cite{QE,romero_abinit_2020,Kuhne2020cp2k,box2025fhiaims_dfpt,kresse_efficient_1996}. 
However, the generality of these implementations varies, as they are typically
limited to specific types of DFT functionals and perturbations.
Manually adding support for additional classes of DFT models or DFT derivatives
can be a daunting task,
resulting in a substantial obstacle
to rapid exploration of novel avenues in materials modeling.

In an attempt to avoid this considerable human effort,
researchers often have to resort to
finite-difference-based approaches,
see for example the comprehensive frameworks for DFT elasticity tensors~\cite{Golesorkhtabar2013ElaStic,de_jong_charting_2015},
phonons~\cite{togo2023phonopy-phono3py},
or other spectroscopic properties~\cite{Bastonero2024}.
However,
finite-difference techniques suffer from well-known deficiencies,
such as their sensitivity to numerical noise
and the need to find an appropriate step size~\cite{LaflammeJanssen2016precise,Golesorkhtabar2013ElaStic}.

This article explores the computation of DFT gradients
employing algorithmic differentiation~(AD) techniques.
AD offers a rigorous mathematical approach to compute derivatives
automatically and accurately~\cite{Griewank2008EDP,Naumann2012TAo}.
Recent advances in general-purpose AD systems%
~\cite{pytorch,jax2018github,Revels2016forwarddiff,Moses2021} have
significantly broadened their applicability in scientific computing~\cite{Baydin2018review,sapienza2024differentiableprogrammingdifferentialequations}.
As we will discuss, AD does not replace DFPT;
rather, it allows to set up and solve appropriate DFPT problems transparently.
This results in a powerful systematic AD-DFPT framework
capable of computing gradients in an end-to-end fashion, across entire DFT workflows.

The use of AD techniques in atomistic modeling is hardly a novelty;
see the broad list of examples related to
machine-learning force fields~\cite{Unke2021,Schmitz2022ad,Langer2023ADheatflux,Gonnheimer2025ad_hessian_mlip},
the differentiable programming of thermodynamical observables%
~\cite{Wang2020,Greener2021,SchwalbeKoda2021,Maliyov2024}, or model
Hamiltonians~\cite{Li2022deeplearning,VargasHernandez2023,Li2024nndft}.
Considering differentiable DFT simulations specifically,
first software packages~\cite{Chen2021,Kasim2022,Zhang2022pyscf,MCasares2024}
with AD capabilities have recently appeared
for simulations employing Gaussian basis sets.
Their ability to compute arbitrary DFT gradients has already enabled
novel approaches to machine-learned DFT functionals
or the inverse design of molecules%
~\cite{Li2021,Kasim2021,Chen2021,Wu2023xcml}.
In plane-wave DFT related settings, recent work focused on AD in orbital-free DFT~\cite{tan2023professAD}
or for implementing direct minimization algorithms~\cite{li2024jrystal}. 

However, a systematic AD treatment in standard plane-wave DFT and DFPT has so far remained an open challenge.
The underlying difficulty stems from the distinct mathematical structure of plane-wave methods:
While Gaussian basis sets result in tractable dense matrices
for the Hamiltonian or the density matrix,
for plane-wave basis sets these central objects are much larger, but structured.
Exploiting this structure appropriately ---
e.g.~via fast Fourier transforms~(FFTs)
and iterative algorithms
--- %
is essential to obtain an efficient code~\cite{kresse_efficient_1996}.
The same strategies are indispensable when computing DFT derivatives.
In other words, the successful use of AD in the plane-wave setting requires
the careful incorporation of the algorithmic insights that underlie decades of DFPT development.

\begin{figure*}
    \centering
       \includegraphics[width=\linewidth]{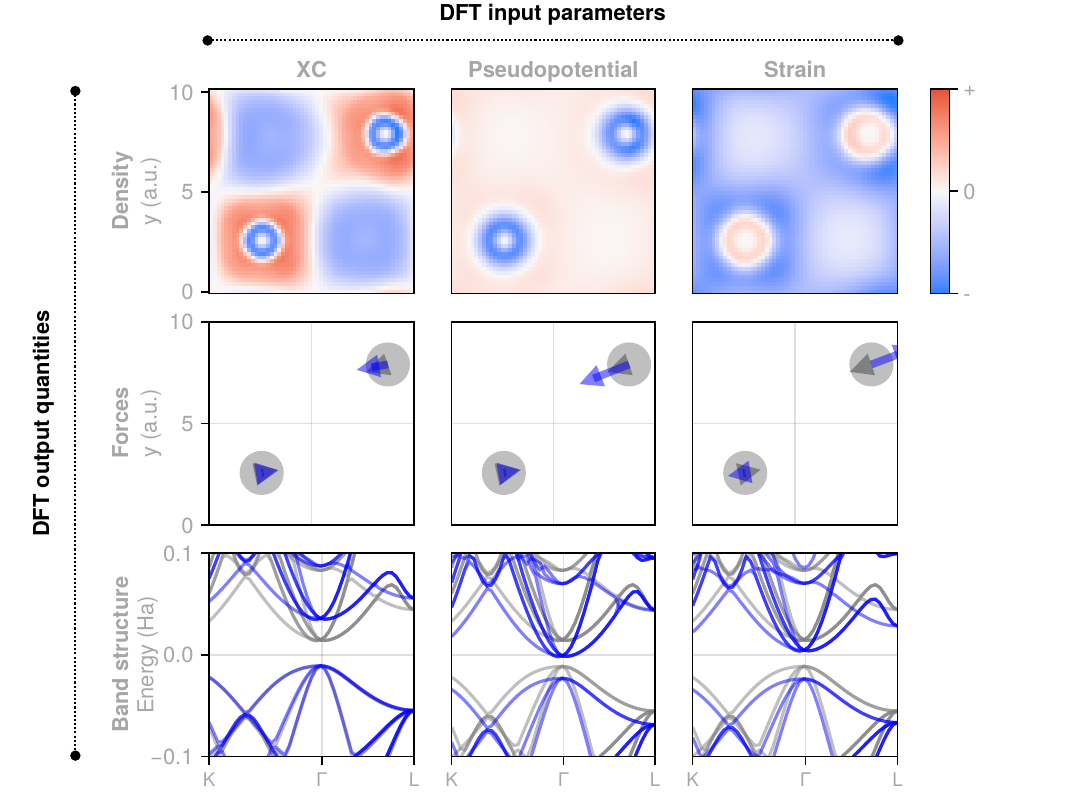}
    \caption{\textbf{Systematic DFT derivatives.}
    Examples of physical quantities (rows) differentiated with respect
    to input parameters (columns),
    illustrating the combinatorial range of quantity-parameter derivatives
    readily accessible with our AD-DFPT framework.
    Quantities are displayed for a silicon unit cell. %
    Densities and non-zero forces are shown along a $z=0$ plane
    and the structure was slightly distorted.
    The parameter-induced changes have been scaled to improve visibility.
    }
    \label{fig:combinatorial}
\end{figure*}

In this work, we present a first AD-DFPT framework for plane-wave DFT.
At the conceptual level, our approach integrates AD with DFPT transparently
and with high generality:
geometric parameters such as strain, and DFT model parameters such as those of
the exchange-correlation (XC) functional or pseudopotentials now all enter on an
equal footing; see~\cref{fig:combinatorial}.
The combinatorial number of possible derivatives
of DFT output quantities versus any of these parameters
thus become readily available for use in materials modeling.

Our framework uses forward-mode AD~\cite{Griewank2008EDP},
where perturbations are propagated forward from input parameters to all
outputs, thus providing a natural generalization of traditional DFPT.
Computationally, the complexity of accumulating explicit derivative
tensors (or gradients) depends on the number of independent input perturbations,
just like in traditional DFPT.
The extension of AD-DFPT to reverse-mode AD, suitable for
high-dimensional gradients arising from loss functions of many input parameters
(such as training highly parametrized machine-learned exchange-correlation functionals), is
left as a promising avenue for future extension.

Practically, we realized our AD-DFPT framework by direct implementation into the Density-Functional Toolkit (DFTK)~\cite{DFTKpaper}, a flexible DFT code written in the Julia programming language.
These efforts were greatly facilitated by the ongoing developments in the Julia community
towards powerful AD tools~\cite{Revels2016forwarddiff,Moses2020,Moses2021,innes2019differentiableprogrammingbridgemachine,frames_white_2025_chainrules,dalle2025commoninterfaceautomaticdifferentiation}.
Moreover, DFTK's simple and tractable code base of only about 10\,000 lines of Julia code
enabled us to readily equip this existing DFT package with AD capabilities.
This is in contrast to previous differentiable DFT software,
which was either written from scratch for the purpose to be differentiable
or represents a hard fork of an existing code base.
We not only believe this integrated development model to be more sustainable in the long run,
but we could already benefit from it during this research:
recent algorithmic advances on solving DFPT problems~\cite{Cances2023,herbst2025efficientkrylovmethodslinear},
which were developed using DFTK, were immediately available to us.

The remainder of this manuscript is structured as follows:
We first provide an overview of the key developments
required to obtain an AD-based framework for
end-to-end differentiable DFT workflows.
We then provide six examples to illustrate
novel research avenues enabled by the framework, namely:
(1)   Computation of elastic constants by applying AD on top of AD,
(2)  engineering a semiconductor band gap,
(3) learning exchange-correlation parameters,
(4)  optimizing pseudopotentials,
(5)   propagating the DFT model error to relaxed geometries,
and
(6)  estimating the error in DFT forces due to the chosen plane-wave cutoff.
Collectively, these highlight how end-to-end differentiation capabilities
turn derivative information into a first‑class asset enhancing
accuracy, design, and reliability in materials modeling.

\section{Results}
\subsection{Algorithmic differentiation framework for plane-wave DFT}
\begin{figure*}
    \centering
    \includegraphics[width=\linewidth]{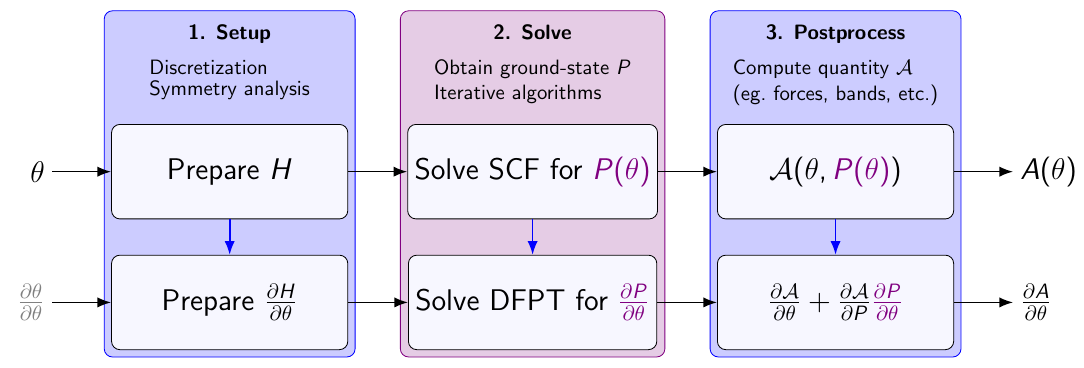}
    \caption{\textbf{End-to-end derivatives in our AD-DFPT framework.} We embed
        plane-wave DFT into a general-purpose AD system, which across the
        entire simulation workflow $A$ (top row) computes the end-to-end
        derivative $\frac{\partial A}{\partial \theta}$ (bottom row).
     Based on forward-mode AD, the full derivative %
     is accumulated starting from the input $\frac{\partial \theta}{\partial\theta}=1$
     and following each primitive computational step in order.
     Here, blue arrows indicate dependencies on intermediate quantities.
     The AD system automatically obtains the Hamiltonian perturbation
     $\frac{\partial H}{\partial \theta}$ entering DFPT,
     as well as the contributions of the postprocessing.
     For the SCF algorithm~\eqref{eq:fixedpoint} we manually define its
     derivative as the matching DFPT algorithm~\eqref{eq:Dyson},
     see details in the main text.
     }
    \label{fig:flowchart}
\end{figure*}

\noindent
An end-to-end differentiable DFT workflow is achieved by making a
DFT code interact seamlessly with a general-purpose AD system. We illustrate
our approach using~\cref{fig:flowchart}, based on the three conceptual
workflow stages: setup, solve, and postprocess.

Entering the setup stage are the simulation parameters $\theta$~(first row of~\cref{fig:flowchart}). 
Depending on context $\theta$ may indicate XC functional coefficients,
the parameters of the pseudopotential model or the system's geometry.
These parameters are needed for the construction
of the plane-wave basis, pseudopotential projectors, and potentials.
Ultimately $\theta$ thus defines the discretized Kohn-Sham Hamiltonian $H(\theta, P)$ and
energy functional $\mathcal{E}(\theta, P)$ as functions of a trial density matrix $P$.
Note that in this section we employ a formulation based on density matrices
(instead of explicitly denoting ground-state density, orbitals and occupations)
for conciseness; in our calculations equivalent orbital-based representations are used instead (see Methods).
In the solve stage, the discretized Hamiltonian and energy functional
are used to determine the
ground state density matrix $P(\theta)$ by iteratively solving the
self-consistent field (SCF) equations.
Finally, the postprocessing stage
evaluates the desired physical quantities, such as total energy,
forces, or band structure, which we indicate by the function $\mathcal{A}$.
Evaluating these quantities $\mathcal{A}$ in turn consumes the self-consistent
density matrix $P(\theta)$, but may also feature an explicit dependency on
simulation parameters $\theta$. 
Considering the workflow in its entirety, from the input parameters to the
predicted DFT quantity, thus defines a function
\begin{align}
    A(\theta) = \mathcal{A}\left(\theta, P(\theta)\right).
\end{align}
The end-to-end derivative of this function follows from the chain rule
\begin{align}
    \label{eq:chainrule}
    \frac{\partial A}{\partial\theta} = \frac{\partial \mathcal{A}}{\partial \theta} + \frac{\partial \mathcal{A}}{\partial P} \frac{\partial P}{\partial \theta}.
\end{align}
Here, the first term represents the explicit dependence of the quantity $A$ on
the parameters, while the second captures the implicit dependence through the
ground-state response $\frac{\partial P}{\partial \theta}$.
In particular the implicit term
is computationally more involved since it requires differentiating the SCF
solution itself.

End-to-end derivatives such as~\cref{eq:chainrule}
can be obtained automatically using modern AD systems.
In a nutshell this is achieved by working directly
on the level of the computer program implementing the workflow $A$,
combining three ingredients:
(1) A library of known differentiation rules for a set of
primitive operations.
Such primitives may range from fine-grained operations,
e.g.~floating-point operation on numbers or matrix arithmetic,
to coarse-grained standard algorithms,
such as FFTs or eigenvalue solvers.
(2) A mechanism to accumulate the full gradient $\frac{\partial A}{\partial \theta}$
from the derivatives of the primitives.
Here, the AD system decomposes the entire workflow $A$ into a sequence
of primitives, applies the tabulated rule to
each, and assembles the full gradient via the chain rule.
(3)
A mechanism for defining new primitives, allowing developers to incorporate domain-specific knowledge into custom differentiation rules.

In our AD-DFPT approach, we use these ingredients to automatically compute the
derivative $\frac{\partial A}{\partial \theta}$ by letting the AD system pass
through the setup, solve and postprocess stages of the
computation, see second row of~\cref{fig:flowchart}.
The setup stage is treated entirely using the system's library of differentiation rules
as well as its accumulation mechanism.
This results in the derivative of the discretized Kohn-Sham Hamiltonian
$\frac{\partial H}{\partial \theta}$.
Following the workflow, the AD system encounters the solve stage.
This stage we define as a custom primitive
\emph{imposing} the derivative of the SCF solver to be evaluated by solving a
matching linear-response (DFPT) problem. As explained below
this yields the response $\frac{\partial P}{\partial \theta}$
from the perturbation $\frac{\partial H}{\partial \theta}$.
Finally, we once again let the AD system differentiate through
the postprocess stage, assembling the desired derivative
as an end result.

Compared to traditional DFPT approaches, AD-DFPT still relies crucially
on a general-purpose linear-response (DFPT) solver, requiring manual
implementation and careful tuning.
However, once this solver is available,
the tedious and error-prone hand-derivation of all possible setup and postprocessing combinations (compare Figure~\ref{fig:combinatorial})
is now completely automated.

To conclude this section, we discuss the custom rule of the solve stage,
that is how to compute $\frac{\partial P}{\partial \theta}$ from $\frac{\partial H}{\partial \theta}$.
For notational simplicity we suppress spin and Brillouin zone sampling in this
discussion and all operators are understood after discretization in a
plane-wave basis of size $N_b$.
In the solve stage we determine the
ground state $P(\theta)$ by minimizing
the free energy $\mathcal{E}$ including the internal Kohn-Sham energy,
the electronic smearing entropy contribution, and the ion-ion electrostatic energy.
Usually this is done by satisfying its first-order stationarity conditions,
the SCF equations
\begin{align}
    \left\{
    \begin{aligned}
        H&(\theta, P)\psi_n = \varepsilon_n\psi_n, \qquad n=1,\ldots,N_b, \\
        P &= \sum_{n=1}^{N_b} f(\varepsilon_n) \psi_n\psi_n^\dagger,
    \end{aligned}
    \right.
  \label{eq:scforbitals}
\end{align}
where the $(\varepsilon_n, \psi_n)$ are $N_b$ orthonormalized eigenpairs of the
Hamiltonian and $f$ is a smearing function enforcing the correct electron count in $P$.
Viewing $f$ as a matrix function acting on the Hamiltonian, we can equivalently write \eqref{eq:scforbitals} as
\begin{align}
    P = f\left(H(\theta, P)\right).
  \label{eq:fixedpoint}
\end{align}
From here, the SCF custom rule is obtained naturally by differentiating with respect to $\theta$.
After rearrangement, this yields
\begin{align}
    \frac{\partial P}{\partial\theta} = \left(1 - \chi_0 K\right)^{-1}\chi_0 \frac{\partial H}{\partial\theta},
    \label{eq:Dyson}
\end{align}
where we use the shorthand ${\chi_0 := \frac{\partial f}{\partial H}}$ and ${K := \frac{\partial H}{\partial P}}$.
Here $\chi_0$ captures the independent susceptibility of the density matrix to
a change in the Hamiltonian, while $K$ collects the dependence of the
Hamiltonian on the density matrix through the nonlinear Hartree and XC terms.
Equation~\eqref{eq:Dyson} is thus a matrix-based formulation
of the Dyson equation to compute the interacting response in DFPT.
It provides us exactly with the linear response problem to be solved
as part of the custom rule of the solve stage.

\newcommand{\secorbitals}{Orbital-based representation for response}%
While the notation of~\eqref{eq:Dyson} is compact, it involves as principal unknowns
density matrices of size $\mathcal{O}(N_b^2)$, which cannot be stored explicitly
given the typical number $N_b$ of basis functions used in plane-wave calculations.
As common in plane-wave DFT we thus rely on
matrix-free formulations and iterative solvers to solve the DFPT problem \eqref{eq:Dyson}
as described in more details in the Methods section \textit{\secorbitals}.

\newcommand{\secelasticity}{Elasticity: Accurate standard properties with minimal human effort}
\subsection{\secelasticity} \label{sec:elastic}

\begin{figure*}
    \centering
    \includegraphics[width=\linewidth]{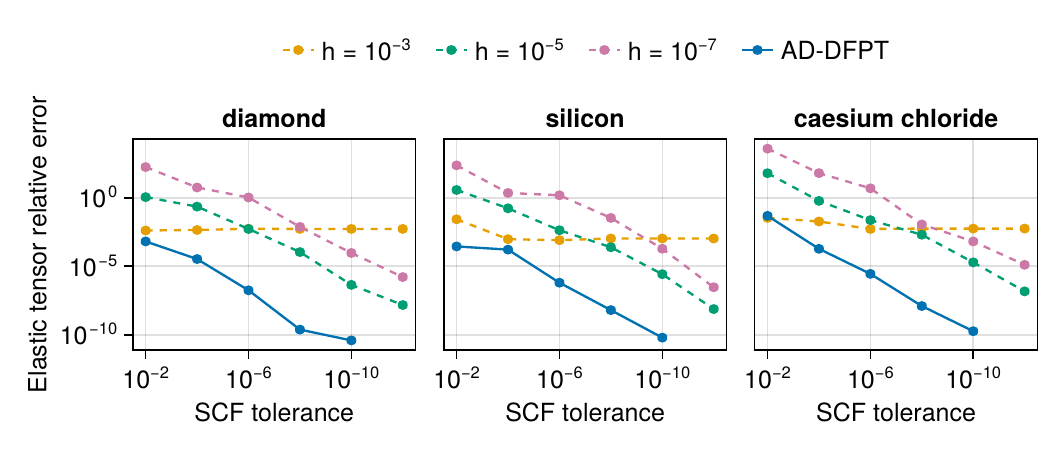}
    \caption{\textbf{Elasticity.} Relative error in the clamped-ion elastic tensor
        ($\|C - C_\text{ref}\|_F / \|C_\text{ref}\|_F$)
        for indicated solids as a function of
        SCF tolerance. The dashed curves correspond to finite-difference
        values obtained on top of
        stresses with step sizes $h$ as indicated in the legend.
        AD-DFPT (solid curve) denotes a direct computation of
        second-order energy derivatives
        within our AD framework. All relative errors are computed with respect to the
        AD-DFPT result at SCF tolerance $10^{-12}$, see Supplementary Table~S1 for the numerical values.
        AD-DFPT proves to agree well with finite differences at tightest SCF tolerances and be the most precise at looser tolerances,
        while finite-difference results deteriorate notably for looser SCF tolerance
        and are sensitive to the step size parameter~$h$.}
    \label{fig:elastic}
\end{figure*}

\noindent
Elastic constants, which characterize the linear response of a material to
strain, are fundamental for predicting mechanical stability, sound velocities,
and thermomechanical behavior.
Traditionally, they are computed either by finite differences (FD)%
~\cite{Golesorkhtabar2013ElaStic,de_jong_charting_2015}
or in the form of providing manual extensions on top of DFPT.
While FD methods require carefully tuned strain increments to balance numerical noise and nonlinear effects,
DFPT-based approaches involve significant implementation efforts~\cite{Baroni1987elastic,Hamann2005metrictensor},
some of which are highly specific to the elastic constants case.

Using our general AD-DFPT framework, elastic constants emerge
naturally and with minimal coding effort.
The starting point is the stress in Voigt notation,
defined from the first derivative of the total energy $E(\eta) = \mathcal{E}(\eta,P(\eta))$
with respect to an applied Voigt strain $\eta$:
\begin{align}
    \label{eqn:strain}
    \sigma(\eta) = \frac{1}{V(\eta)} \frac{\partial \mathcal{E}}{\partial \eta},
\end{align}
where $V(\eta)$ is the volume of the strained unit cell, and we have used the
Hellmann-Feynman theorem. Minimizing the total energy, the equilibrium strain
$\eta^\star$ induces zero stress. At equilibrium, the elastic stiffness tensor
is defined as
\begin{align}
    \label{eqn:elestic}
    C = \frac{\partial\sigma}{\partial\eta}\bigg\vert_{\eta^\star}.
\end{align}
In DFTK, Hellmann-Feynman stresses are obtained in the postprocessing stage
by implementing literally~\cref{eqn:strain}, directly as a call to
the AD system, see~\cite[\texttt{src/postprocess/stresses.jl}]{herbst_2025_15793538}.
With our AD-DFPT framework elastic constants follow
immediately from a second invocation of the AD system,
computing the end-to-end derivative of the stress $\sigma$
by literally implementing \eqref{eqn:elestic}.
No hand-coded second derivatives are required.

As shown in~\cref{fig:elastic}, our approach inherits the robustness and
precision of DFPT. For three solids spanning a range of mechanical hardness,
the precision of elastic constants is benchmarked as a function of SCF
convergence tolerance.
At the tighest tolerance, AD-DFPT agrees well with finite differences, see Supplementary Table~S1 for the numerical values.
At the loosest SCF tolerances considered, our AD-DFPT approach proves to be the most
precise.
In contrast, finite-difference results show an increased dependence
on SCF tolerances: large steps lead to non-convergent error curves as other
error sources dominate, while very small steps amplify the SCF noise and
degrade precision.
This makes AD-DFPT techniques the preferred approach
when considering the trade-off between human implementation time
and derivative accuracy.

\newcommand{\secinv}{Inverse materials design}
\subsection{\secinv}
\begin{figure*}[!ht]
    \centering
    \includegraphics[width=\linewidth]{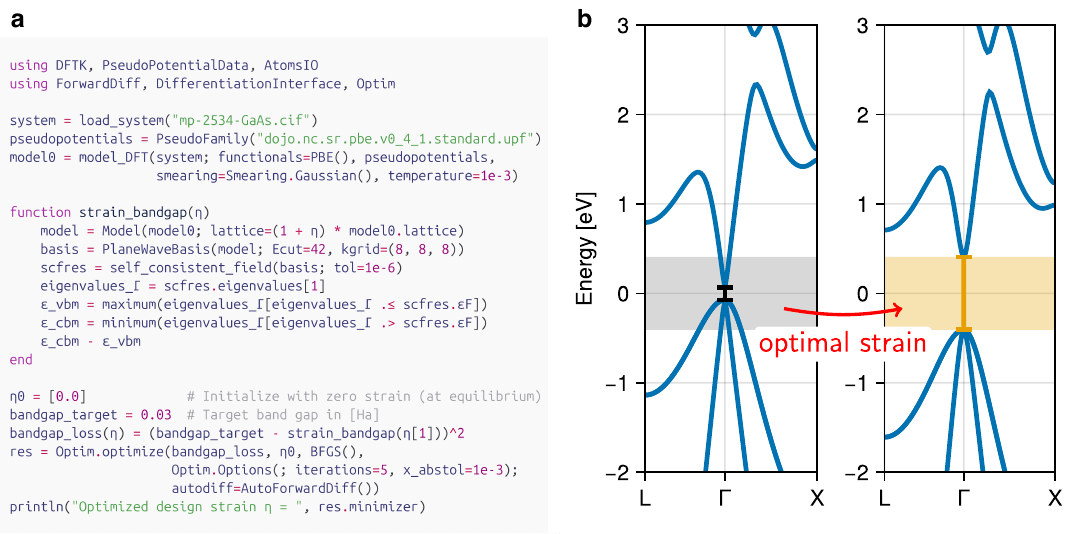}
    \caption{\textbf{\secinv.}
        \textbf{a} Minimal code example tuning the band gap of bulk GaAs with respect
        to volumetric strain.
        \textbf{b} Band structure of GaAs with the band gap before (left)
        and after (right) minimizing $\Lgap$.
        Energies are shown relative to the middle of the band gap. The optimizer internally invokes automatic differentiation to compute the required gradient, without requiring user intervention.
    }
    \label{fig:inversedesigngaas}
\end{figure*}
\noindent
A successful approach in computational materials discovery is inverse materials design.
In contrast to the usual, forward direction to estimate the functional properties of
a material given its structure, this approach does the reverse:
starting from a desired set of properties it seeks those atomistic structures satisfying these properties most closely%
~\cite{Franceschetti1999,Zunger2018,Noh2020}.
For example, when fine-tuning carrier mobilities or lifetimes in semiconductor devices
one usually seeks materials with specific band characteristics,
e.g. particular band gaps or band valleys.
To illustrate how a differentiable DFT code can be beneficial in this endeavor
we will consider a toy example,
namely the fine tuning of the band gap of bulk gallium arsenide (GaAs)
by applying a bulk volumetric strain $\eta$. %
This example is inspired by the remarkable successes of strain engineering
in the context of proposing better-suited optoelectronic devices
~\cite{%
Manasevit1982,%
Shi2019,%
Sun2007,%
Ponce2019%
}.

Mathematically, we can formulate this problem as the minimization of a loss function such as
\begin{align}
    \label{eqn:gaploss}
    \Lgap(\eta) = \left(E_g^\text{target} - E_g(\eta) \right)^2,
\end{align}
which measures the discrepancy of a predicted band gap $E_g(\eta)$ under strain $\eta$
from the targeted value $E_g^\text{target}$.
In our example we will simply obtain $E_g(\eta)$ using a Perdew-Burke-Ernzerhof (PBE)~\cite{Perdew1996} calculation on strained GaAs,
see the function \texttt{strain\_bandgap} in Figure~\ref{fig:inversedesigngaas}a.
In traditional DFT codes, obtaining the gradient of $E_g$ and thus the gradient
of $\Lgap$ is challenging, such that naive grid search techniques
or other derivative-free methods are still commonly employed
--- making inverse design problems an expensive endeavor in general.

However, employing an end-to-end
differentiable DFT code such as DFTK enables to compute
the gradient $\partial \Lgap / \partial\eta$ directly,
such that we can
use a classic algorithm like Broyden–Fletcher–Goldfarb–Shanno (BFGS)~\cite{Nocedal2006Chapter6quasinewton}
to rapidly minimize the loss.
\cref{fig:inversedesigngaas} demonstrates this on our GaAs example, where
the optimization achieves the target band gap in just three BFGS iterations.
For completeness, we also provide the full user code required to obtain this
result in \cref{fig:inversedesigngaas}a.
Notably, in this example the optimizer transparently triggers the AD-based
computation of the required gradient of the \texttt{bandgap\_loss} function,
thus enabling even novice users
to perform gradient-based inverse design with minimal
boilerplate.

\newcommand{\secxc}{Learning the exchange-correlation functional}
\subsection{\secxc} \label{sec:finetuning}
\begin{figure*}
    \centering
    \includegraphics[width=\linewidth]{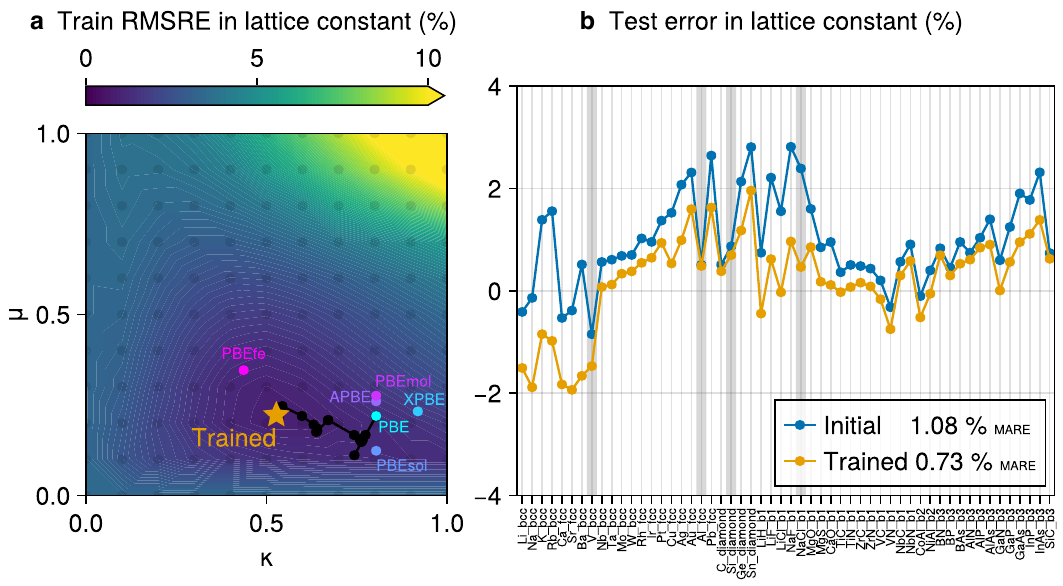}
    \caption{\textbf{\secxc.} Experimental lattice constants are targeted for solids in the Sol58LC dataset~\cite{lundgaard_mbeef-vdw_2016}. \textbf{a} 
    The training loss landscape in two parameters $\mu,\kappa$ of the PBE functional~\cite{Perdew1996}
    is visualized by an exhaustive grid search of the root mean squared relative error (RMSRE), along with several variants from the literature~\cite{Perdew2008pbesol,delCampo2012pbemol,xu2004xpbe,Constantin2011apbe,Sarmiento2015pbefe}.
    The efficient trajectory of the AD-DFPT-enabled optimization is shown in black. \textbf{b} Relative lattice constant errors for solids in the test set. The train set of Si, Al, V, NaCl is indicated in gray. Fine-tuning improves agreement on average across the dataset, though some metals (e.g. Li, Na, and even V) show overcompensation.
    }
    \label{fig:xcfinetuning}
\end{figure*}
\noindent
The data-driven construction of exchange-correlation (XC) functionals
has a long history, ranging from early
semi-empirical fits of functionals to reference data
to more recently increasingly sophisticated machine learning~(ML) strategies%
~\cite{Mortensen2005,Snyder2012,Bogojeski2020,Li2021,Kirkpatrick2021,Kasim2021,Bystrom2022,Bystrom2024gaps,luise2025accurate,Dick2020neuralxc,Cuierrier2021,Polak2025realspaceML}.
Considering materials modeling,
ML approaches have considered fitting
to reference data
such as atomization energies~\cite{Mortensen2005,Wellendorff2012beefvdw,Wellendorff2014,Aldegunde2016},
lattice constants~\cite{Kovacs2022}
and band gaps~\cite{Borlido2020,Bystrom2024gaps}.
In most cases, these fitting procedures are indirect:
functional parameters are optimized
while keeping the electronic density fixed to
the density of a baseline functional such as PBE.
Moreover, equilibrium properties are often
approximated from energy-volume curve fitting around
fixed pre-relaxed structures.
These approximations are motivated by cost, but they obscure how parameter
changes propagate through the full simulation pipeline,
leading to possibly suboptimal fits.

Our differentiable DFT framework enables fully self-consistent, gradient-based
optimization of XC parameters against bulk observables, treating both the
electronic ground state and the relaxed geometry as differentiable functions of
the functional parameters $\theta$. For example, to fit lattice constants, we
define a loss function over a dataset of materials:
\begin{align}
    L_\textrm{xc}(\theta) = \frac{1}{N} \sum_{i=1}^N &\left(\frac{a^\star(x_i ,\theta) - a^\text{expt}_i}{a^\text{expt}_i}\right)^2,\\
    \text{where}\quad 
    a^\star(x_i ,\theta) &= \arg\min_{a}\; E(x_i,\theta,a),\\
    E(x_i,\theta,a) &= \min_P\; \mathcal{E}(x_i,\theta,a, P),
    \label{eq:latticeconstloss}
\end{align}
$a^\text{expt}_i$ is the experimental lattice constant
and $a^\star(x_i ,\theta)$ is obtained by
geometry optimization for material $x_i$ at XC parameters $\theta$. This setup involves two levels of implicit differentiation: one for the SCF solution and one for the geometry optimization. To handle the latter, we wrap the lattice optimization in a custom differentiation rule that applies implicit differentiation to the optimality condition. This mirrors how we imposed DFPT as the derivative of the SCF solution: internally, the chain rule propagates through stress evaluation, which triggers the DFPT response. The required implicit derivative of the equilibrium lattice constant with respect to functional parameters takes the form
\begin{align}
    \frac{\partial a^\star}{\partial\theta} = -\left(\frac{\partial^2 E}{\partial a^2}\right)^{-1} \left(\frac{\partial^2 E}{\partial\theta \partial a}\right)
    \label{eq:latticeconstantgradient}
\end{align}
evaluated at the converged structure. Using this expression, the full loss gradient $\frac{\partial L_\textrm{xc}}{\partial \theta}$ is composed automatically within our framework.

In~\cref{fig:xcfinetuning}, we start from the PBE functional and optimize
two of its parameters (see details in the Methods section)
on four lattice constants selected from the Sol58LC benchmark set~\cite{lundgaard_mbeef-vdw_2016}.
The fine-tuned functional reduces the root mean squared
relative error compared to several standard PBE variants, with all predictions
obtained through fully self-consistent calculations. Some metals exhibit
overcompensation, including vanadium even though it was included in training.
The optimizer has to balance competing regimes, reflecting the limited
flexibility of the chosen parametrization.
Such trade-offs are well established in the construction of semilocal
functionals, for instance between the accuracy of PBE for molecules and
that of PBEsol for solids~\cite{Perdew2008pbesol},
both generalized gradient approximations (GGA).
We remark that such trade-offs can be improved
by using more expressive forms
such as meta-GGAs~\cite{perdew2009revtpss,Kovacs2022,Lebeda2024lak},
which is however beyond the scope of this work.
Overall,
this example illustrates how AD
enables systematic, targeted exploration of functional refinements in a fully
self-consistent setting.

\newcommand{\secpseudo}{Property-driven pseudopotential optimization}
\subsection{\secpseudo}
\begin{figure*}[ht]
    \centering
    \includegraphics[width=\linewidth]{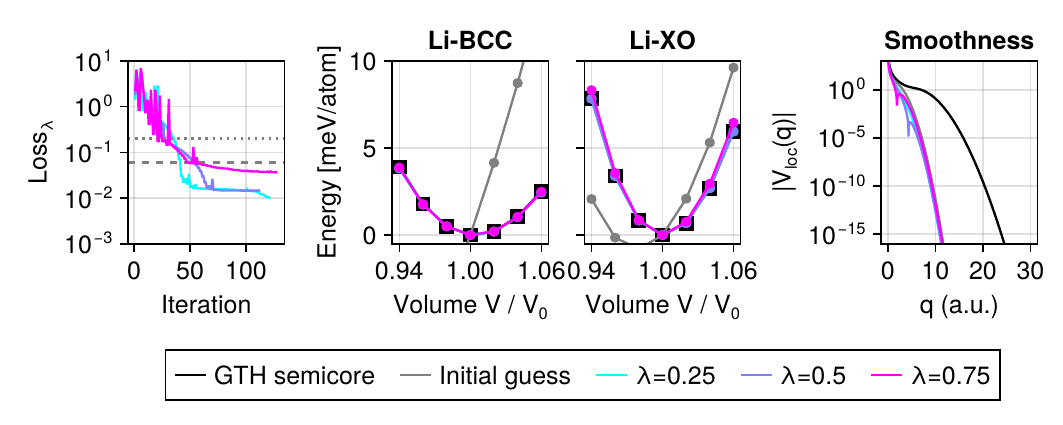}
    \caption{
        \textbf{\secpseudo.}
    A valence-only pseudopotential for $\text{Li}$ is trained against
    energy-volume curves of the more expensive semicore pseudopotential.
    The rightmost plot provides the decay
        of the Fourier components of the resulting local potential $V_\text{loc}$
        wrt.~$q$, the magnitude of the wave vector.
    The parameter $\lambda$ controls the relative weight of the unary and oxide loss
    terms as detailed in the Methods section.}
    \label{fig:psptuning}
\end{figure*}
\noindent
Pseudopotentials~(PSPs) are essential for the efficiency of plane-wave DFT,
yet they can introduce errors of similar magnitude to that
of the XC functional~\cite{Bosoni2023}.
Such pseudopotential errors can become especially large
if a PSP is employed in combination with a functional,
that differs from the functional used to fit the PSP
parameters in the first place~\cite{Mazdziarz2024incompatiblepseudo}.
Despite ongoing efforts to automate validation and
benchmarking~\cite{Lejaeghere2016,Prandini2018,Bosoni2023}, the PSP fitting process
itself remains largely manual or based on derivative-free
optimization~\cite{hamann_optimized_2013,SG15,VanSetten2018,SHOJAEI2023multiobjectiveONCV}.

Here, our AD-DFPT framework provides new opportunities
for pseudopotential generation.
We demonstrate these briefly using the example of fitting
a pseudopotential for lithium,
a lightweight element where the choice of
pseudopotential strongly affects smoothness
and transferability. Specifically, we train a valence-only one-electron
pseudopotential to reproduce the energy-volume curves of a more accurate,
three-electron semicore pseudopotential across two structurally distinct
compounds: elemental Li in the body-centered cubic (BCC) phase and LiO in a rocksalt structure.
Note that this semicore potential is only a stand-in
to avoid the additional complexity of performing an all-electron calculation.

\cref{fig:psptuning} summarizes the optimization result: we observe substantial
improvement in agreement with the reference, while preserving the smoothness of
the potential required for rapid convergence of the discretization. To quantify
agreement, we minimize a composite loss function combining normalized
energy-volume curve errors across both compounds. Since the loss depends
only on total energies, the required gradients with respect to pseudopotential
parameters are computed efficiently using the Hellmann-Feynman theorem.

Our AD-DFPT framework thus enables a gradient-based fitting
of PSP parameters directly employing bulk DFT observables in our loss function.

\newcommand{\secuq}{Propagating XC functional uncertainty}
\subsection{\secuq}\label{sec:forwarduq}
\noindent
In recent years a number of approaches have been developed
to provide statistical uncertainty estimates in the parameters of the XC functional.
Examples are the Bayesian error estimation functional (BEEF) family~\cite{Mortensen2005,Wellendorff2012beefvdw},
approaches based on Bayesian linear regression~\cite{Aldegunde2016}
or mixtures of established functionals~\cite{hansen_uncertainty-aware_2025}.
Provided such an error-aware DFT functional is chosen,
the built-in parameter uncertainty
estimate ought to be propagated to physical predictions,
such as equilibrium geometry and lattice constants.
For this purpose,
previous work relied on sampling-based ensemble propagation methods, combined
with additional approximations such as non-self-consistent calculations and
equation of state fitting%
~\cite{Mortensen2005,Aldegunde2016,hansen_uncertainty-aware_2025}.

\begin{figure}
    \centering
    \includegraphics[width=\linewidth]{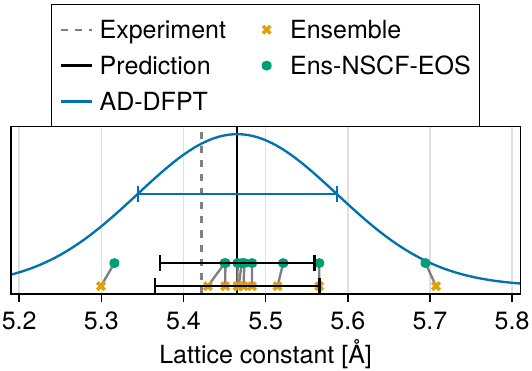}
    \caption{
        \textbf{\secuq.}
    Predictive uncertainty in the relaxed lattice constant of silicon
    obtained by propagating the parameter uncertainty of the BEEF
    functional~\cite{Mortensen2005} forward using three methods:
    full ensemble (10 geometry relaxations, orange),
    an ensemble based on non-self-consistent equation-of-state fits (Ens-NSCF-EOS, green),
    and a linearized analytic approximation
    using end-to-end differentiation and AD-DFPT (blue).
    The latter
    relies only on a single linearization around the mean parameters,
    avoiding the need for a choice of ensemble size or further approximations.}
    \label{fig:beef-latticeconst}
\end{figure}

The ability to compute end-to-end derivatives in our AD-DFPT framework
provides a new ingredient for such uncertainty propagation tasks,
namely to linearize entire computational workflows.
We demonstrate this in the blue curve in \cref{fig:beef-latticeconst},
which displays the distribution of lattice constants
resulting from propagating the uncertainty encoded in
the BEEF parameters forward through
both the SCF and the geometry optimization.

Specifically,
if $a^\star(\theta)$ denotes the optimal lattice constant depending
on the BEEF parameter value $\theta$,
a linearization around the mean parameter $\theta_0$ yields
\begin{align}
    a^\star(\theta) \approx a^\star(\theta_0) + J \cdot (\theta - \theta_0).
\end{align}
Here, $J = \frac{\partial a^\star}{\partial \theta}\vert_{\theta_0}$ is the same
total derivative as~\cref{eq:latticeconstantgradient}, readily computed
by applying our AD-DFPT framework.
Since the BEEF posterior for $\theta$ is modeled as a Gaussian distribution,
applying a linear pushforward approximation again yields an analytic Gaussian
$\mathcal{N}(a^\star(\theta_0), J\Sigma J^\top)$ for the uncertainty in $a^\star$,
where $\theta_0$ is the mean and $\Sigma$ the covariance of the BEEF posterior.

Notably, AD-based linearization
provides a viable alternative to ensemble propagation methods.
In these latter kinds of methods,
$N$ sets of XC parameters are sampled from the BEEF posterior
and each is considered independently.
A full ensemble therefore requires $N$ separate
geometry relaxations, which in practice is prohibitively costly.
To reduce cost,
previous works~\cite{Mortensen2005,Aldegunde2016,hansen_uncertainty-aware_2025} have often
employed an approximate procedure that we denote here
as Ens-NSCF-EOS. Instead of re-relaxing the structure for each sample,
Ens-NSCF-EOS performs an equation-of-state fit over seven fixed volumes around the
mean-parameter equilibrium, using non-self-consistent energy evaluations based
on the mean-parameter density and orbitals.

As shown in~\cref{fig:beef-latticeconst}, our AD-based linearization yields results
comparable to an ensemble of 10 independent geometry relaxations (orange), and
its Ens-NSCF-EOS approximation (green). Unsurprisingly, the error bars from
such small ensembles are not converged.
While agreement improves when increasing the
ensemble size to $30$ (not shown in the figure),
this indicates the challenge of selecting an ensemble size,
which gives good results, but remains computationally feasible.

In contrast, the AD-DFPT linearization for uncertainty propagation avoids tuning choices such as ensemble size as well as any further
approximations by equation-of-state fits or non-self-consistent evaluations. Instead, it requires only a single relaxation and derivative
evaluation,
and it extends mechanically to any quantity available in the
end-to-end differentiable workflow.

\newcommand{\secpwerror}{Estimation of the plane-wave basis error}
\subsection{\secpwerror}
\begin{figure*}
    \centering
    \includegraphics[width=\linewidth]{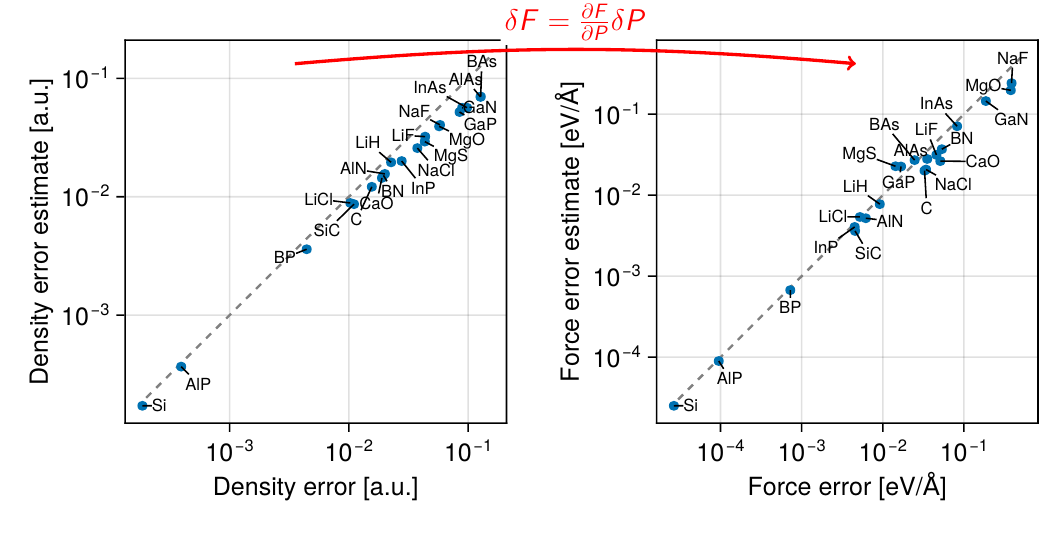}
    \caption{\textbf{\secpwerror}. Propagation of plane-wave discretization error estimates from the density to the forces using AD. The estimates, following the method from~\cite{Cances2022}, approximate the error in the density and forces due to the low energy cutoff $E_\text{cut}$ of $\SI{20}{\hartree}$ used for all solids. They correlate well with reference values. The density error metric is the integrated absolute density difference over the unit cell, normalized by the number of electrons. The force error metric is the largest force difference magnitude, i.e. $\max_{\text{atom }i} \Vert\Delta F_i\Vert_2$.}
    \label{fig:pw-error-estimate}
\end{figure*}
\noindent
Choosing the kinetic energy cutoff $E_\mathrm{cut}$ for plane-wave computations
remains a difficult tradeoff between accuracy and cost with
the ideal cutoff depending on the precise system,
the computed properties, and the desired accuracy.
Recommended cutoffs from pseudopotential
libraries cannot always reflect these nuances.
At the same time performing explicit convergence studies for each simulation
greatly increases the overall cost of the simulation.
Without a cheap and precise way to control the discretization error,
one often has no other option but to resort to running simulations
with an increased $E_\mathrm{cut}$ to ensure convergence.
Especially when performing massive data generation
over a large dataset of material structures this drives up the total computational cost
considerably~\cite{mazitov2025petmaduniversalinteratomicpotential}.

Yet, reliable error estimates have received increasing interest in the
mathematical community of DFT.
In particular, Cancès et al. have proposed a strategy~\cite{Cances2022} to
estimate the DFT plane-wave discretization error. This strategy computes an
error estimate that is specific to the system and property of interest, making
it a promising avenue.
Based on a standard SCF and the resulting initial density matrix $P$,
the approach results in a perturbative correction $\delta P$
which approximates the plane-wave discretization error in $P$.
Here, the key contribution of \cite{Cances2022} is an efficient algorithm,
such that the effect of increasing the basis
is approximately captured, but
without having to perform the full SCF in a larger basis.

Coupling this approach with our
differentiable DFT framework enables
the propagation of the error estimate $\delta P$ to quantities of interest such
as the interatomic forces $F$.
In fact, early versions of our AD framework were already used in~\cite{Cances2022}.
By linearization of the force computation $F(P)$,
the discretization error estimate in the forces is:
\begin{align}
    \label{eqn:forceprop}
    \delta F \approx \frac{\partial F}{\partial P} \delta P.
\end{align}
This quantity is readily computed from $P$ and $\delta P$ with a single forward-mode AD pass.
Notably, the perturbation $\delta P$ in this equation
is computed following~\cite{Cances2022},
which already involves solving some approximate Dyson equation.
Thus, no additional solution of \eqref{eq:Dyson} is required as part of the AD procedure as we only use AD to compute
$\frac{\partial F}{\partial P}$.

We illustrate this technique on a range of 21 insulating bulk solids from the
Sol58LC dataset, using a fixed plane-wave kinetic energy cutoff of
$E_\mathrm{cut} = \SI{20}{\hartree}$.
Of note, this value is lower than the recommended cutoff for most solids.
As such, we expect a large system-dependent discretization error in the converged
density and the derived forces,
which we would hope to capture with the error estimate procedure.
Indeed, as~\cref{fig:pw-error-estimate} summarizes,
we obtain an excellent correlation between the error estimate and the reference
error (obtained by comparing against a high-$E_\text{cut}$ computation).
In particular, not only is the
electronic density error accurately estimated (left),
but so is the error in the force (right),
computing using \eqref{eqn:forceprop}.

Hand-implementing unusual derivatives such as $\frac{\partial F}{\partial P}$ is
already a tedious and error-prone process for experienced code developers,
but can become an insurmountable obstacle
for practitioners
testing such error-estimation strategies.
Our differentiable plane-wave DFT code
seamlessly provides such derivatives
and in this way furthers the development of error estimation strategies:
by computing appropriate derivatives
propagating the density error estimates
to other quantities of interest
becomes readily feasible.

\section{Discussion}
\noindent
In this article we presented the AD-DFPT framework for DFT gradient computation,
an accurate and automated approach
to compute end-to-end derivatives across the entire DFT workflow.
By integrating algorithmic differentiation~(AD) techniques
with classic density-functional perturbation theory~(DFPT),
it shares the favorable robustness and accuracy properties of
established DFPT-based methods,
while extending their reach to arbitrary input parameters and postprocessing quantities.
In particular it avoids common pitfalls of finite-difference techniques,
such as the complex interplay between the optimal step size
and numerical details such as plane-wave cutoff or SCF tolerance.
This combination  opens the door
to high-precision gradient-based DFT workflows
even for researchers who are not DFT implementation experts.

We demonstrated multiple emerging research opportunities,
covering a wide range of tasks, such as inverse materials design,
uncertainty quantification of DFT simulations,
or the learning of improved functionals or pseudopotentials.
Importantly, with an AD-DFPT implementation at
hand we were able to push the state of the art
and directly
target relaxed self-consistent material properties as
reference data,
instead of relying on surrogate losses or fixed-density approximations.
Given the increasing interest in integrating machine learning approaches
directly within the formulation of DFT models,
our developments provide an important and timely foundation
to support such efforts.

While AD provides a systematic route to differentiation,
its application does not eliminate the inherent mathematical and
numerical challenges of a physical model as complex as DFT.
We emphasize this point on three practical issues.
First, for specific expressions the
standard derivative rules of the AD system
do not always yield a numerically stable derivative.
Such cases can be solved
by providing a custom rule with a hand-coded derivative implementation, or
preferably by modifying the initial undifferentiated code to enhance stability;
see the example of the Fermi-Dirac function in the Supplementary Information.
Second, some DFT quantities are not always differentiable.
An example is the band gap, since its definition involves minima and maxima
over eigenvalues which are non-differentiable at crossings.
In such cases, both numerical and algorithmic differentiation techniques
will still compute \textit{some value} for the derivative without
an indication that the numerical result may be unreliable.
In this regard,
various solutions (e.g.~appropriate smoothening
of the differentiated function~\cite[Chapter~5]{blondel_elements_2024})
have been suggested in the AD community~\cite[Chapter~14]{Griewank2008EDP},
which should be explored in the context of AD-DFPT
in future work.
Third, the crystal symmetry analysis~\cite{togo_spglib_2024}
used by DFT codes to reduce the effective number of $k$-points
needs to be adapted to additionally consider
crystal perturbations.
Indeed, geometric perturbations of a crystal typically break some of its symmetries,
which must then be filtered out;
see an illustration in Supplementary Figure S1.
An automated algorithm is work in progress.

The current implementation of AD-DFPT in DFTK
fully supports forward-mode propagation of general
parameter derivatives through
DFT computations involving norm-conserving pseudopotentials,
generalized gradient approximation (GGA) functionals
and no spin polarization.
While considering other models or spin-polarized systems
requires additional implementation effort,
such extensions integrate well into the presented AD-DFPT framework.
In particular, automatically generating the derivatives of the setup and
postprocess stages in~\cref{fig:flowchart} is not limited to our current
setting and generalizes straightforwardly to other functionals and models.
Going beyond GGA functionals, for example, is primarily limited
by the capabilities of the underlying DFPT solver.
Developing a reliable and efficient DFPT solver capable of treating meta-GGA
or hybrid-DFT functionals, remains a considerable technical challenge and poses
an important avenue for future research
to extend the applicability of AD-DFPT methods.
Finally, our forward-mode implementation prepares the ground
towards more challenging reverse-mode AD techniques,
enabling as an outlook
the efficient optimization in
high-dimensional parameter spaces, for example when training deep-learning
exchange-correlation functionals.

Looking ahead to the nested workflows common in materials simulation, packages for higher-level tasks such as geometry optimization or molecular dynamics could provide their own differentiation rules, relying on the underlying DFT code to compute parameter derivatives transparently. This would achieve a clear separation of concerns and make differentiation of such workflows naturally composable.

In summary,
AD-DFPT brings plane-wave DFT in line with the general progress towards
differentiable programming in scientific computing.
As exemplified the resulting end-to-end differentiable
workflows have the potential to significantly enhance
materials design, error control,
and systematic learning of model parameters in computational materials science.

\section{Methods}
\noindent
In the following we provide details on the implementation of
the AD-DFPT framework in DFTK,
followed by an outline of the computational setup of each example.

\subsection{\secorbitals}
\noindent
The density-matrix-based formulation
of Equation~\eqref{eq:Dyson} is convenient
to illustrate the mathematical structure of DFPT.
However,
manipulating objects such as dense Hamiltonians or density matrices
(with $\mathcal{O}(N_b^2)$ storage cost) is prohibitive
for basis sets with a large number $N_b$ of functions, such as plane-waves.
In our computations we therefore follow the standard approach
to employ sparse representations in terms of Kohn-Sham orbitals
as well as iterative techniques for solving response problems.
We sketch these briefly in this section.

Consider the non-differentiated form of DFPT,
that is the SCF problem \eqref{eq:fixedpoint}.
To avoid the $\mathcal{O}(N_b^2)$ storage cost of Hamiltonians or density matrices
when solving the Kohn-Sham equations~\eqref{eq:scforbitals}
one employs iterative methods.
The density matrix is represented
implicitly as a truncated set of $N$ partially
occupied Kohn-Sham orbitals $\psi_n$ and occupations $f_n$ with $N \ll N_b$,
thus in $\mathcal{O}(N_b)$ storage.
In turn the Hamiltonian is represented as a sum of three sparse terms
with again $\mathcal{O}(N_b)$ storage:
a diagonal kinetic term,
the low-rank nonlocal pseudopotential projectors,
and real-space local potentials applied via FFT-based convolutions~\cite{kresse_efficient_1996}.

Taking the derivative of such structured representations %
retains this structure. Therefore sparse $\mathcal{O}(N_b)$ representations
for the corresponding perturbations $\delta P$ and $\delta H$
can similarly be constructed and carried through all
stages of the DFT workflow of \cref{fig:flowchart}.
More precisely any Hamiltonian perturbation $\delta H$ can be
expressed through perturbations of its sparse terms and
any admissible (i.e.~representable in the $N$-truncated subspace)
density matrix perturbation $\delta P$ can in turn be parametrized
by $N$ pairs $(\delta\psi_n, \delta f_n)$
of perturbations to its orbitals and occupations
\begin{align}
    \delta P = \sum_{n=1}^N \delta f_n \psi_n\psi_n^\dagger + f_n \left(\delta\psi_n\psi_n^\dagger + \psi_n\delta\psi_n^\dagger\right).
\end{align}
We remark that the choice of $\delta\psi_n$ and $\delta f_n$
to represent $\delta P$ is not unique, which in DFTK
is fixed according to the \textit{minimal gauge} described in~\cite{Cances2023}.

Based on this orbital-based representation for $\delta P$
the response in~\cref{eq:Dyson} is solved iteratively
using the inexact generalized minimal residual (GMRES) algorithm of~\cite{herbst2025efficientkrylovmethodslinear},
which only requires matrix-vector products of $(1 - \chi_0 K)$ with trial vectors.
For the latter each application of $\chi_0$ to a trial Hamiltonian perturbation
$\delta H$ computes the orbital responses $\delta \psi_n$ as a sum of two ingredients:
a sum-over-states formula
for the occupied contributions
and the iterative solution of the
Sternheimer equations for the unoccupied contributions,
using a tailored
preconditioned conjugate gradient algorithm~\cite{Cances2023}.
This reflects a broader design principle: iterative algorithms for
linearized response problems parallel those used in the underlying nonlinear problem.
In our case, the nested iterative structure of Anderson-accelerated SCF iteration with
an inner LOBPCG (locally optimal block preconditioned conjugate gradient~\cite{knyazev2001lobpcg}) eigensolver
carries over to a GMRES iteration (the linear analogue of Anderson acceleration~\cite{walker2011anderson_gmres})
with an inner conjugate gradient algorithm for the Sternheimer equations.

\subsection{Integration of forward-mode AD within DFTK}
\noindent
For our AD-DFPT framework in DFTK we employ
the Julia forward-mode AD package ForwardDiff~\cite{Revels2016forwarddiff},
which provides AD capabilities using a technique known as operator-overloading AD.
In a nutshell ForwardDiff defines a new number type, the dual number.
When used instead of a standard floating-point number,
it changes a function's behavior by computing simultaneously its value and derivatives, thus in line with executing Figure~\ref{fig:flowchart} from left to right in both rows simultaneously.
For standard primitive functions%
~(e.g.~multiplication, addition, trigonometric functions, LAPACK-based linear algebra~\cite{laug}),
ForwardDiff provides differentiation rules based on such dual numbers
as well as appropriate chain rule expressions to compose derivative results.

Relying on Julia's multiple dispatch mechanism, DFTK is generic in the
employed floating-point type, which includes dual numbers.
The largest part of the code base is thus made differentiable simply by switching
the floating-point type from normal floating-point numbers to
dual numbers when the gradient
of a DFT quantity should be computed.
The only exceptions are cases where a custom rule should be employed for
differentiation instead of a decomposition down to standard primitives.

In ForwardDiff such custom rules
can be provided by overloading a function with a method that is specialized
to arguments of the dual number type.
On top of our standard \texttt{self\_consistent\_field} function to execute
the first row of the solve stage of~\cref{fig:flowchart} we therefore also provide
a special \texttt{self\_consistent\_field} function for dual numbers in DFTK.
Whenever ForwardDiff attempts to compute the derivative of the solve stage
then this function in DFTK will be called,
enabling us to extract $\frac{\partial H}{\partial \theta}$ from the AD system,
solve the Dyson equation \eqref{eq:Dyson} using our DFPT solver
and re-inject the solution $\frac{\partial P}{\partial \theta}$ back to ForwardDiff.

\subsection{\secelasticity}
\noindent
For all calculations we employ the PBE functional~\cite{Perdew1996}, PseudoDojo pseudopotentials~\cite{VanSetten2018},
Gaussian smearing of $10^{-3}$ $\text{Ha}$ and a $8\times8\times 8$ $k$-mesh.
The plane-wave cutoff was chosen following the \textit{normal} recommendations~\cite{VanSetten2018}.
Crystal structures for diamond, silicon (diamond structure), and caesium chloride are generated from the Atomic Simulation Environment~\cite{ase-paper},
and are relaxed before computing elastic constants.
For the computation of the elasticity tensor only a single strain pattern of ${\eta = (1,0,0,1,0,0)}$ has been used,
which recovers all $(C_{11}, C_{12}, C_{12}, C_{44}, 0, 0)$ in our cubic crystals.
Any crystal symmetries of the unstrained crystal, which would be broken by this perturbation,
are removed
during all computations.
For the finite-difference computations we employ a central formula.
The SCF tolerance corresponds to the $L_2$ error in the density.

\subsection{\secinv}
\noindent
The fully self-contained code for the inverse design example is included in~\cref{fig:inversedesigngaas}.

\subsection{\secxc}
\noindent
In this example we
optimize the two free parameters $\kappa$ and $\mu$ in the PBE
exchange enhancement factor~\cite{Perdew1996}
\begin{align}
    F_X(s) = 1 + \kappa - \kappa / (1 + \mu s^2)
\end{align}
where $s$ is the reduced density gradient.
The reference experimental lattice constants including zero-point corrections
are taken from the Sol58LC data set~\cite{lundgaard_mbeef-vdw_2016}.
The outermost parameter optimization loop to minimize~\cref{eq:latticeconstloss}
employs BFGS as implemented in~\cite{mogensen2018optim} with a backtracking linesearch
and implicit parameter gradients derived according to~\cref{eq:latticeconstantgradient}.
DFT calculations use a Gaussian smearing of $10^{-3}$ Ha and
PseudoDojo pseudopotentials~\cite{VanSetten2018}.
We follow the \textit{normal} recommendations~\cite{VanSetten2018}
for the plane-wave basis cutoffs, giving a range from $18$ to $49$ Ha.
Additionally, kinetic energy cutoff smearing is used~\cite{Cances2023modif} to
enforce a smooth lattice relaxation.
Note that the $k$-grid, ensuring a maximal $k$-point spacing of
$0.15\,\text{\AA}^{-1}$, is determined once for each solid at the initial step of the
optimization and held fixed afterwards.

\subsection{\secpseudo}
\noindent
We optimize six selected parameters of a valence-only Li pseudopotential in the Goedecker-Teter-Hutter (GTH) parametrization~\cite{Goedecker1996}.
The initial guess is the ``LDA'' version, while the semicore reference is from the PBE table~\cite{krack_pseudopotentials_2005,Kuhne2020cp2k}.
Reference structures of Li-BCC and LiO rocksalt (``Li-XO``) are taken from~\cite{Bosoni2023}, corresponding to the relaxed all-electron PBE structures. For each compound, we generate energy-volume curves using seven uniformly spaced volumes in the range $\pm 6\%$ around the reference.
We impose a normalized energy-volume loss function per compound, adapted from the recent $\varepsilon$-metric~\cite{Bosoni2023}:
\begin{align}
    \varepsilon = \sqrt{\frac{\sum_i (E_i^a - E_i^b)^2}{\sqrt{\sum_i (E_i^a -  \overline{E^a})^2 \sum_i (E_i^b -  \overline{E^b})^2}}}.
\end{align}
Here, $E^a_i=E^a(V_i)$ and $E^b_i=E_b(V_i)$ are total energy differences relative to the energy at the reference volume $V_0$, at volume $V_i$ from models $a$ and $b$, respectively, and $\overline{E^a},\overline{E^b}$ are their volume-averaged energies. The overall training loss is defined as a weighted sum over the two compounds as
$L_\text{pseudo}(\theta) = \lambda \varepsilon_\text{Li-BCC}(\theta) + (1-\lambda) \varepsilon_\text{LiO}(\theta)$,
where $\theta$ now denote the pseudopotential parameters and $\lambda$ controls the balance between the two training structures.
The outermost parameter optimization loop uses BFGS as implemented
in~\cite{mogensen2018optim} with a backtracking linesearch
and total energy parameter gradients computed using the Hellmann-Feynman theorem.
DFT calculations use Gaussian smearing of $0.00225$ Ha with a moderate $8\times8\times8$ $k$-mesh consistently for both reference and prediction, comparable to recent pseudopotential benchmarks~\cite{garrity_pseudopotentials_2014,VanSetten2018,Prandini2018}.
The plane-wave basis cutoff used is 120 Ha for semicore Li and O, and 20 Ha for valence-only Li.

\subsection{\secuq}
\noindent
Numerical parameters are identical to the example \textit{\secxc}, except for the
choice of DFT functional. Here, we employ the BEEF exchange-correlation
functional with the parameters reported in~\cite{Mortensen2005}. The silicon
geometry used has been optimized tightly using the mean XC parameters before
considering the uncertainty propagation. For the Ens-NSCF-EOS ensemble,
total energies are recomputed on a fixed grid of seven volumes symmetrically
spaced within $\pm 6\%$ around the equilibrium volume, using the mean-parameter
density and orbitals. The equilibrium lattice constant for each sampled
parameter is then extracted by fitting a Birch-Murnaghan equation of state. For
the AD-based linear pushforward, the derivative of the relaxed lattice constant
with respect to the XC parameters is computed via implicit differentiation,
using a single geometry optimization and derivative computation
at the mean parameters.

\subsection{\secpwerror}
\noindent
We use the PBE functional~\cite{Perdew1996}, the PseudoDojo pseudopotentials, a
uniform plane-wave cutoff of $\SI{20}{\hartree}$ and a minimal $k$-spacing of
\SI{0.15}{\per\angstrom}. For each system, the first atom is displaced compared
to the equilibrium structure such that the largest interatomic force magnitude
is around \SI{0.5}{\eV\per\angstrom}. The error estimates are computed
following~\cite{Cances2022} using a reference cutoff
$E_\text{cut,ref} = 1.5\times$ the \textit{high}
recommendation for the pseudopotentials~\cite{VanSetten2018}.
The reference error is obtained by
comparison against an expensive reference SCF computation for each system,
using $E_\text{cut,ref}$ as the plane-wave cutoff and otherwise the same parameters.

\section{Data availability}
\noindent
All data necessary to reproduce the experiments and plots are included in the code repository.

\section{Code availability}
\noindent
Our AD-DFPT framework is directly included in DFTK v0.7.16 at \url{https://dftk.org}. An archived copy is available at \cite{herbst_2025_15793538}.
All code to reproduce calculations and plots is included in the public repository at \url{https://github.com/niklasschmitz/ad-dfpt}, archived at~\cite{schmitz_2025_17084313}.

\bibliography{literature}

\begin{thebibliography}{111}%
\makeatletter
\providecommand \@ifxundefined [1]{%
 \@ifx{#1\undefined}
}%
\providecommand \@ifnum [1]{%
 \ifnum #1\expandafter \@firstoftwo
 \else \expandafter \@secondoftwo
 \fi
}%
\providecommand \@ifx [1]{%
 \ifx #1\expandafter \@firstoftwo
 \else \expandafter \@secondoftwo
 \fi
}%
\providecommand \natexlab [1]{#1}%
\providecommand \enquote  [1]{``#1''}%
\providecommand \bibnamefont  [1]{#1}%
\providecommand \bibfnamefont [1]{#1}%
\providecommand \citenamefont [1]{#1}%
\providecommand \href@noop [0]{\@secondoftwo}%
\providecommand \href [0]{\begingroup \@sanitize@url \@href}%
\providecommand \@href[1]{\@@startlink{#1}\@@href}%
\providecommand \@@href[1]{\endgroup#1\@@endlink}%
\providecommand \@sanitize@url [0]{\catcode `\\12\catcode `\$12\catcode
  `\&12\catcode `\#12\catcode `\^12\catcode `\_12\catcode `\%12\relax}%
\providecommand \@@startlink[1]{}%
\providecommand \@@endlink[0]{}%
\providecommand \url  [0]{\begingroup\@sanitize@url \@url }%
\providecommand \@url [1]{\endgroup\@href {#1}{\urlprefix }}%
\providecommand \urlprefix  [0]{URL }%
\providecommand \Eprint [0]{\href }%
\providecommand \doibase [0]{https://doi.org/}%
\providecommand \selectlanguage [0]{\@gobble}%
\providecommand \bibinfo  [0]{\@secondoftwo}%
\providecommand \bibfield  [0]{\@secondoftwo}%
\providecommand \translation [1]{[#1]}%
\providecommand \BibitemOpen [0]{}%
\providecommand \bibitemStop [0]{}%
\providecommand \bibitemNoStop [0]{.\EOS\space}%
\providecommand \EOS [0]{\spacefactor3000\relax}%
\providecommand \BibitemShut  [1]{\csname bibitem#1\endcsname}%
\let\auto@bib@innerbib\@empty
\bibitem [{\citenamefont {Hohenberg}\ and\ \citenamefont
  {Kohn}(1964)}]{HohenbergKohn1964}%
  \BibitemOpen
  \bibfield  {author} {\bibinfo {author} {\bibfnamefont {P.}~\bibnamefont
  {Hohenberg}}\ and\ \bibinfo {author} {\bibfnamefont {W.}~\bibnamefont
  {Kohn}},\ }\bibfield  {title} {\bibinfo {title} {Inhomogeneous electron
  gas},\ }\href {https://doi.org/10.1103/PhysRev.136.B864} {\bibfield
  {journal} {\bibinfo  {journal} {Phys. Rev.}\ }\textbf {\bibinfo {volume}
  {136}},\ \bibinfo {pages} {B864} (\bibinfo {year} {1964})}\BibitemShut
  {NoStop}%
\bibitem [{\citenamefont {Kohn}\ and\ \citenamefont
  {Sham}(1965)}]{KohnSham1965}%
  \BibitemOpen
  \bibfield  {author} {\bibinfo {author} {\bibfnamefont {W.}~\bibnamefont
  {Kohn}}\ and\ \bibinfo {author} {\bibfnamefont {L.~J.}\ \bibnamefont
  {Sham}},\ }\bibfield  {title} {\bibinfo {title} {Self-consistent equations
  including exchange and correlation effects},\ }\href
  {https://doi.org/10.1103/PhysRev.140.A1133} {\bibfield  {journal} {\bibinfo
  {journal} {Phys. Rev.}\ }\textbf {\bibinfo {volume} {140}},\ \bibinfo {pages}
  {A1133} (\bibinfo {year} {1965})}\BibitemShut {NoStop}%
\bibitem [{\citenamefont {Baroni}\ \emph {et~al.}(2001)\citenamefont {Baroni},
  \citenamefont {de~Gironcoli}, \citenamefont {Dal~Corso},\ and\ \citenamefont
  {Giannozzi}}]{Baroni2001}%
  \BibitemOpen
  \bibfield  {author} {\bibinfo {author} {\bibfnamefont {S.}~\bibnamefont
  {Baroni}}, \bibinfo {author} {\bibfnamefont {S.}~\bibnamefont
  {de~Gironcoli}}, \bibinfo {author} {\bibfnamefont {A.}~\bibnamefont
  {Dal~Corso}},\ and\ \bibinfo {author} {\bibfnamefont {P.}~\bibnamefont
  {Giannozzi}},\ }\bibfield  {title} {\bibinfo {title} {Phonons and related
  crystal properties from density-functional perturbation theory},\ }\href
  {https://doi.org/10.1103/revmodphys.73.515} {\bibfield  {journal} {\bibinfo
  {journal} {Reviews of Modern Physics}\ }\textbf {\bibinfo {volume} {73}},\
  \bibinfo {pages} {515} (\bibinfo {year} {2001})}\BibitemShut {NoStop}%
\bibitem [{\citenamefont {Gonze}\ and\ \citenamefont
  {Lee}(1997)}]{Gonze1997dynamicalmatrices}%
  \BibitemOpen
  \bibfield  {author} {\bibinfo {author} {\bibfnamefont {X.}~\bibnamefont
  {Gonze}}\ and\ \bibinfo {author} {\bibfnamefont {C.}~\bibnamefont {Lee}},\
  }\bibfield  {title} {\bibinfo {title} {Dynamical matrices, born effective
  charges, dielectric permittivity tensors, and interatomic force constants
  from density-functional perturbation theory},\ }\href
  {https://doi.org/10.1103/PhysRevB.55.10355} {\bibfield  {journal} {\bibinfo
  {journal} {Phys. Rev. B}\ }\textbf {\bibinfo {volume} {55}},\ \bibinfo
  {pages} {10355} (\bibinfo {year} {1997})}\BibitemShut {NoStop}%
\bibitem [{\citenamefont {Baroni}\ \emph {et~al.}(1987)\citenamefont {Baroni},
  \citenamefont {Giannozzi},\ and\ \citenamefont {Testa}}]{Baroni1987elastic}%
  \BibitemOpen
  \bibfield  {author} {\bibinfo {author} {\bibfnamefont {S.}~\bibnamefont
  {Baroni}}, \bibinfo {author} {\bibfnamefont {P.}~\bibnamefont {Giannozzi}},\
  and\ \bibinfo {author} {\bibfnamefont {A.}~\bibnamefont {Testa}},\ }\bibfield
   {title} {\bibinfo {title} {Elastic constants of crystals from
  linear-response theory},\ }\href
  {https://doi.org/10.1103/PhysRevLett.59.2662} {\bibfield  {journal} {\bibinfo
   {journal} {Phys. Rev. Lett.}\ }\textbf {\bibinfo {volume} {59}},\ \bibinfo
  {pages} {2662} (\bibinfo {year} {1987})}\BibitemShut {NoStop}%
\bibitem [{\citenamefont {Gonze}(1997)}]{Gonze1997conjugategradientdfpt}%
  \BibitemOpen
  \bibfield  {author} {\bibinfo {author} {\bibfnamefont {X.}~\bibnamefont
  {Gonze}},\ }\bibfield  {title} {\bibinfo {title} {First-principles responses
  of solids to atomic displacements and homogeneous electric fields:
  Implementation of a conjugate-gradient algorithm},\ }\href
  {https://doi.org/10.1103/PhysRevB.55.10337} {\bibfield  {journal} {\bibinfo
  {journal} {Phys. Rev. B}\ }\textbf {\bibinfo {volume} {55}},\ \bibinfo
  {pages} {10337} (\bibinfo {year} {1997})}\BibitemShut {NoStop}%
\bibitem [{\citenamefont {Hamann}\ \emph {et~al.}(2005)\citenamefont {Hamann},
  \citenamefont {Wu}, \citenamefont {Rabe},\ and\ \citenamefont
  {Vanderbilt}}]{Hamann2005metrictensor}%
  \BibitemOpen
  \bibfield  {author} {\bibinfo {author} {\bibfnamefont {D.~R.}\ \bibnamefont
  {Hamann}}, \bibinfo {author} {\bibfnamefont {X.}~\bibnamefont {Wu}}, \bibinfo
  {author} {\bibfnamefont {K.~M.}\ \bibnamefont {Rabe}},\ and\ \bibinfo
  {author} {\bibfnamefont {D.}~\bibnamefont {Vanderbilt}},\ }\bibfield  {title}
  {\bibinfo {title} {Metric tensor formulation of strain in density-functional
  perturbation theory},\ }\href {https://doi.org/10.1103/PhysRevB.71.035117}
  {\bibfield  {journal} {\bibinfo  {journal} {Phys. Rev. B}\ }\textbf {\bibinfo
  {volume} {71}},\ \bibinfo {pages} {035117} (\bibinfo {year}
  {2005})}\BibitemShut {NoStop}%
\bibitem [{\citenamefont {Wu}\ \emph {et~al.}(2005)\citenamefont {Wu},
  \citenamefont {Vanderbilt},\ and\ \citenamefont
  {Hamann}}]{Wu2005systematicdfpt}%
  \BibitemOpen
  \bibfield  {author} {\bibinfo {author} {\bibfnamefont {X.}~\bibnamefont
  {Wu}}, \bibinfo {author} {\bibfnamefont {D.}~\bibnamefont {Vanderbilt}},\
  and\ \bibinfo {author} {\bibfnamefont {D.~R.}\ \bibnamefont {Hamann}},\
  }\bibfield  {title} {\bibinfo {title} {Systematic treatment of displacements,
  strains, and electric fields in density-functional perturbation theory},\
  }\href {https://doi.org/10.1103/PhysRevB.72.035105} {\bibfield  {journal}
  {\bibinfo  {journal} {Phys. Rev. B}\ }\textbf {\bibinfo {volume} {72}},\
  \bibinfo {pages} {035105} (\bibinfo {year} {2005})}\BibitemShut {NoStop}%
\bibitem [{\citenamefont {Gajdo\ifmmode~\check{s}\else \v{s}\fi{}}\ \emph
  {et~al.}(2006)\citenamefont {Gajdo\ifmmode~\check{s}\else \v{s}\fi{}},
  \citenamefont {Hummer}, \citenamefont {Kresse}, \citenamefont
  {Furthm\"uller},\ and\ \citenamefont {Bechstedt}}]{gajdos2006vaspdfpt}%
  \BibitemOpen
  \bibfield  {author} {\bibinfo {author} {\bibfnamefont {M.}~\bibnamefont
  {Gajdo\ifmmode~\check{s}\else \v{s}\fi{}}}, \bibinfo {author} {\bibfnamefont
  {K.}~\bibnamefont {Hummer}}, \bibinfo {author} {\bibfnamefont
  {G.}~\bibnamefont {Kresse}}, \bibinfo {author} {\bibfnamefont
  {J.}~\bibnamefont {Furthm\"uller}},\ and\ \bibinfo {author} {\bibfnamefont
  {F.}~\bibnamefont {Bechstedt}},\ }\bibfield  {title} {\bibinfo {title}
  {Linear optical properties in the projector-augmented wave methodology},\
  }\href {https://doi.org/10.1103/PhysRevB.73.045112} {\bibfield  {journal}
  {\bibinfo  {journal} {Phys. Rev. B}\ }\textbf {\bibinfo {volume} {73}},\
  \bibinfo {pages} {045112} (\bibinfo {year} {2006})}\BibitemShut {NoStop}%
\bibitem [{\citenamefont {Laflamme~Janssen}\ \emph {et~al.}(2016)\citenamefont
  {Laflamme~Janssen}, \citenamefont {Gillet}, \citenamefont {Ponc\'e},
  \citenamefont {Martin}, \citenamefont {Torrent},\ and\ \citenamefont
  {Gonze}}]{LaflammeJanssen2016precise}%
  \BibitemOpen
  \bibfield  {author} {\bibinfo {author} {\bibfnamefont {J.}~\bibnamefont
  {Laflamme~Janssen}}, \bibinfo {author} {\bibfnamefont {Y.}~\bibnamefont
  {Gillet}}, \bibinfo {author} {\bibfnamefont {S.}~\bibnamefont {Ponc\'e}},
  \bibinfo {author} {\bibfnamefont {A.}~\bibnamefont {Martin}}, \bibinfo
  {author} {\bibfnamefont {M.}~\bibnamefont {Torrent}},\ and\ \bibinfo {author}
  {\bibfnamefont {X.}~\bibnamefont {Gonze}},\ }\bibfield  {title} {\bibinfo
  {title} {Precise effective masses from density functional perturbation
  theory},\ }\href {https://doi.org/10.1103/PhysRevB.93.205147} {\bibfield
  {journal} {\bibinfo  {journal} {Phys. Rev. B}\ }\textbf {\bibinfo {volume}
  {93}},\ \bibinfo {pages} {205147} (\bibinfo {year} {2016})}\BibitemShut
  {NoStop}%
\bibitem [{\citenamefont {Canc\`{e}s}\ \emph
  {et~al.}(2023{\natexlab{a}})\citenamefont {Canc\`{e}s}, \citenamefont
  {Herbst}, \citenamefont {Kemlin}, \citenamefont {Levitt},\ and\ \citenamefont
  {Stamm}}]{Cances2023}%
  \BibitemOpen
  \bibfield  {author} {\bibinfo {author} {\bibfnamefont {E.}~\bibnamefont
  {Canc\`{e}s}}, \bibinfo {author} {\bibfnamefont {M.~F.}\ \bibnamefont
  {Herbst}}, \bibinfo {author} {\bibfnamefont {G.}~\bibnamefont {Kemlin}},
  \bibinfo {author} {\bibfnamefont {A.}~\bibnamefont {Levitt}},\ and\ \bibinfo
  {author} {\bibfnamefont {B.}~\bibnamefont {Stamm}},\ }\bibfield  {title}
  {\bibinfo {title} {Numerical stability and efficiency of response property
  calculations in density functional theory},\ }\href
  {https://doi.org/10.1007/s11005-023-01645-3} {\bibfield  {journal} {\bibinfo
  {journal} {Letters in Mathematical Physics}\ }\textbf {\bibinfo {volume}
  {113}},\ \bibinfo {pages} {21} (\bibinfo {year}
  {2023}{\natexlab{a}})}\BibitemShut {NoStop}%
\bibitem [{\citenamefont {Gonze}\ \emph {et~al.}(2024)\citenamefont {Gonze},
  \citenamefont {Rostami},\ and\ \citenamefont
  {Tantardini}}]{gonze_variational_2024}%
  \BibitemOpen
  \bibfield  {author} {\bibinfo {author} {\bibfnamefont {X.}~\bibnamefont
  {Gonze}}, \bibinfo {author} {\bibfnamefont {S.}~\bibnamefont {Rostami}},\
  and\ \bibinfo {author} {\bibfnamefont {C.}~\bibnamefont {Tantardini}},\
  }\bibfield  {title} {\bibinfo {title} {Variational {Density} {Functional}
  {Perturbation} {Theory} for {Metals}},\ }\href
  {https://doi.org/10.1103/PhysRevB.109.014317} {\bibfield  {journal} {\bibinfo
   {journal} {Phys. Rev. B}\ }\textbf {\bibinfo {volume} {109}},\ \bibinfo
  {pages} {014317} (\bibinfo {year} {2024})}\BibitemShut {NoStop}%
\bibitem [{\citenamefont {Giannozzi}\ \emph {et~al.}(2020)\citenamefont
  {Giannozzi}, \citenamefont {Baseggio}, \citenamefont {Bonf\`{a}},
  \citenamefont {Brunato}, \citenamefont {Car}, \citenamefont {Carnimeo},
  \citenamefont {Cavazzoni}, \citenamefont {de~Gironcoli}, \citenamefont
  {Delugas}, \citenamefont {Ferrari~Ruffino}, \citenamefont {Ferretti},
  \citenamefont {Marzari}, \citenamefont {Timrov}, \citenamefont {Urru},\ and\
  \citenamefont {Baroni}}]{QE}%
  \BibitemOpen
  \bibfield  {author} {\bibinfo {author} {\bibfnamefont {P.}~\bibnamefont
  {Giannozzi}}, \bibinfo {author} {\bibfnamefont {O.}~\bibnamefont {Baseggio}},
  \bibinfo {author} {\bibfnamefont {P.}~\bibnamefont {Bonf\`{a}}}, \bibinfo
  {author} {\bibfnamefont {D.}~\bibnamefont {Brunato}}, \bibinfo {author}
  {\bibfnamefont {R.}~\bibnamefont {Car}}, \bibinfo {author} {\bibfnamefont
  {I.}~\bibnamefont {Carnimeo}}, \bibinfo {author} {\bibfnamefont
  {C.}~\bibnamefont {Cavazzoni}}, \bibinfo {author} {\bibfnamefont
  {S.}~\bibnamefont {de~Gironcoli}}, \bibinfo {author} {\bibfnamefont
  {P.}~\bibnamefont {Delugas}}, \bibinfo {author} {\bibfnamefont
  {F.}~\bibnamefont {Ferrari~Ruffino}}, \bibinfo {author} {\bibfnamefont
  {A.}~\bibnamefont {Ferretti}}, \bibinfo {author} {\bibfnamefont
  {N.}~\bibnamefont {Marzari}}, \bibinfo {author} {\bibfnamefont
  {I.}~\bibnamefont {Timrov}}, \bibinfo {author} {\bibfnamefont
  {A.}~\bibnamefont {Urru}},\ and\ \bibinfo {author} {\bibfnamefont
  {S.}~\bibnamefont {Baroni}},\ }\bibfield  {title} {\bibinfo {title}
  {{Quantum} {ESPRESSO} toward the exascale},\ }\href
  {https://doi.org/10.1063/5.0005082} {\bibfield  {journal} {\bibinfo
  {journal} {The Journal of Chemical Physics}\ }\textbf {\bibinfo {volume}
  {152}},\ \bibinfo {pages} {154105} (\bibinfo {year} {2020})}\BibitemShut
  {NoStop}%
\bibitem [{\citenamefont {Romero}\ \emph {et~al.}(2020)\citenamefont {Romero},
  \citenamefont {Allan}, \citenamefont {Amadon}, \citenamefont {Antonius},
  \citenamefont {Applencourt}, \citenamefont {Baguet}, \citenamefont {Bieder},
  \citenamefont {Bottin}, \citenamefont {Bouchet}, \citenamefont {Bousquet},
  \citenamefont {Bruneval}, \citenamefont {Brunin}, \citenamefont {Caliste},
  \citenamefont {Côté}, \citenamefont {Denier}, \citenamefont {Dreyer},
  \citenamefont {Ghosez}, \citenamefont {Giantomassi}, \citenamefont {Gillet},
  \citenamefont {Gingras}, \citenamefont {Hamann}, \citenamefont {Hautier},
  \citenamefont {Jollet}, \citenamefont {Jomard}, \citenamefont {Martin},
  \citenamefont {Miranda}, \citenamefont {Naccarato}, \citenamefont {Petretto},
  \citenamefont {Pike}, \citenamefont {Planes}, \citenamefont {Prokhorenko},
  \citenamefont {Rangel}, \citenamefont {Ricci}, \citenamefont {Rignanese},
  \citenamefont {Royo}, \citenamefont {Stengel}, \citenamefont {Torrent},
  \citenamefont {van Setten}, \citenamefont {Van~Troeye}, \citenamefont
  {Verstraete}, \citenamefont {Wiktor}, \citenamefont {Zwanziger},\ and\
  \citenamefont {Gonze}}]{romero_abinit_2020}%
  \BibitemOpen
  \bibfield  {author} {\bibinfo {author} {\bibfnamefont {A.~H.}\ \bibnamefont
  {Romero}}, \bibinfo {author} {\bibfnamefont {D.~C.}\ \bibnamefont {Allan}},
  \bibinfo {author} {\bibfnamefont {B.}~\bibnamefont {Amadon}}, \bibinfo
  {author} {\bibfnamefont {G.}~\bibnamefont {Antonius}}, \bibinfo {author}
  {\bibfnamefont {T.}~\bibnamefont {Applencourt}}, \bibinfo {author}
  {\bibfnamefont {L.}~\bibnamefont {Baguet}}, \bibinfo {author} {\bibfnamefont
  {J.}~\bibnamefont {Bieder}}, \bibinfo {author} {\bibfnamefont
  {F.}~\bibnamefont {Bottin}}, \bibinfo {author} {\bibfnamefont
  {J.}~\bibnamefont {Bouchet}}, \bibinfo {author} {\bibfnamefont
  {E.}~\bibnamefont {Bousquet}}, \bibinfo {author} {\bibfnamefont
  {F.}~\bibnamefont {Bruneval}}, \bibinfo {author} {\bibfnamefont
  {G.}~\bibnamefont {Brunin}}, \bibinfo {author} {\bibfnamefont
  {D.}~\bibnamefont {Caliste}}, \bibinfo {author} {\bibfnamefont
  {M.}~\bibnamefont {Côté}}, \bibinfo {author} {\bibfnamefont
  {J.}~\bibnamefont {Denier}}, \bibinfo {author} {\bibfnamefont
  {C.}~\bibnamefont {Dreyer}}, \bibinfo {author} {\bibfnamefont
  {P.}~\bibnamefont {Ghosez}}, \bibinfo {author} {\bibfnamefont
  {M.}~\bibnamefont {Giantomassi}}, \bibinfo {author} {\bibfnamefont
  {Y.}~\bibnamefont {Gillet}}, \bibinfo {author} {\bibfnamefont
  {O.}~\bibnamefont {Gingras}}, \bibinfo {author} {\bibfnamefont {D.~R.}\
  \bibnamefont {Hamann}}, \bibinfo {author} {\bibfnamefont {G.}~\bibnamefont
  {Hautier}}, \bibinfo {author} {\bibfnamefont {F.}~\bibnamefont {Jollet}},
  \bibinfo {author} {\bibfnamefont {G.}~\bibnamefont {Jomard}}, \bibinfo
  {author} {\bibfnamefont {A.}~\bibnamefont {Martin}}, \bibinfo {author}
  {\bibfnamefont {H.~P.~C.}\ \bibnamefont {Miranda}}, \bibinfo {author}
  {\bibfnamefont {F.}~\bibnamefont {Naccarato}}, \bibinfo {author}
  {\bibfnamefont {G.}~\bibnamefont {Petretto}}, \bibinfo {author}
  {\bibfnamefont {N.~A.}\ \bibnamefont {Pike}}, \bibinfo {author}
  {\bibfnamefont {V.}~\bibnamefont {Planes}}, \bibinfo {author} {\bibfnamefont
  {S.}~\bibnamefont {Prokhorenko}}, \bibinfo {author} {\bibfnamefont
  {T.}~\bibnamefont {Rangel}}, \bibinfo {author} {\bibfnamefont
  {F.}~\bibnamefont {Ricci}}, \bibinfo {author} {\bibfnamefont {G.-M.}\
  \bibnamefont {Rignanese}}, \bibinfo {author} {\bibfnamefont {M.}~\bibnamefont
  {Royo}}, \bibinfo {author} {\bibfnamefont {M.}~\bibnamefont {Stengel}},
  \bibinfo {author} {\bibfnamefont {M.}~\bibnamefont {Torrent}}, \bibinfo
  {author} {\bibfnamefont {M.~J.}\ \bibnamefont {van Setten}}, \bibinfo
  {author} {\bibfnamefont {B.}~\bibnamefont {Van~Troeye}}, \bibinfo {author}
  {\bibfnamefont {M.~J.}\ \bibnamefont {Verstraete}}, \bibinfo {author}
  {\bibfnamefont {J.}~\bibnamefont {Wiktor}}, \bibinfo {author} {\bibfnamefont
  {J.~W.}\ \bibnamefont {Zwanziger}},\ and\ \bibinfo {author} {\bibfnamefont
  {X.}~\bibnamefont {Gonze}},\ }\bibfield  {title} {\bibinfo {title} {{ABINIT}:
  {Overview} and focus on selected capabilities},\ }\href
  {https://doi.org/10.1063/1.5144261} {\bibfield  {journal} {\bibinfo
  {journal} {The Journal of Chemical Physics}\ }\textbf {\bibinfo {volume}
  {152}},\ \bibinfo {pages} {124102} (\bibinfo {year} {2020})}\BibitemShut
  {NoStop}%
\bibitem [{\citenamefont {Kühne}\ \emph {et~al.}(2020)\citenamefont {Kühne},
  \citenamefont {Iannuzzi}, \citenamefont {Del~Ben}, \citenamefont {Rybkin},
  \citenamefont {Seewald}, \citenamefont {Stein}, \citenamefont {Laino},
  \citenamefont {Khaliullin}, \citenamefont {Schütt}, \citenamefont
  {Schiffmann}, \citenamefont {Golze}, \citenamefont {Wilhelm}, \citenamefont
  {Chulkov}, \citenamefont {Bani-Hashemian}, \citenamefont {Weber},
  \citenamefont {Borštnik}, \citenamefont {Taillefumier}, \citenamefont
  {Jakobovits}, \citenamefont {Lazzaro}, \citenamefont {Pabst}, \citenamefont
  {Müller}, \citenamefont {Schade}, \citenamefont {Guidon}, \citenamefont
  {Andermatt}, \citenamefont {Holmberg}, \citenamefont {Schenter},
  \citenamefont {Hehn}, \citenamefont {Bussy}, \citenamefont {Belleflamme},
  \citenamefont {Tabacchi}, \citenamefont {Glöß}, \citenamefont {Lass},
  \citenamefont {Bethune}, \citenamefont {Mundy}, \citenamefont {Plessl},
  \citenamefont {Watkins}, \citenamefont {VandeVondele}, \citenamefont
  {Krack},\ and\ \citenamefont {Hutter}}]{Kuhne2020cp2k}%
  \BibitemOpen
  \bibfield  {author} {\bibinfo {author} {\bibfnamefont {T.~D.}\ \bibnamefont
  {Kühne}}, \bibinfo {author} {\bibfnamefont {M.}~\bibnamefont {Iannuzzi}},
  \bibinfo {author} {\bibfnamefont {M.}~\bibnamefont {Del~Ben}}, \bibinfo
  {author} {\bibfnamefont {V.~V.}\ \bibnamefont {Rybkin}}, \bibinfo {author}
  {\bibfnamefont {P.}~\bibnamefont {Seewald}}, \bibinfo {author} {\bibfnamefont
  {F.}~\bibnamefont {Stein}}, \bibinfo {author} {\bibfnamefont
  {T.}~\bibnamefont {Laino}}, \bibinfo {author} {\bibfnamefont {R.~Z.}\
  \bibnamefont {Khaliullin}}, \bibinfo {author} {\bibfnamefont
  {O.}~\bibnamefont {Schütt}}, \bibinfo {author} {\bibfnamefont
  {F.}~\bibnamefont {Schiffmann}}, \bibinfo {author} {\bibfnamefont
  {D.}~\bibnamefont {Golze}}, \bibinfo {author} {\bibfnamefont
  {J.}~\bibnamefont {Wilhelm}}, \bibinfo {author} {\bibfnamefont
  {S.}~\bibnamefont {Chulkov}}, \bibinfo {author} {\bibfnamefont {M.~H.}\
  \bibnamefont {Bani-Hashemian}}, \bibinfo {author} {\bibfnamefont
  {V.}~\bibnamefont {Weber}}, \bibinfo {author} {\bibfnamefont
  {U.}~\bibnamefont {Borštnik}}, \bibinfo {author} {\bibfnamefont
  {M.}~\bibnamefont {Taillefumier}}, \bibinfo {author} {\bibfnamefont {A.~S.}\
  \bibnamefont {Jakobovits}}, \bibinfo {author} {\bibfnamefont
  {A.}~\bibnamefont {Lazzaro}}, \bibinfo {author} {\bibfnamefont
  {H.}~\bibnamefont {Pabst}}, \bibinfo {author} {\bibfnamefont
  {T.}~\bibnamefont {Müller}}, \bibinfo {author} {\bibfnamefont
  {R.}~\bibnamefont {Schade}}, \bibinfo {author} {\bibfnamefont
  {M.}~\bibnamefont {Guidon}}, \bibinfo {author} {\bibfnamefont
  {S.}~\bibnamefont {Andermatt}}, \bibinfo {author} {\bibfnamefont
  {N.}~\bibnamefont {Holmberg}}, \bibinfo {author} {\bibfnamefont {G.~K.}\
  \bibnamefont {Schenter}}, \bibinfo {author} {\bibfnamefont {A.}~\bibnamefont
  {Hehn}}, \bibinfo {author} {\bibfnamefont {A.}~\bibnamefont {Bussy}},
  \bibinfo {author} {\bibfnamefont {F.}~\bibnamefont {Belleflamme}}, \bibinfo
  {author} {\bibfnamefont {G.}~\bibnamefont {Tabacchi}}, \bibinfo {author}
  {\bibfnamefont {A.}~\bibnamefont {Glöß}}, \bibinfo {author} {\bibfnamefont
  {M.}~\bibnamefont {Lass}}, \bibinfo {author} {\bibfnamefont {I.}~\bibnamefont
  {Bethune}}, \bibinfo {author} {\bibfnamefont {C.~J.}\ \bibnamefont {Mundy}},
  \bibinfo {author} {\bibfnamefont {C.}~\bibnamefont {Plessl}}, \bibinfo
  {author} {\bibfnamefont {M.}~\bibnamefont {Watkins}}, \bibinfo {author}
  {\bibfnamefont {J.}~\bibnamefont {VandeVondele}}, \bibinfo {author}
  {\bibfnamefont {M.}~\bibnamefont {Krack}},\ and\ \bibinfo {author}
  {\bibfnamefont {J.}~\bibnamefont {Hutter}},\ }\bibfield  {title} {\bibinfo
  {title} {{CP2K}: An electronic structure and molecular dynamics software
  package - {Quickstep}: Efficient and accurate electronic structure
  calculations},\ }\href {https://doi.org/10.1063/5.0007045} {\bibfield
  {journal} {\bibinfo  {journal} {The Journal of Chemical Physics}\ }\textbf
  {\bibinfo {volume} {152}},\ \bibinfo {pages} {194103} (\bibinfo {year}
  {2020})}\BibitemShut {NoStop}%
\bibitem [{\citenamefont {Box}\ \emph {et~al.}(2025)\citenamefont {Box},
  \citenamefont {Maurer}, \citenamefont {Shang}, \citenamefont {Scheffler},
  \citenamefont {Blum}, \citenamefont {Carbogno},\ and\ \citenamefont
  {Rossi}}]{box2025fhiaims_dfpt}%
  \BibitemOpen
  \bibfield  {author} {\bibinfo {author} {\bibfnamefont {C.~L.}\ \bibnamefont
  {Box}}, \bibinfo {author} {\bibfnamefont {R.~J.}\ \bibnamefont {Maurer}},
  \bibinfo {author} {\bibfnamefont {H.}~\bibnamefont {Shang}}, \bibinfo
  {author} {\bibfnamefont {M.}~\bibnamefont {Scheffler}}, \bibinfo {author}
  {\bibfnamefont {V.}~\bibnamefont {Blum}}, \bibinfo {author} {\bibfnamefont
  {C.}~\bibnamefont {Carbogno}},\ and\ \bibinfo {author} {\bibfnamefont
  {M.}~\bibnamefont {Rossi}},\ }\href {https://arxiv.org/abs/2501.16091}
  {\bibinfo {title} {Density-functional perturbation theory with numeric
  atom-centered orbitals}} (\bibinfo {year} {2025}),\ \Eprint
  {https://arxiv.org/abs/2501.16091} {arXiv:2501.16091 [cond-mat.mtrl-sci]}
  \BibitemShut {NoStop}%
\bibitem [{\citenamefont {Kresse}\ and\ \citenamefont
  {Furthmüller}(1996)}]{kresse_efficient_1996}%
  \BibitemOpen
  \bibfield  {author} {\bibinfo {author} {\bibfnamefont {G.}~\bibnamefont
  {Kresse}}\ and\ \bibinfo {author} {\bibfnamefont {J.}~\bibnamefont
  {Furthmüller}},\ }\bibfield  {title} {\bibinfo {title} {Efficient iterative
  schemes for ab initio total-energy calculations using a plane-wave basis
  set},\ }\href {https://doi.org/10.1103/PhysRevB.54.11169} {\bibfield
  {journal} {\bibinfo  {journal} {Phys. Rev. B}\ }\textbf {\bibinfo {volume}
  {54}},\ \bibinfo {pages} {11169} (\bibinfo {year} {1996})}\BibitemShut
  {NoStop}%
\bibitem [{\citenamefont {Golesorkhtabar}\ \emph {et~al.}(2013)\citenamefont
  {Golesorkhtabar}, \citenamefont {Pavone}, \citenamefont {Spitaler},
  \citenamefont {Puschnig},\ and\ \citenamefont
  {Draxl}}]{Golesorkhtabar2013ElaStic}%
  \BibitemOpen
  \bibfield  {author} {\bibinfo {author} {\bibfnamefont {R.}~\bibnamefont
  {Golesorkhtabar}}, \bibinfo {author} {\bibfnamefont {P.}~\bibnamefont
  {Pavone}}, \bibinfo {author} {\bibfnamefont {J.}~\bibnamefont {Spitaler}},
  \bibinfo {author} {\bibfnamefont {P.}~\bibnamefont {Puschnig}},\ and\
  \bibinfo {author} {\bibfnamefont {C.}~\bibnamefont {Draxl}},\ }\bibfield
  {title} {\bibinfo {title} {{ElaStic}: A tool for calculating second-order
  elastic constants from first principles},\ }\href
  {https://doi.org/https://doi.org/10.1016/j.cpc.2013.03.010} {\bibfield
  {journal} {\bibinfo  {journal} {Computer Physics Communications}\ }\textbf
  {\bibinfo {volume} {184}},\ \bibinfo {pages} {1861} (\bibinfo {year}
  {2013})}\BibitemShut {NoStop}%
\bibitem [{\citenamefont {de~Jong}\ \emph {et~al.}(2015)\citenamefont
  {de~Jong}, \citenamefont {Chen}, \citenamefont {Angsten}, \citenamefont
  {Jain}, \citenamefont {Notestine}, \citenamefont {Gamst}, \citenamefont
  {Sluiter}, \citenamefont {Krishna~Ande}, \citenamefont {van~der Zwaag},
  \citenamefont {Plata}, \citenamefont {Toher}, \citenamefont {Curtarolo},
  \citenamefont {Ceder}, \citenamefont {Persson},\ and\ \citenamefont
  {Asta}}]{de_jong_charting_2015}%
  \BibitemOpen
  \bibfield  {author} {\bibinfo {author} {\bibfnamefont {M.}~\bibnamefont
  {de~Jong}}, \bibinfo {author} {\bibfnamefont {W.}~\bibnamefont {Chen}},
  \bibinfo {author} {\bibfnamefont {T.}~\bibnamefont {Angsten}}, \bibinfo
  {author} {\bibfnamefont {A.}~\bibnamefont {Jain}}, \bibinfo {author}
  {\bibfnamefont {R.}~\bibnamefont {Notestine}}, \bibinfo {author}
  {\bibfnamefont {A.}~\bibnamefont {Gamst}}, \bibinfo {author} {\bibfnamefont
  {M.}~\bibnamefont {Sluiter}}, \bibinfo {author} {\bibfnamefont
  {C.}~\bibnamefont {Krishna~Ande}}, \bibinfo {author} {\bibfnamefont
  {S.}~\bibnamefont {van~der Zwaag}}, \bibinfo {author} {\bibfnamefont {J.~J.}\
  \bibnamefont {Plata}}, \bibinfo {author} {\bibfnamefont {C.}~\bibnamefont
  {Toher}}, \bibinfo {author} {\bibfnamefont {S.}~\bibnamefont {Curtarolo}},
  \bibinfo {author} {\bibfnamefont {G.}~\bibnamefont {Ceder}}, \bibinfo
  {author} {\bibfnamefont {K.~A.}\ \bibnamefont {Persson}},\ and\ \bibinfo
  {author} {\bibfnamefont {M.}~\bibnamefont {Asta}},\ }\bibfield  {title}
  {\bibinfo {title} {Charting the complete elastic properties of inorganic
  crystalline compounds},\ }\href {https://doi.org/10.1038/sdata.2015.9}
  {\bibfield  {journal} {\bibinfo  {journal} {Scientific Data}\ }\textbf
  {\bibinfo {volume} {2}},\ \bibinfo {pages} {150009} (\bibinfo {year}
  {2015})}\BibitemShut {NoStop}%
\bibitem [{\citenamefont {Togo}\ \emph {et~al.}(2023)\citenamefont {Togo},
  \citenamefont {Chaput}, \citenamefont {Tadano},\ and\ \citenamefont
  {Tanaka}}]{togo2023phonopy-phono3py}%
  \BibitemOpen
  \bibfield  {author} {\bibinfo {author} {\bibfnamefont {A.}~\bibnamefont
  {Togo}}, \bibinfo {author} {\bibfnamefont {L.}~\bibnamefont {Chaput}},
  \bibinfo {author} {\bibfnamefont {T.}~\bibnamefont {Tadano}},\ and\ \bibinfo
  {author} {\bibfnamefont {I.}~\bibnamefont {Tanaka}},\ }\bibfield  {title}
  {\bibinfo {title} {Implementation strategies in phonopy and phono3py},\
  }\href {https://doi.org/10.1088/1361-648X/acd831} {\bibfield  {journal}
  {\bibinfo  {journal} {J. Phys. Condens. Matter}\ }\textbf {\bibinfo {volume}
  {35}},\ \bibinfo {pages} {353001} (\bibinfo {year} {2023})}\BibitemShut
  {NoStop}%
\bibitem [{\citenamefont {Bastonero}\ and\ \citenamefont
  {Marzari}(2024)}]{Bastonero2024}%
  \BibitemOpen
  \bibfield  {author} {\bibinfo {author} {\bibfnamefont {L.}~\bibnamefont
  {Bastonero}}\ and\ \bibinfo {author} {\bibfnamefont {N.}~\bibnamefont
  {Marzari}},\ }\bibfield  {title} {\bibinfo {title} {Automated all-functionals
  infrared and raman spectra},\ }\href
  {https://doi.org/10.1038/s41524-024-01236-3} {\bibfield  {journal} {\bibinfo
  {journal} {npj Computational Materials}\ }\textbf {\bibinfo {volume} {10}},\
  \bibinfo {pages} {55} (\bibinfo {year} {2024})}\BibitemShut {NoStop}%
\bibitem [{\citenamefont {Griewank}\ and\ \citenamefont
  {Walther}(2008)}]{Griewank2008EDP}%
  \BibitemOpen
  \bibfield  {author} {\bibinfo {author} {\bibfnamefont {A.}~\bibnamefont
  {Griewank}}\ and\ \bibinfo {author} {\bibfnamefont {A.}~\bibnamefont
  {Walther}},\ }\href {https://doi.org/10.1137/1.9780898717761} {\emph
  {\bibinfo {title} {Evaluating Derivatives: {P}rinciples and Techniques of
  Algorithmic Differentiation}}},\ \bibinfo {edition} {2nd}\ ed.,\ \bibinfo
  {series} {Other Titles in Applied Mathematics}\ No.\ \bibinfo {number} {105}\
  (\bibinfo  {publisher} {SIAM},\ \bibinfo {year} {2008})\BibitemShut {NoStop}%
\bibitem [{\citenamefont {Naumann}(2012)}]{Naumann2012TAo}%
  \BibitemOpen
  \bibfield  {author} {\bibinfo {author} {\bibfnamefont {U.}~\bibnamefont
  {Naumann}},\ }\href {https://doi.org/10.1137/1.9781611972078} {\emph
  {\bibinfo {title} {The Art of Differentiating Computer Programs. An
  Introduction to Algorithmic Differentiation.}}},\ \bibinfo {series}
  {Software, Environments, and Tools}\ No.\ \bibinfo {number} {SE24}\ (\bibinfo
   {publisher} {SIAM},\ \bibinfo {year} {2012})\BibitemShut {NoStop}%
\bibitem [{\citenamefont {Paszke}\ \emph {et~al.}(2019)\citenamefont {Paszke},
  \citenamefont {Gross}, \citenamefont {Massa}, \citenamefont {Lerer},
  \citenamefont {Bradbury}, \citenamefont {Chanan}, \citenamefont {Killeen},
  \citenamefont {Lin}, \citenamefont {Gimelshein}, \citenamefont {Antiga},
  \citenamefont {Desmaison}, \citenamefont {K\"{o}pf}, \citenamefont {Yang},
  \citenamefont {DeVito}, \citenamefont {Raison}, \citenamefont {Tejani},
  \citenamefont {Chilamkurthy}, \citenamefont {Steiner}, \citenamefont {Fang},
  \citenamefont {Bai},\ and\ \citenamefont {Chintala}}]{pytorch}%
  \BibitemOpen
  \bibfield  {author} {\bibinfo {author} {\bibfnamefont {A.}~\bibnamefont
  {Paszke}}, \bibinfo {author} {\bibfnamefont {S.}~\bibnamefont {Gross}},
  \bibinfo {author} {\bibfnamefont {F.}~\bibnamefont {Massa}}, \bibinfo
  {author} {\bibfnamefont {A.}~\bibnamefont {Lerer}}, \bibinfo {author}
  {\bibfnamefont {J.}~\bibnamefont {Bradbury}}, \bibinfo {author}
  {\bibfnamefont {G.}~\bibnamefont {Chanan}}, \bibinfo {author} {\bibfnamefont
  {T.}~\bibnamefont {Killeen}}, \bibinfo {author} {\bibfnamefont
  {Z.}~\bibnamefont {Lin}}, \bibinfo {author} {\bibfnamefont {N.}~\bibnamefont
  {Gimelshein}}, \bibinfo {author} {\bibfnamefont {L.}~\bibnamefont {Antiga}},
  \bibinfo {author} {\bibfnamefont {A.}~\bibnamefont {Desmaison}}, \bibinfo
  {author} {\bibfnamefont {A.}~\bibnamefont {K\"{o}pf}}, \bibinfo {author}
  {\bibfnamefont {E.}~\bibnamefont {Yang}}, \bibinfo {author} {\bibfnamefont
  {Z.}~\bibnamefont {DeVito}}, \bibinfo {author} {\bibfnamefont
  {M.}~\bibnamefont {Raison}}, \bibinfo {author} {\bibfnamefont
  {A.}~\bibnamefont {Tejani}}, \bibinfo {author} {\bibfnamefont
  {S.}~\bibnamefont {Chilamkurthy}}, \bibinfo {author} {\bibfnamefont
  {B.}~\bibnamefont {Steiner}}, \bibinfo {author} {\bibfnamefont
  {L.}~\bibnamefont {Fang}}, \bibinfo {author} {\bibfnamefont {J.}~\bibnamefont
  {Bai}},\ and\ \bibinfo {author} {\bibfnamefont {S.}~\bibnamefont
  {Chintala}},\ }\bibinfo {title} {{PyTorch}: an imperative style,
  high-performance deep learning library},\ in\ \href@noop {} {\emph {\bibinfo
  {booktitle} {Proceedings of the 33rd International Conference on Neural
  Information Processing Systems}}}\ (\bibinfo  {publisher} {Curran Associates
  Inc.},\ \bibinfo {address} {Red Hook, NY, USA},\ \bibinfo {year}
  {2019})\BibitemShut {NoStop}%
\bibitem [{\citenamefont {Bradbury}\ \emph {et~al.}(2018)\citenamefont
  {Bradbury}, \citenamefont {Frostig}, \citenamefont {Hawkins}, \citenamefont
  {Johnson}, \citenamefont {Leary}, \citenamefont {Maclaurin}, \citenamefont
  {Necula}, \citenamefont {Paszke}, \citenamefont {Vander{P}las}, \citenamefont
  {Wanderman-{M}ilne},\ and\ \citenamefont {Zhang}}]{jax2018github}%
  \BibitemOpen
  \bibfield  {author} {\bibinfo {author} {\bibfnamefont {J.}~\bibnamefont
  {Bradbury}}, \bibinfo {author} {\bibfnamefont {R.}~\bibnamefont {Frostig}},
  \bibinfo {author} {\bibfnamefont {P.}~\bibnamefont {Hawkins}}, \bibinfo
  {author} {\bibfnamefont {M.~J.}\ \bibnamefont {Johnson}}, \bibinfo {author}
  {\bibfnamefont {C.}~\bibnamefont {Leary}}, \bibinfo {author} {\bibfnamefont
  {D.}~\bibnamefont {Maclaurin}}, \bibinfo {author} {\bibfnamefont
  {G.}~\bibnamefont {Necula}}, \bibinfo {author} {\bibfnamefont
  {A.}~\bibnamefont {Paszke}}, \bibinfo {author} {\bibfnamefont
  {J.}~\bibnamefont {Vander{P}las}}, \bibinfo {author} {\bibfnamefont
  {S.}~\bibnamefont {Wanderman-{M}ilne}},\ and\ \bibinfo {author}
  {\bibfnamefont {Q.}~\bibnamefont {Zhang}},\ }\href
  {http://github.com/jax-ml/jax} {\bibinfo {title} {{JAX}: composable
  transformations of {P}ython+{N}um{P}y programs}} (\bibinfo {year}
  {2018})\BibitemShut {NoStop}%
\bibitem [{\citenamefont {Revels}\ \emph {et~al.}(2016)\citenamefont {Revels},
  \citenamefont {Lubin},\ and\ \citenamefont
  {Papamarkou}}]{Revels2016forwarddiff}%
  \BibitemOpen
  \bibfield  {author} {\bibinfo {author} {\bibfnamefont {J.}~\bibnamefont
  {Revels}}, \bibinfo {author} {\bibfnamefont {M.}~\bibnamefont {Lubin}},\ and\
  \bibinfo {author} {\bibfnamefont {T.}~\bibnamefont {Papamarkou}},\ }\href
  {https://arxiv.org/abs/1607.07892} {\bibinfo {title} {Forward-mode automatic
  differentiation in {Julia}}} (\bibinfo {year} {2016}),\ \Eprint
  {https://arxiv.org/abs/1607.07892} {arXiv:1607.07892 [cs.MS]} \BibitemShut
  {NoStop}%
\bibitem [{\citenamefont {Moses}\ \emph {et~al.}(2021)\citenamefont {Moses},
  \citenamefont {Churavy}, \citenamefont {Paehler}, \citenamefont
  {H\"{u}ckelheim}, \citenamefont {Narayanan}, \citenamefont {Schanen},\ and\
  \citenamefont {Doerfert}}]{Moses2021}%
  \BibitemOpen
  \bibfield  {author} {\bibinfo {author} {\bibfnamefont {W.~S.}\ \bibnamefont
  {Moses}}, \bibinfo {author} {\bibfnamefont {V.}~\bibnamefont {Churavy}},
  \bibinfo {author} {\bibfnamefont {L.}~\bibnamefont {Paehler}}, \bibinfo
  {author} {\bibfnamefont {J.}~\bibnamefont {H\"{u}ckelheim}}, \bibinfo
  {author} {\bibfnamefont {S.~H.~K.}\ \bibnamefont {Narayanan}}, \bibinfo
  {author} {\bibfnamefont {M.}~\bibnamefont {Schanen}},\ and\ \bibinfo {author}
  {\bibfnamefont {J.}~\bibnamefont {Doerfert}},\ }\bibfield  {title} {\bibinfo
  {title} {Reverse-mode automatic differentiation and optimization of {GPU}
  kernels via enzyme},\ }in\ \href {https://doi.org/10.1145/3458817.3476165}
  {\emph {\bibinfo {booktitle} {SC '21: The International Conference for High
  Performance Computing, Networking, Storage and Analysis}}}\ (\bibinfo
  {publisher} {ACM},\ \bibinfo {year} {2021})\BibitemShut {NoStop}%
\bibitem [{\citenamefont {Baydin}\ \emph {et~al.}(2018)\citenamefont {Baydin},
  \citenamefont {Pearlmutter}, \citenamefont {Radul},\ and\ \citenamefont
  {Siskind}}]{Baydin2018review}%
  \BibitemOpen
  \bibfield  {author} {\bibinfo {author} {\bibfnamefont {A.~G.}\ \bibnamefont
  {Baydin}}, \bibinfo {author} {\bibfnamefont {B.~A.}\ \bibnamefont
  {Pearlmutter}}, \bibinfo {author} {\bibfnamefont {A.~A.}\ \bibnamefont
  {Radul}},\ and\ \bibinfo {author} {\bibfnamefont {J.~M.}\ \bibnamefont
  {Siskind}},\ }\bibfield  {title} {\bibinfo {title} {Automatic differentiation
  in machine learning: a survey},\ }\href
  {http://jmlr.org/papers/v18/17-468.html} {\bibfield  {journal} {\bibinfo
  {journal} {Journal of Machine Learning Research}\ }\textbf {\bibinfo {volume}
  {18}},\ \bibinfo {pages} {1} (\bibinfo {year} {2018})}\BibitemShut {NoStop}%
\bibitem [{\citenamefont {Sapienza}\ \emph {et~al.}(2024)\citenamefont
  {Sapienza}, \citenamefont {Bolibar}, \citenamefont {Schäfer}, \citenamefont
  {Groenke}, \citenamefont {Pal}, \citenamefont {Boussange}, \citenamefont
  {Heimbach}, \citenamefont {Hooker}, \citenamefont {Pérez}, \citenamefont
  {Persson},\ and\ \citenamefont
  {Rackauckas}}]{sapienza2024differentiableprogrammingdifferentialequations}%
  \BibitemOpen
  \bibfield  {author} {\bibinfo {author} {\bibfnamefont {F.}~\bibnamefont
  {Sapienza}}, \bibinfo {author} {\bibfnamefont {J.}~\bibnamefont {Bolibar}},
  \bibinfo {author} {\bibfnamefont {F.}~\bibnamefont {Schäfer}}, \bibinfo
  {author} {\bibfnamefont {B.}~\bibnamefont {Groenke}}, \bibinfo {author}
  {\bibfnamefont {A.}~\bibnamefont {Pal}}, \bibinfo {author} {\bibfnamefont
  {V.}~\bibnamefont {Boussange}}, \bibinfo {author} {\bibfnamefont
  {P.}~\bibnamefont {Heimbach}}, \bibinfo {author} {\bibfnamefont
  {G.}~\bibnamefont {Hooker}}, \bibinfo {author} {\bibfnamefont
  {F.}~\bibnamefont {Pérez}}, \bibinfo {author} {\bibfnamefont {P.-O.}\
  \bibnamefont {Persson}},\ and\ \bibinfo {author} {\bibfnamefont
  {C.}~\bibnamefont {Rackauckas}},\ }\href {https://arxiv.org/abs/2406.09699}
  {\bibinfo {title} {Differentiable programming for differential equations: A
  review}} (\bibinfo {year} {2024}),\ \Eprint
  {https://arxiv.org/abs/2406.09699} {arXiv:2406.09699 [math.NA]} \BibitemShut
  {NoStop}%
\bibitem [{\citenamefont {Unke}\ \emph {et~al.}(2021)\citenamefont {Unke},
  \citenamefont {Chmiela}, \citenamefont {Sauceda}, \citenamefont {Gastegger},
  \citenamefont {Poltavsky}, \citenamefont {Sch\"{u}tt}, \citenamefont
  {Tkatchenko},\ and\ \citenamefont {M\"{u}ller}}]{Unke2021}%
  \BibitemOpen
  \bibfield  {author} {\bibinfo {author} {\bibfnamefont {O.~T.}\ \bibnamefont
  {Unke}}, \bibinfo {author} {\bibfnamefont {S.}~\bibnamefont {Chmiela}},
  \bibinfo {author} {\bibfnamefont {H.~E.}\ \bibnamefont {Sauceda}}, \bibinfo
  {author} {\bibfnamefont {M.}~\bibnamefont {Gastegger}}, \bibinfo {author}
  {\bibfnamefont {I.}~\bibnamefont {Poltavsky}}, \bibinfo {author}
  {\bibfnamefont {K.~T.}\ \bibnamefont {Sch\"{u}tt}}, \bibinfo {author}
  {\bibfnamefont {A.}~\bibnamefont {Tkatchenko}},\ and\ \bibinfo {author}
  {\bibfnamefont {K.-R.}\ \bibnamefont {M\"{u}ller}},\ }\bibfield  {title}
  {\bibinfo {title} {Machine learning force fields},\ }\href
  {https://doi.org/10.1021/acs.chemrev.0c01111} {\bibfield  {journal} {\bibinfo
   {journal} {Chemical Reviews}\ }\textbf {\bibinfo {volume} {121}},\ \bibinfo
  {pages} {10142} (\bibinfo {year} {2021})}\BibitemShut {NoStop}%
\bibitem [{\citenamefont {Schmitz}\ \emph {et~al.}(2022)\citenamefont
  {Schmitz}, \citenamefont {M{\"u}ller},\ and\ \citenamefont
  {Chmiela}}]{Schmitz2022ad}%
  \BibitemOpen
  \bibfield  {author} {\bibinfo {author} {\bibfnamefont {N.~F.}\ \bibnamefont
  {Schmitz}}, \bibinfo {author} {\bibfnamefont {K.-R.}\ \bibnamefont
  {M{\"u}ller}},\ and\ \bibinfo {author} {\bibfnamefont {S.}~\bibnamefont
  {Chmiela}},\ }\bibfield  {title} {\bibinfo {title} {Algorithmic
  differentiation for automated modeling of machine learned force fields},\
  }\href {https://doi.org/10.1021/acs.jpclett.2c02632} {\bibfield  {journal}
  {\bibinfo  {journal} {The Journal of Physical Chemistry Letters}\ }\textbf
  {\bibinfo {volume} {13}},\ \bibinfo {pages} {10183} (\bibinfo {year}
  {2022})}\BibitemShut {NoStop}%
\bibitem [{\citenamefont {Langer}\ \emph {et~al.}(2023)\citenamefont {Langer},
  \citenamefont {Frank},\ and\ \citenamefont {Knoop}}]{Langer2023ADheatflux}%
  \BibitemOpen
  \bibfield  {author} {\bibinfo {author} {\bibfnamefont {M.~F.}\ \bibnamefont
  {Langer}}, \bibinfo {author} {\bibfnamefont {J.~T.}\ \bibnamefont {Frank}},\
  and\ \bibinfo {author} {\bibfnamefont {F.}~\bibnamefont {Knoop}},\ }\bibfield
   {title} {\bibinfo {title} {Stress and heat flux via automatic
  differentiation},\ }\href {https://doi.org/10.1063/5.0155760} {\bibfield
  {journal} {\bibinfo  {journal} {The Journal of Chemical Physics}\ }\textbf
  {\bibinfo {volume} {159}},\ \bibinfo {pages} {174105} (\bibinfo {year}
  {2023})}\BibitemShut {NoStop}%
\bibitem [{\citenamefont {G{\"o}nnheimer}\ \emph {et~al.}(2025)\citenamefont
  {G{\"o}nnheimer}, \citenamefont {Reuter},\ and\ \citenamefont
  {Margraf}}]{Gonnheimer2025ad_hessian_mlip}%
  \BibitemOpen
  \bibfield  {author} {\bibinfo {author} {\bibfnamefont {N.}~\bibnamefont
  {G{\"o}nnheimer}}, \bibinfo {author} {\bibfnamefont {K.}~\bibnamefont
  {Reuter}},\ and\ \bibinfo {author} {\bibfnamefont {J.~T.}\ \bibnamefont
  {Margraf}},\ }\bibfield  {title} {\bibinfo {title} {Beyond numerical
  hessians: Higher-order derivatives for machine learning interatomic
  potentials via automatic differentiation},\ }\href
  {https://doi.org/10.1021/acs.jctc.4c01790} {\bibfield  {journal} {\bibinfo
  {journal} {Journal of Chemical Theory and Computation}\ }\textbf {\bibinfo
  {volume} {21}},\ \bibinfo {pages} {4742} (\bibinfo {year}
  {2025})}\BibitemShut {NoStop}%
\bibitem [{\citenamefont {Wang}\ \emph {et~al.}(2019)\citenamefont {Wang},
  \citenamefont {Axelrod},\ and\ \citenamefont
  {G{\'o}mez-Bombarelli}}]{Wang2020}%
  \BibitemOpen
  \bibfield  {author} {\bibinfo {author} {\bibfnamefont {W.}~\bibnamefont
  {Wang}}, \bibinfo {author} {\bibfnamefont {S.}~\bibnamefont {Axelrod}},\ and\
  \bibinfo {author} {\bibfnamefont {R.}~\bibnamefont {G{\'o}mez-Bombarelli}},\
  }\bibfield  {title} {\bibinfo {title} {Differentiable molecular simulations
  for control and learning},\ }in\ \href
  {https://openreview.net/forum?id=YDNzrQRsu} {\emph {\bibinfo {booktitle}
  {ICLR 2020 Workshop on Integration of Deep Neural Models and Differential
  Equations}}}\ (\bibinfo {year} {2019})\BibitemShut {NoStop}%
\bibitem [{\citenamefont {Greener}\ and\ \citenamefont
  {Jones}(2021)}]{Greener2021}%
  \BibitemOpen
  \bibfield  {author} {\bibinfo {author} {\bibfnamefont {J.~G.}\ \bibnamefont
  {Greener}}\ and\ \bibinfo {author} {\bibfnamefont {D.~T.}\ \bibnamefont
  {Jones}},\ }\bibfield  {title} {\bibinfo {title} {Differentiable molecular
  simulation can learn all the parameters in a coarse-grained force field for
  proteins},\ }\href {https://doi.org/10.1371/journal.pone.0256990} {\bibfield
  {journal} {\bibinfo  {journal} {PLOS ONE}\ }\textbf {\bibinfo {volume}
  {16}},\ \bibinfo {pages} {e0256990} (\bibinfo {year} {2021})}\BibitemShut
  {NoStop}%
\bibitem [{\citenamefont {Schwalbe-Koda}\ \emph {et~al.}(2021)\citenamefont
  {Schwalbe-Koda}, \citenamefont {Tan},\ and\ \citenamefont
  {G\'{o}mez-Bombarelli}}]{SchwalbeKoda2021}%
  \BibitemOpen
  \bibfield  {author} {\bibinfo {author} {\bibfnamefont {D.}~\bibnamefont
  {Schwalbe-Koda}}, \bibinfo {author} {\bibfnamefont {A.~R.}\ \bibnamefont
  {Tan}},\ and\ \bibinfo {author} {\bibfnamefont {R.}~\bibnamefont
  {G\'{o}mez-Bombarelli}},\ }\bibfield  {title} {\bibinfo {title}
  {Differentiable sampling of molecular geometries with uncertainty-based
  adversarial attacks},\ }\href {https://doi.org/10.1038/s41467-021-25342-8}
  {\bibfield  {journal} {\bibinfo  {journal} {Nature Communications}\ }\textbf
  {\bibinfo {volume} {12}},\ \bibinfo {pages} {5104} (\bibinfo {year}
  {2021})}\BibitemShut {NoStop}%
\bibitem [{\citenamefont {Maliyov}\ \emph {et~al.}(2025)\citenamefont
  {Maliyov}, \citenamefont {Grigorev},\ and\ \citenamefont
  {Swinburne}}]{Maliyov2024}%
  \BibitemOpen
  \bibfield  {author} {\bibinfo {author} {\bibfnamefont {I.}~\bibnamefont
  {Maliyov}}, \bibinfo {author} {\bibfnamefont {P.}~\bibnamefont {Grigorev}},\
  and\ \bibinfo {author} {\bibfnamefont {T.~D.}\ \bibnamefont {Swinburne}},\
  }\bibfield  {title} {\bibinfo {title} {Exploring parameter dependence of
  atomic minima with implicit differentiation},\ }\href
  {https://doi.org/10.1038/s41524-024-01506-0} {\bibfield  {journal} {\bibinfo
  {journal} {npj Computational Materials}\ }\textbf {\bibinfo {volume} {11}},\
  \bibinfo {pages} {22} (\bibinfo {year} {2025})}\BibitemShut {NoStop}%
\bibitem [{\citenamefont {Li}\ \emph {et~al.}(2022)\citenamefont {Li},
  \citenamefont {Wang}, \citenamefont {Zou}, \citenamefont {Ye}, \citenamefont
  {Xu}, \citenamefont {Gong}, \citenamefont {Duan},\ and\ \citenamefont
  {Xu}}]{Li2022deeplearning}%
  \BibitemOpen
  \bibfield  {author} {\bibinfo {author} {\bibfnamefont {H.}~\bibnamefont
  {Li}}, \bibinfo {author} {\bibfnamefont {Z.}~\bibnamefont {Wang}}, \bibinfo
  {author} {\bibfnamefont {N.}~\bibnamefont {Zou}}, \bibinfo {author}
  {\bibfnamefont {M.}~\bibnamefont {Ye}}, \bibinfo {author} {\bibfnamefont
  {R.}~\bibnamefont {Xu}}, \bibinfo {author} {\bibfnamefont {X.}~\bibnamefont
  {Gong}}, \bibinfo {author} {\bibfnamefont {W.}~\bibnamefont {Duan}},\ and\
  \bibinfo {author} {\bibfnamefont {Y.}~\bibnamefont {Xu}},\ }\bibfield
  {title} {\bibinfo {title} {Deep-learning density functional theory
  hamiltonian for efficient ab initio electronic-structure calculation},\
  }\href {https://doi.org/10.1038/s43588-022-00265-6} {\bibfield  {journal}
  {\bibinfo  {journal} {Nature Computational Science}\ }\textbf {\bibinfo
  {volume} {2}},\ \bibinfo {pages} {367} (\bibinfo {year} {2022})}\BibitemShut
  {NoStop}%
\bibitem [{\citenamefont {Vargas–Hernández}\ \emph
  {et~al.}(2023)\citenamefont {Vargas–Hernández}, \citenamefont {Jorner},
  \citenamefont {Pollice},\ and\ \citenamefont
  {Aspuru–Guzik}}]{VargasHernandez2023}%
  \BibitemOpen
  \bibfield  {author} {\bibinfo {author} {\bibfnamefont {R.~A.}\ \bibnamefont
  {Vargas–Hernández}}, \bibinfo {author} {\bibfnamefont {K.}~\bibnamefont
  {Jorner}}, \bibinfo {author} {\bibfnamefont {R.}~\bibnamefont {Pollice}},\
  and\ \bibinfo {author} {\bibfnamefont {A.}~\bibnamefont {Aspuru–Guzik}},\
  }\bibfield  {title} {\bibinfo {title} {Inverse molecular design and parameter
  optimization with hückel theory using automatic differentiation},\ }\href
  {https://doi.org/10.1063/5.0137103} {\bibfield  {journal} {\bibinfo
  {journal} {The Journal of Chemical Physics}\ }\textbf {\bibinfo {volume}
  {158}},\ \bibinfo {pages} {104801} (\bibinfo {year} {2023})}\BibitemShut
  {NoStop}%
\bibitem [{\citenamefont {Li}\ \emph {et~al.}(2024{\natexlab{a}})\citenamefont
  {Li}, \citenamefont {Tang}, \citenamefont {Chen}, \citenamefont {Sun},
  \citenamefont {Zhao}, \citenamefont {Li}, \citenamefont {Tao}, \citenamefont
  {Yuan}, \citenamefont {Duan},\ and\ \citenamefont {Xu}}]{Li2024nndft}%
  \BibitemOpen
  \bibfield  {author} {\bibinfo {author} {\bibfnamefont {Y.}~\bibnamefont
  {Li}}, \bibinfo {author} {\bibfnamefont {Z.}~\bibnamefont {Tang}}, \bibinfo
  {author} {\bibfnamefont {Z.}~\bibnamefont {Chen}}, \bibinfo {author}
  {\bibfnamefont {M.}~\bibnamefont {Sun}}, \bibinfo {author} {\bibfnamefont
  {B.}~\bibnamefont {Zhao}}, \bibinfo {author} {\bibfnamefont {H.}~\bibnamefont
  {Li}}, \bibinfo {author} {\bibfnamefont {H.}~\bibnamefont {Tao}}, \bibinfo
  {author} {\bibfnamefont {Z.}~\bibnamefont {Yuan}}, \bibinfo {author}
  {\bibfnamefont {W.}~\bibnamefont {Duan}},\ and\ \bibinfo {author}
  {\bibfnamefont {Y.}~\bibnamefont {Xu}},\ }\bibfield  {title} {\bibinfo
  {title} {Neural-network density functional theory based on variational energy
  minimization},\ }\href {https://doi.org/10.1103/PhysRevLett.133.076401}
  {\bibfield  {journal} {\bibinfo  {journal} {Phys. Rev. Lett.}\ }\textbf
  {\bibinfo {volume} {133}},\ \bibinfo {pages} {076401} (\bibinfo {year}
  {2024}{\natexlab{a}})}\BibitemShut {NoStop}%
\bibitem [{\citenamefont {Chen}\ \emph {et~al.}(2021)\citenamefont {Chen},
  \citenamefont {Zhang}, \citenamefont {Wang},\ and\ \citenamefont
  {E}}]{Chen2021}%
  \BibitemOpen
  \bibfield  {author} {\bibinfo {author} {\bibfnamefont {Y.}~\bibnamefont
  {Chen}}, \bibinfo {author} {\bibfnamefont {L.}~\bibnamefont {Zhang}},
  \bibinfo {author} {\bibfnamefont {H.}~\bibnamefont {Wang}},\ and\ \bibinfo
  {author} {\bibfnamefont {W.}~\bibnamefont {E}},\ }\bibfield  {title}
  {\bibinfo {title} {{DeePKS}: A comprehensive data-driven approach toward
  chemically accurate density functional theory},\ }\href
  {https://doi.org/10.1021/acs.jctc.0c00872} {\bibfield  {journal} {\bibinfo
  {journal} {Journal of Chemical Theory and Computation}\ }\textbf {\bibinfo
  {volume} {17}},\ \bibinfo {pages} {170} (\bibinfo {year} {2021})}\BibitemShut
  {NoStop}%
\bibitem [{\citenamefont {Kasim}\ \emph {et~al.}(2022)\citenamefont {Kasim},
  \citenamefont {Lehtola},\ and\ \citenamefont {Vinko}}]{Kasim2022}%
  \BibitemOpen
  \bibfield  {author} {\bibinfo {author} {\bibfnamefont {M.~F.}\ \bibnamefont
  {Kasim}}, \bibinfo {author} {\bibfnamefont {S.}~\bibnamefont {Lehtola}},\
  and\ \bibinfo {author} {\bibfnamefont {S.~M.}\ \bibnamefont {Vinko}},\
  }\bibfield  {title} {\bibinfo {title} {{DQC}: A {Python} program package for
  differentiable quantum chemistry},\ }\href
  {https://doi.org/10.1063/5.0076202} {\bibfield  {journal} {\bibinfo
  {journal} {The Journal of Chemical Physics}\ }\textbf {\bibinfo {volume}
  {156}},\ \bibinfo {pages} {084801} (\bibinfo {year} {2022})}\BibitemShut
  {NoStop}%
\bibitem [{\citenamefont {Zhang}\ and\ \citenamefont
  {Chan}(2022)}]{Zhang2022pyscf}%
  \BibitemOpen
  \bibfield  {author} {\bibinfo {author} {\bibfnamefont {X.}~\bibnamefont
  {Zhang}}\ and\ \bibinfo {author} {\bibfnamefont {G.~K.-L.}\ \bibnamefont
  {Chan}},\ }\bibfield  {title} {\bibinfo {title} {Differentiable quantum
  chemistry with {PySCF} for molecules and materials at the mean-field level
  and beyond},\ }\href {https://doi.org/10.1063/5.0118200} {\bibfield
  {journal} {\bibinfo  {journal} {The Journal of Chemical Physics}\ }\textbf
  {\bibinfo {volume} {157}},\ \bibinfo {pages} {204801} (\bibinfo {year}
  {2022})}\BibitemShut {NoStop}%
\bibitem [{\citenamefont {M.~Casares}\ \emph {et~al.}(2024)\citenamefont
  {M.~Casares}, \citenamefont {Baker}, \citenamefont {Medvidović},
  \citenamefont {Reis},\ and\ \citenamefont {Arrazola}}]{MCasares2024}%
  \BibitemOpen
  \bibfield  {author} {\bibinfo {author} {\bibfnamefont {P.~A.}\ \bibnamefont
  {M.~Casares}}, \bibinfo {author} {\bibfnamefont {J.~S.}\ \bibnamefont
  {Baker}}, \bibinfo {author} {\bibfnamefont {M.}~\bibnamefont {Medvidović}},
  \bibinfo {author} {\bibfnamefont {R.~d.}\ \bibnamefont {Reis}},\ and\
  \bibinfo {author} {\bibfnamefont {J.~M.}\ \bibnamefont {Arrazola}},\
  }\bibfield  {title} {\bibinfo {title} {{GradDFT}. a software library for
  machine learning enhanced density functional theory},\ }\href
  {https://doi.org/10.1063/5.0181037} {\bibfield  {journal} {\bibinfo
  {journal} {The Journal of Chemical Physics}\ }\textbf {\bibinfo {volume}
  {160}},\ \bibinfo {pages} {062501} (\bibinfo {year} {2024})}\BibitemShut
  {NoStop}%
\bibitem [{\citenamefont {Li}\ \emph {et~al.}(2021)\citenamefont {Li},
  \citenamefont {Hoyer}, \citenamefont {Pederson} \emph {et~al.}}]{Li2021}%
  \BibitemOpen
  \bibfield  {author} {\bibinfo {author} {\bibfnamefont {L.}~\bibnamefont
  {Li}}, \bibinfo {author} {\bibfnamefont {S.}~\bibnamefont {Hoyer}}, \bibinfo
  {author} {\bibfnamefont {R.}~\bibnamefont {Pederson}}, \emph {et~al.},\
  }\bibfield  {title} {\bibinfo {title} {{Kohn}-{Sham} equations as
  regularizer: Building prior knowledge into machine-learned physics},\ }\href
  {https://doi.org/10.1103/physrevlett.126.036401} {\bibfield  {journal}
  {\bibinfo  {journal} {Physical Review Letters}\ }\textbf {\bibinfo {volume}
  {126}},\ \bibinfo {pages} {036401} (\bibinfo {year} {2021})}\BibitemShut
  {NoStop}%
\bibitem [{\citenamefont {Kasim}\ and\ \citenamefont
  {Vinko}(2021)}]{Kasim2021}%
  \BibitemOpen
  \bibfield  {author} {\bibinfo {author} {\bibfnamefont {M.~F.}\ \bibnamefont
  {Kasim}}\ and\ \bibinfo {author} {\bibfnamefont {S.~M.}\ \bibnamefont
  {Vinko}},\ }\bibfield  {title} {\bibinfo {title} {Learning the
  exchange-correlation functional from nature with fully differentiable density
  functional theory},\ }\href {https://doi.org/10.1103/PhysRevLett.127.126403}
  {\bibfield  {journal} {\bibinfo  {journal} {Phys. Rev. Lett.}\ }\textbf
  {\bibinfo {volume} {127}},\ \bibinfo {pages} {126403} (\bibinfo {year}
  {2021})}\BibitemShut {NoStop}%
\bibitem [{\citenamefont {Wu}\ \emph {et~al.}(2023)\citenamefont {Wu},
  \citenamefont {Pun}, \citenamefont {Zheng},\ and\ \citenamefont
  {Chen}}]{Wu2023xcml}%
  \BibitemOpen
  \bibfield  {author} {\bibinfo {author} {\bibfnamefont {J.}~\bibnamefont
  {Wu}}, \bibinfo {author} {\bibfnamefont {S.-M.}\ \bibnamefont {Pun}},
  \bibinfo {author} {\bibfnamefont {X.}~\bibnamefont {Zheng}},\ and\ \bibinfo
  {author} {\bibfnamefont {G.}~\bibnamefont {Chen}},\ }\bibfield  {title}
  {\bibinfo {title} {Construct exchange-correlation functional via machine
  learning},\ }\href {https://doi.org/10.1063/5.0150587} {\bibfield  {journal}
  {\bibinfo  {journal} {The Journal of Chemical Physics}\ }\textbf {\bibinfo
  {volume} {159}},\ \bibinfo {pages} {090901} (\bibinfo {year}
  {2023})}\BibitemShut {NoStop}%
\bibitem [{\citenamefont {Tan}\ \emph {et~al.}(2023)\citenamefont {Tan},
  \citenamefont {Pickard},\ and\ \citenamefont {Witt}}]{tan2023professAD}%
  \BibitemOpen
  \bibfield  {author} {\bibinfo {author} {\bibfnamefont {C.~W.}\ \bibnamefont
  {Tan}}, \bibinfo {author} {\bibfnamefont {C.~J.}\ \bibnamefont {Pickard}},\
  and\ \bibinfo {author} {\bibfnamefont {W.~C.}\ \bibnamefont {Witt}},\
  }\bibfield  {title} {\bibinfo {title} {Automatic differentiation for
  orbital-free density functional theory},\ }\href
  {https://doi.org/10.1063/5.0138429} {\bibfield  {journal} {\bibinfo
  {journal} {The Journal of Chemical Physics}\ }\textbf {\bibinfo {volume}
  {158}},\ \bibinfo {pages} {124801} (\bibinfo {year} {2023})}\BibitemShut
  {NoStop}%
\bibitem [{\citenamefont {Li}\ \emph {et~al.}(2024{\natexlab{b}})\citenamefont
  {Li}, \citenamefont {Shi}, \citenamefont {Dale}, \citenamefont {Vignale},\
  and\ \citenamefont {Lin}}]{li2024jrystal}%
  \BibitemOpen
  \bibfield  {author} {\bibinfo {author} {\bibfnamefont {T.}~\bibnamefont
  {Li}}, \bibinfo {author} {\bibfnamefont {Z.}~\bibnamefont {Shi}}, \bibinfo
  {author} {\bibfnamefont {S.~G.}\ \bibnamefont {Dale}}, \bibinfo {author}
  {\bibfnamefont {G.}~\bibnamefont {Vignale}},\ and\ \bibinfo {author}
  {\bibfnamefont {M.}~\bibnamefont {Lin}},\ }\bibfield  {title} {\bibinfo
  {title} {{Jrystal}: A {JAX}-based differentiable density functional theory
  framework for materials},\ }in\ \href@noop {} {\emph {\bibinfo {booktitle}
  {Machine Learning and the Physical Sciences Workshop at NeurIPS 2024}}}\
  (\bibinfo {year} {2024})\BibitemShut {NoStop}%
\bibitem [{\citenamefont {Herbst}\ \emph {et~al.}(2021)\citenamefont {Herbst},
  \citenamefont {Levitt},\ and\ \citenamefont {Cancès}}]{DFTKpaper}%
  \BibitemOpen
  \bibfield  {author} {\bibinfo {author} {\bibfnamefont {M.~F.}\ \bibnamefont
  {Herbst}}, \bibinfo {author} {\bibfnamefont {A.}~\bibnamefont {Levitt}},\
  and\ \bibinfo {author} {\bibfnamefont {E.}~\bibnamefont {Cancès}},\
  }\bibfield  {title} {\bibinfo {title} {{DFTK}: A {Julian} approach for
  simulating electrons in solids},\ }\href
  {https://doi.org/10.21105/jcon.00069} {\bibfield  {journal} {\bibinfo
  {journal} {Proceedings of the JuliaCon Conference}\ }\textbf {\bibinfo
  {volume} {3}},\ \bibinfo {pages} {69} (\bibinfo {year} {2021})}\BibitemShut
  {NoStop}%
\bibitem [{\citenamefont {Moses}\ and\ \citenamefont
  {Churavy}(2020)}]{Moses2020}%
  \BibitemOpen
  \bibfield  {author} {\bibinfo {author} {\bibfnamefont {W.}~\bibnamefont
  {Moses}}\ and\ \bibinfo {author} {\bibfnamefont {V.}~\bibnamefont
  {Churavy}},\ }\bibfield  {title} {\bibinfo {title} {Instead of rewriting
  foreign code for machine learning, automatically synthesize fast gradients},\
  }in\ \href
  {https://proceedings.neurips.cc/paper\_files/paper/2020/file/9332c513ef44b682e9347822c2e457ac-Paper.pdf}
  {\emph {\bibinfo {booktitle} {Advances in Neural Information Processing
  Systems}}},\ Vol.~\bibinfo {volume} {33},\ \bibinfo {editor} {edited by\
  \bibinfo {editor} {\bibfnamefont {H.}~\bibnamefont {Larochelle}}, \bibinfo
  {editor} {\bibfnamefont {M.}~\bibnamefont {Ranzato}}, \bibinfo {editor}
  {\bibfnamefont {R.}~\bibnamefont {Hadsell}}, \bibinfo {editor} {\bibfnamefont
  {M.}~\bibnamefont {Balcan}},\ and\ \bibinfo {editor} {\bibfnamefont
  {H.}~\bibnamefont {Lin}}}\ (\bibinfo  {publisher} {Curran Associates, Inc.},\
  \bibinfo {year} {2020})\ pp.\ \bibinfo {pages} {12472--12485}\BibitemShut
  {NoStop}%
\bibitem [{\citenamefont {Innes}\ \emph {et~al.}(2019)\citenamefont {Innes},
  \citenamefont {Edelman}, \citenamefont {Fischer}, \citenamefont {Rackauckas},
  \citenamefont {Saba}, \citenamefont {Shah},\ and\ \citenamefont
  {Tebbutt}}]{innes2019differentiableprogrammingbridgemachine}%
  \BibitemOpen
  \bibfield  {author} {\bibinfo {author} {\bibfnamefont {M.}~\bibnamefont
  {Innes}}, \bibinfo {author} {\bibfnamefont {A.}~\bibnamefont {Edelman}},
  \bibinfo {author} {\bibfnamefont {K.}~\bibnamefont {Fischer}}, \bibinfo
  {author} {\bibfnamefont {C.}~\bibnamefont {Rackauckas}}, \bibinfo {author}
  {\bibfnamefont {E.}~\bibnamefont {Saba}}, \bibinfo {author} {\bibfnamefont
  {V.~B.}\ \bibnamefont {Shah}},\ and\ \bibinfo {author} {\bibfnamefont
  {W.}~\bibnamefont {Tebbutt}},\ }\href {https://arxiv.org/abs/1907.07587}
  {\bibinfo {title} {A differentiable programming system to bridge machine
  learning and scientific computing}} (\bibinfo {year} {2019}),\ \Eprint
  {https://arxiv.org/abs/1907.07587} {arXiv:1907.07587 [cs.PL]} \BibitemShut
  {NoStop}%
\bibitem [{\citenamefont {White}\ \emph {et~al.}(2025)\citenamefont {White},
  \citenamefont {Abbott}, \citenamefont {Revels}, \citenamefont {Zgubic},
  \citenamefont {Axen}, \citenamefont {Arslan}, \citenamefont {Schaub},
  \citenamefont {Robinson}, \citenamefont {Ma}, \citenamefont {Heim},
  \citenamefont {Rackauckas}, \citenamefont {Sam}, \citenamefont {Widmann},
  \citenamefont {Dhingra}, \citenamefont {Tebbutt}, \citenamefont {Schmitz},
  \citenamefont {Protter}, \citenamefont {Lucibello}, \citenamefont {Fischer},
  \citenamefont {Sajko}, \citenamefont {Heintzmann}, \citenamefont
  {frankschae}, \citenamefont {Noack}, \citenamefont {Smirnov}, \citenamefont
  {Zhabinski}, \citenamefont {Finnegan}, \citenamefont {mattBrzezinski},\ and\
  \citenamefont {Ling}}]{frames_white_2025_chainrules}%
  \BibitemOpen
  \bibfield  {author} {\bibinfo {author} {\bibfnamefont {F.}~\bibnamefont
  {White}}, \bibinfo {author} {\bibfnamefont {M.}~\bibnamefont {Abbott}},
  \bibinfo {author} {\bibfnamefont {J.}~\bibnamefont {Revels}}, \bibinfo
  {author} {\bibfnamefont {M.}~\bibnamefont {Zgubic}}, \bibinfo {author}
  {\bibfnamefont {S.}~\bibnamefont {Axen}}, \bibinfo {author} {\bibfnamefont
  {A.}~\bibnamefont {Arslan}}, \bibinfo {author} {\bibfnamefont {S.~D.}\
  \bibnamefont {Schaub}}, \bibinfo {author} {\bibfnamefont {N.}~\bibnamefont
  {Robinson}}, \bibinfo {author} {\bibfnamefont {Y.}~\bibnamefont {Ma}},
  \bibinfo {author} {\bibfnamefont {N.}~\bibnamefont {Heim}}, \bibinfo {author}
  {\bibfnamefont {C.}~\bibnamefont {Rackauckas}}, \bibinfo {author}
  {\bibnamefont {Sam}}, \bibinfo {author} {\bibfnamefont {D.}~\bibnamefont
  {Widmann}}, \bibinfo {author} {\bibfnamefont {G.}~\bibnamefont {Dhingra}},
  \bibinfo {author} {\bibfnamefont {W.}~\bibnamefont {Tebbutt}}, \bibinfo
  {author} {\bibfnamefont {N.}~\bibnamefont {Schmitz}}, \bibinfo {author}
  {\bibfnamefont {M.}~\bibnamefont {Protter}}, \bibinfo {author} {\bibfnamefont
  {C.}~\bibnamefont {Lucibello}}, \bibinfo {author} {\bibfnamefont
  {K.}~\bibnamefont {Fischer}}, \bibinfo {author} {\bibfnamefont
  {N.}~\bibnamefont {Sajko}}, \bibinfo {author} {\bibfnamefont
  {R.}~\bibnamefont {Heintzmann}}, \bibinfo {author} {\bibnamefont
  {frankschae}}, \bibinfo {author} {\bibfnamefont {A.}~\bibnamefont {Noack}},
  \bibinfo {author} {\bibfnamefont {A.}~\bibnamefont {Smirnov}}, \bibinfo
  {author} {\bibfnamefont {A.}~\bibnamefont {Zhabinski}}, \bibinfo {author}
  {\bibfnamefont {R.}~\bibnamefont {Finnegan}}, \bibinfo {author} {\bibnamefont
  {mattBrzezinski}},\ and\ \bibinfo {author} {\bibfnamefont {J.}~\bibnamefont
  {Ling}},\ }\href {https://doi.org/10.5281/zenodo.15658875} {\bibinfo {title}
  {{JuliaDiff}/{ChainRules}.jl: v1.72.5}} (\bibinfo {year} {2025})\BibitemShut
  {NoStop}%
\bibitem [{\citenamefont {Dalle}\ and\ \citenamefont
  {Hill}(2025)}]{dalle2025commoninterfaceautomaticdifferentiation}%
  \BibitemOpen
  \bibfield  {author} {\bibinfo {author} {\bibfnamefont {G.}~\bibnamefont
  {Dalle}}\ and\ \bibinfo {author} {\bibfnamefont {A.}~\bibnamefont {Hill}},\
  }\href {https://arxiv.org/abs/2505.05542} {\bibinfo {title} {A common
  interface for automatic differentiation}} (\bibinfo {year} {2025}),\ \Eprint
  {https://arxiv.org/abs/2505.05542} {arXiv:2505.05542 [cs.MS]} \BibitemShut
  {NoStop}%
\bibitem [{\citenamefont {Herbst}\ and\ \citenamefont
  {Sun}(2025)}]{herbst2025efficientkrylovmethodslinear}%
  \BibitemOpen
  \bibfield  {author} {\bibinfo {author} {\bibfnamefont {M.~F.}\ \bibnamefont
  {Herbst}}\ and\ \bibinfo {author} {\bibfnamefont {B.}~\bibnamefont {Sun}},\
  }\href {https://arxiv.org/abs/2505.02319} {\bibinfo {title} {Efficient
  {Krylov} methods for linear response in plane-wave electronic structure
  calculations}} (\bibinfo {year} {2025}),\ \Eprint
  {https://arxiv.org/abs/2505.02319} {arXiv:2505.02319 [math.NA]} \BibitemShut
  {NoStop}%
\bibitem [{\citenamefont {Herbst}\ \emph {et~al.}(2025)\citenamefont {Herbst},
  \citenamefont {Levitt},\ and\ \citenamefont
  {Cancès}}]{herbst_2025_15793538}%
  \BibitemOpen
  \bibfield  {author} {\bibinfo {author} {\bibfnamefont {M.~F.}\ \bibnamefont
  {Herbst}}, \bibinfo {author} {\bibfnamefont {A.}~\bibnamefont {Levitt}},\
  and\ \bibinfo {author} {\bibfnamefont {E.}~\bibnamefont {Cancès}},\ }\href
  {https://doi.org/10.5281/zenodo.15793538} {\bibinfo {title} {{DFTK}: The
  {Density}-functional toolkit, v0.7.16}} (\bibinfo {year} {2025}),\ \Eprint
  {https://arxiv.org/abs/15793538} {zenodo:15793538} \BibitemShut {NoStop}%
\bibitem [{\citenamefont {Franceschetti}\ and\ \citenamefont
  {Zunger}(1999)}]{Franceschetti1999}%
  \BibitemOpen
  \bibfield  {author} {\bibinfo {author} {\bibfnamefont {A.}~\bibnamefont
  {Franceschetti}}\ and\ \bibinfo {author} {\bibfnamefont {A.}~\bibnamefont
  {Zunger}},\ }\bibfield  {title} {\bibinfo {title} {The inverse band-structure
  problem of finding an atomic configuration with given electronic
  properties},\ }\href {https://doi.org/10.1038/46995} {\bibfield  {journal}
  {\bibinfo  {journal} {Nature}\ }\textbf {\bibinfo {volume} {402}},\ \bibinfo
  {pages} {60} (\bibinfo {year} {1999})}\BibitemShut {NoStop}%
\bibitem [{\citenamefont {Zunger}(2018)}]{Zunger2018}%
  \BibitemOpen
  \bibfield  {author} {\bibinfo {author} {\bibfnamefont {A.}~\bibnamefont
  {Zunger}},\ }\bibfield  {title} {\bibinfo {title} {Inverse design in search
  of materials with target functionalities},\ }\href
  {https://doi.org/10.1038/s41570-018-0121} {\bibfield  {journal} {\bibinfo
  {journal} {Nature Reviews Chemistry}\ }\textbf {\bibinfo {volume} {2}},\
  \bibinfo {pages} {0121} (\bibinfo {year} {2018})}\BibitemShut {NoStop}%
\bibitem [{\citenamefont {Noh}\ \emph {et~al.}(2020)\citenamefont {Noh},
  \citenamefont {Gu}, \citenamefont {Kim},\ and\ \citenamefont
  {Jung}}]{Noh2020}%
  \BibitemOpen
  \bibfield  {author} {\bibinfo {author} {\bibfnamefont {J.}~\bibnamefont
  {Noh}}, \bibinfo {author} {\bibfnamefont {G.~H.}\ \bibnamefont {Gu}},
  \bibinfo {author} {\bibfnamefont {S.}~\bibnamefont {Kim}},\ and\ \bibinfo
  {author} {\bibfnamefont {Y.}~\bibnamefont {Jung}},\ }\bibfield  {title}
  {\bibinfo {title} {Machine-enabled inverse design of inorganic solid
  materials: promises and challenges},\ }\href
  {https://doi.org/10.1039/d0sc00594k} {\bibfield  {journal} {\bibinfo
  {journal} {Chemical Science}\ }\textbf {\bibinfo {volume} {11}},\ \bibinfo
  {pages} {4871} (\bibinfo {year} {2020})}\BibitemShut {NoStop}%
\bibitem [{\citenamefont {Manasevit}\ \emph {et~al.}(1982)\citenamefont
  {Manasevit}, \citenamefont {Gergis},\ and\ \citenamefont
  {Jones}}]{Manasevit1982}%
  \BibitemOpen
  \bibfield  {author} {\bibinfo {author} {\bibfnamefont {H.~M.}\ \bibnamefont
  {Manasevit}}, \bibinfo {author} {\bibfnamefont {I.~S.}\ \bibnamefont
  {Gergis}},\ and\ \bibinfo {author} {\bibfnamefont {A.~B.}\ \bibnamefont
  {Jones}},\ }\bibfield  {title} {\bibinfo {title} {Electron mobility
  enhancement in epitaxial multilayer {Si-Si1-xGex} alloy films on (100)
  {Si}},\ }\href {https://doi.org/10.1063/1.93533} {\bibfield  {journal}
  {\bibinfo  {journal} {Applied Physics Letters}\ }\textbf {\bibinfo {volume}
  {41}},\ \bibinfo {pages} {464} (\bibinfo {year} {1982})}\BibitemShut
  {NoStop}%
\bibitem [{\citenamefont {Shi}\ \emph {et~al.}(2019)\citenamefont {Shi},
  \citenamefont {Tsymbalov}, \citenamefont {Dao}, \citenamefont {Suresh},
  \citenamefont {Shapeev},\ and\ \citenamefont {Li}}]{Shi2019}%
  \BibitemOpen
  \bibfield  {author} {\bibinfo {author} {\bibfnamefont {Z.}~\bibnamefont
  {Shi}}, \bibinfo {author} {\bibfnamefont {E.}~\bibnamefont {Tsymbalov}},
  \bibinfo {author} {\bibfnamefont {M.}~\bibnamefont {Dao}}, \bibinfo {author}
  {\bibfnamefont {S.}~\bibnamefont {Suresh}}, \bibinfo {author} {\bibfnamefont
  {A.}~\bibnamefont {Shapeev}},\ and\ \bibinfo {author} {\bibfnamefont
  {J.}~\bibnamefont {Li}},\ }\bibfield  {title} {\bibinfo {title} {Deep elastic
  strain engineering of bandgap through machine learning},\ }\href
  {https://doi.org/10.1073/pnas.1818555116} {\bibfield  {journal} {\bibinfo
  {journal} {Proceedings of the National Academy of Sciences}\ }\textbf
  {\bibinfo {volume} {116}},\ \bibinfo {pages} {4117} (\bibinfo {year}
  {2019})}\BibitemShut {NoStop}%
\bibitem [{\citenamefont {Sun}\ \emph {et~al.}(2007)\citenamefont {Sun},
  \citenamefont {Thompson},\ and\ \citenamefont {Nishida}}]{Sun2007}%
  \BibitemOpen
  \bibfield  {author} {\bibinfo {author} {\bibfnamefont {Y.}~\bibnamefont
  {Sun}}, \bibinfo {author} {\bibfnamefont {S.~E.}\ \bibnamefont {Thompson}},\
  and\ \bibinfo {author} {\bibfnamefont {T.}~\bibnamefont {Nishida}},\
  }\bibfield  {title} {\bibinfo {title} {Physics of strain effects in
  semiconductors and metal-oxide-semiconductor field-effect transistors},\
  }\href {https://doi.org/10.1063/1.2730561} {\bibfield  {journal} {\bibinfo
  {journal} {Journal of Applied Physics}\ }\textbf {\bibinfo {volume} {101}},\
  \bibinfo {pages} {104503} (\bibinfo {year} {2007})}\BibitemShut {NoStop}%
\bibitem [{\citenamefont {Ponc\'e}\ \emph {et~al.}(2019)\citenamefont
  {Ponc\'e}, \citenamefont {Jena},\ and\ \citenamefont {Giustino}}]{Ponce2019}%
  \BibitemOpen
  \bibfield  {author} {\bibinfo {author} {\bibfnamefont {S.}~\bibnamefont
  {Ponc\'e}}, \bibinfo {author} {\bibfnamefont {D.}~\bibnamefont {Jena}},\ and\
  \bibinfo {author} {\bibfnamefont {F.}~\bibnamefont {Giustino}},\ }\bibfield
  {title} {\bibinfo {title} {Hole mobility of strained gan from first
  principles},\ }\href {https://doi.org/10.1103/PhysRevB.100.085204} {\bibfield
   {journal} {\bibinfo  {journal} {Phys. Rev. B}\ }\textbf {\bibinfo {volume}
  {100}},\ \bibinfo {pages} {085204} (\bibinfo {year} {2019})}\BibitemShut
  {NoStop}%
\bibitem [{\citenamefont {Perdew}\ \emph {et~al.}(1996)\citenamefont {Perdew},
  \citenamefont {Burke},\ and\ \citenamefont {Ernzerhof}}]{Perdew1996}%
  \BibitemOpen
  \bibfield  {author} {\bibinfo {author} {\bibfnamefont {J.~P.}\ \bibnamefont
  {Perdew}}, \bibinfo {author} {\bibfnamefont {K.}~\bibnamefont {Burke}},\ and\
  \bibinfo {author} {\bibfnamefont {M.}~\bibnamefont {Ernzerhof}},\ }\bibfield
  {title} {\bibinfo {title} {Generalized gradient approximation made simple},\
  }\href {https://doi.org/10.1103/PhysRevLett.77.3865} {\bibfield  {journal}
  {\bibinfo  {journal} {Phys. Rev. Lett.}\ }\textbf {\bibinfo {volume} {77}},\
  \bibinfo {pages} {3865} (\bibinfo {year} {1996})}\BibitemShut {NoStop}%
\bibitem [{\citenamefont {Nocedal}\ and\ \citenamefont
  {Wright}(2006)}]{Nocedal2006Chapter6quasinewton}%
  \BibitemOpen
  \bibfield  {author} {\bibinfo {author} {\bibfnamefont {J.}~\bibnamefont
  {Nocedal}}\ and\ \bibinfo {author} {\bibfnamefont {S.~J.}\ \bibnamefont
  {Wright}},\ }\bibinfo {title} {Quasi-{Newton} methods},\ in\ \href
  {https://doi.org/10.1007/978-0-387-40065-5_6} {\emph {\bibinfo {booktitle}
  {Numerical Optimization}}}\ (\bibinfo  {publisher} {Springer},\ \bibinfo
  {address} {New York, NY},\ \bibinfo {year} {2006})\ pp.\ \bibinfo {pages}
  {135--163}\BibitemShut {NoStop}%
\bibitem [{\citenamefont {Lundgaard}\ \emph {et~al.}(2016)\citenamefont
  {Lundgaard}, \citenamefont {Wellendorff}, \citenamefont {Voss}, \citenamefont
  {Jacobsen},\ and\ \citenamefont {Bligaard}}]{lundgaard_mbeef-vdw_2016}%
  \BibitemOpen
  \bibfield  {author} {\bibinfo {author} {\bibfnamefont {K.~T.}\ \bibnamefont
  {Lundgaard}}, \bibinfo {author} {\bibfnamefont {J.}~\bibnamefont
  {Wellendorff}}, \bibinfo {author} {\bibfnamefont {J.}~\bibnamefont {Voss}},
  \bibinfo {author} {\bibfnamefont {K.~W.}\ \bibnamefont {Jacobsen}},\ and\
  \bibinfo {author} {\bibfnamefont {T.}~\bibnamefont {Bligaard}},\ }\bibfield
  {title} {\bibinfo {title} {{mBEEF}-{vdW}: {Robust} fitting of error
  estimation density functionals},\ }\href
  {https://doi.org/10.1103/PhysRevB.93.235162} {\bibfield  {journal} {\bibinfo
  {journal} {Phys. Rev. B}\ }\textbf {\bibinfo {volume} {93}},\ \bibinfo
  {pages} {235162} (\bibinfo {year} {2016})}\BibitemShut {NoStop}%
\bibitem [{\citenamefont {Perdew}\ \emph {et~al.}(2008)\citenamefont {Perdew},
  \citenamefont {Ruzsinszky}, \citenamefont {Csonka}, \citenamefont {Vydrov},
  \citenamefont {Scuseria}, \citenamefont {Constantin}, \citenamefont {Zhou},\
  and\ \citenamefont {Burke}}]{Perdew2008pbesol}%
  \BibitemOpen
  \bibfield  {author} {\bibinfo {author} {\bibfnamefont {J.~P.}\ \bibnamefont
  {Perdew}}, \bibinfo {author} {\bibfnamefont {A.}~\bibnamefont {Ruzsinszky}},
  \bibinfo {author} {\bibfnamefont {G.~I.}\ \bibnamefont {Csonka}}, \bibinfo
  {author} {\bibfnamefont {O.~A.}\ \bibnamefont {Vydrov}}, \bibinfo {author}
  {\bibfnamefont {G.~E.}\ \bibnamefont {Scuseria}}, \bibinfo {author}
  {\bibfnamefont {L.~A.}\ \bibnamefont {Constantin}}, \bibinfo {author}
  {\bibfnamefont {X.}~\bibnamefont {Zhou}},\ and\ \bibinfo {author}
  {\bibfnamefont {K.}~\bibnamefont {Burke}},\ }\bibfield  {title} {\bibinfo
  {title} {Restoring the density-gradient expansion for exchange in solids and
  surfaces},\ }\href {https://doi.org/10.1103/PhysRevLett.100.136406}
  {\bibfield  {journal} {\bibinfo  {journal} {Phys. Rev. Lett.}\ }\textbf
  {\bibinfo {volume} {100}},\ \bibinfo {pages} {136406} (\bibinfo {year}
  {2008})}\BibitemShut {NoStop}%
\bibitem [{\citenamefont {del Campo}\ \emph {et~al.}(2012)\citenamefont {del
  Campo}, \citenamefont {Gázquez}, \citenamefont {Trickey},\ and\
  \citenamefont {Vela}}]{delCampo2012pbemol}%
  \BibitemOpen
  \bibfield  {author} {\bibinfo {author} {\bibfnamefont {J.~M.}\ \bibnamefont
  {del Campo}}, \bibinfo {author} {\bibfnamefont {J.~L.}\ \bibnamefont
  {Gázquez}}, \bibinfo {author} {\bibfnamefont {S.~B.}\ \bibnamefont
  {Trickey}},\ and\ \bibinfo {author} {\bibfnamefont {A.}~\bibnamefont
  {Vela}},\ }\bibfield  {title} {\bibinfo {title} {Non-empirical improvement of
  {PBE} and its hybrid {PBE0} for general description of molecular
  properties},\ }\href {https://doi.org/10.1063/1.3691197} {\bibfield
  {journal} {\bibinfo  {journal} {The Journal of Chemical Physics}\ }\textbf
  {\bibinfo {volume} {136}},\ \bibinfo {pages} {104108} (\bibinfo {year}
  {2012})}\BibitemShut {NoStop}%
\bibitem [{\citenamefont {Xu}\ and\ \citenamefont
  {Goddard}(2004)}]{xu2004xpbe}%
  \BibitemOpen
  \bibfield  {author} {\bibinfo {author} {\bibfnamefont {X.}~\bibnamefont
  {Xu}}\ and\ \bibinfo {author} {\bibfnamefont {I.}~\bibnamefont {Goddard},
  \bibfnamefont {William~A.}},\ }\bibfield  {title} {\bibinfo {title} {The
  extended perdew-burke-ernzerhof functional with improved accuracy for
  thermodynamic and electronic properties of molecular systems},\ }\href
  {https://doi.org/10.1063/1.1771632} {\bibfield  {journal} {\bibinfo
  {journal} {The Journal of Chemical Physics}\ }\textbf {\bibinfo {volume}
  {121}},\ \bibinfo {pages} {4068} (\bibinfo {year} {2004})}\BibitemShut
  {NoStop}%
\bibitem [{\citenamefont {Constantin}\ \emph {et~al.}(2011)\citenamefont
  {Constantin}, \citenamefont {Fabiano}, \citenamefont {Laricchia},\ and\
  \citenamefont {Della~Sala}}]{Constantin2011apbe}%
  \BibitemOpen
  \bibfield  {author} {\bibinfo {author} {\bibfnamefont {L.~A.}\ \bibnamefont
  {Constantin}}, \bibinfo {author} {\bibfnamefont {E.}~\bibnamefont {Fabiano}},
  \bibinfo {author} {\bibfnamefont {S.}~\bibnamefont {Laricchia}},\ and\
  \bibinfo {author} {\bibfnamefont {F.}~\bibnamefont {Della~Sala}},\ }\bibfield
   {title} {\bibinfo {title} {Semiclassical neutral atom as a reference system
  in density functional theory},\ }\href
  {https://doi.org/10.1103/PhysRevLett.106.186406} {\bibfield  {journal}
  {\bibinfo  {journal} {Phys. Rev. Lett.}\ }\textbf {\bibinfo {volume} {106}},\
  \bibinfo {pages} {186406} (\bibinfo {year} {2011})}\BibitemShut {NoStop}%
\bibitem [{\citenamefont {Sarmiento-P{\'e}rez}\ \emph
  {et~al.}(2015)\citenamefont {Sarmiento-P{\'e}rez}, \citenamefont {Botti},\
  and\ \citenamefont {Marques}}]{Sarmiento2015pbefe}%
  \BibitemOpen
  \bibfield  {author} {\bibinfo {author} {\bibfnamefont {R.}~\bibnamefont
  {Sarmiento-P{\'e}rez}}, \bibinfo {author} {\bibfnamefont {S.}~\bibnamefont
  {Botti}},\ and\ \bibinfo {author} {\bibfnamefont {M.~A.~L.}\ \bibnamefont
  {Marques}},\ }\bibfield  {title} {\bibinfo {title} {Optimized exchange and
  correlation semilocal functional for the calculation of energies of
  formation},\ }\href {https://doi.org/10.1021/acs.jctc.5b00529} {\bibfield
  {journal} {\bibinfo  {journal} {Journal of Chemical Theory and Computation}\
  }\textbf {\bibinfo {volume} {11}},\ \bibinfo {pages} {3844} (\bibinfo {year}
  {2015})}\BibitemShut {NoStop}%
\bibitem [{\citenamefont {Mortensen}\ \emph {et~al.}(2005)\citenamefont
  {Mortensen}, \citenamefont {Kaasbjerg}, \citenamefont {Frederiksen},
  \citenamefont {N\o{}rskov}, \citenamefont {Sethna},\ and\ \citenamefont
  {Jacobsen}}]{Mortensen2005}%
  \BibitemOpen
  \bibfield  {author} {\bibinfo {author} {\bibfnamefont {J.~J.}\ \bibnamefont
  {Mortensen}}, \bibinfo {author} {\bibfnamefont {K.}~\bibnamefont
  {Kaasbjerg}}, \bibinfo {author} {\bibfnamefont {S.~L.}\ \bibnamefont
  {Frederiksen}}, \bibinfo {author} {\bibfnamefont {J.~K.}\ \bibnamefont
  {N\o{}rskov}}, \bibinfo {author} {\bibfnamefont {J.~P.}\ \bibnamefont
  {Sethna}},\ and\ \bibinfo {author} {\bibfnamefont {K.~W.}\ \bibnamefont
  {Jacobsen}},\ }\bibfield  {title} {\bibinfo {title} {Bayesian error
  estimation in density-functional theory},\ }\href
  {https://doi.org/10.1103/PhysRevLett.95.216401} {\bibfield  {journal}
  {\bibinfo  {journal} {Phys. Rev. Lett.}\ }\textbf {\bibinfo {volume} {95}},\
  \bibinfo {pages} {216401} (\bibinfo {year} {2005})}\BibitemShut {NoStop}%
\bibitem [{\citenamefont {Snyder}\ \emph {et~al.}(2012)\citenamefont {Snyder},
  \citenamefont {Rupp}, \citenamefont {Hansen}, \citenamefont {M\"uller},\ and\
  \citenamefont {Burke}}]{Snyder2012}%
  \BibitemOpen
  \bibfield  {author} {\bibinfo {author} {\bibfnamefont {J.~C.}\ \bibnamefont
  {Snyder}}, \bibinfo {author} {\bibfnamefont {M.}~\bibnamefont {Rupp}},
  \bibinfo {author} {\bibfnamefont {K.}~\bibnamefont {Hansen}}, \bibinfo
  {author} {\bibfnamefont {K.-R.}\ \bibnamefont {M\"uller}},\ and\ \bibinfo
  {author} {\bibfnamefont {K.}~\bibnamefont {Burke}},\ }\bibfield  {title}
  {\bibinfo {title} {Finding density functionals with machine learning},\
  }\href {https://doi.org/10.1103/PhysRevLett.108.253002} {\bibfield  {journal}
  {\bibinfo  {journal} {Phys. Rev. Lett.}\ }\textbf {\bibinfo {volume} {108}},\
  \bibinfo {pages} {253002} (\bibinfo {year} {2012})}\BibitemShut {NoStop}%
\bibitem [{\citenamefont {Bogojeski}\ \emph {et~al.}(2020)\citenamefont
  {Bogojeski}, \citenamefont {Vogt-Maranto}, \citenamefont {Tuckerman},
  \citenamefont {M{\"u}ller},\ and\ \citenamefont {Burke}}]{Bogojeski2020}%
  \BibitemOpen
  \bibfield  {author} {\bibinfo {author} {\bibfnamefont {M.}~\bibnamefont
  {Bogojeski}}, \bibinfo {author} {\bibfnamefont {L.}~\bibnamefont
  {Vogt-Maranto}}, \bibinfo {author} {\bibfnamefont {M.~E.}\ \bibnamefont
  {Tuckerman}}, \bibinfo {author} {\bibfnamefont {K.-R.}\ \bibnamefont
  {M{\"u}ller}},\ and\ \bibinfo {author} {\bibfnamefont {K.}~\bibnamefont
  {Burke}},\ }\bibfield  {title} {\bibinfo {title} {Quantum chemical accuracy
  from density functional approximations via machine learning},\ }\href
  {https://doi.org/10.1038/s41467-020-19093-1} {\bibfield  {journal} {\bibinfo
  {journal} {Nature Communications}\ }\textbf {\bibinfo {volume} {11}},\
  \bibinfo {pages} {5223} (\bibinfo {year} {2020})}\BibitemShut {NoStop}%
\bibitem [{\citenamefont {Kirkpatrick}\ \emph {et~al.}(2021)\citenamefont
  {Kirkpatrick}, \citenamefont {McMorrow}, \citenamefont {Turban},
  \citenamefont {Gaunt}, \citenamefont {Spencer}, \citenamefont {Matthews},
  \citenamefont {Obika}, \citenamefont {Thiry}, \citenamefont {Fortunato},
  \citenamefont {Pfau}, \citenamefont {Castellanos}, \citenamefont {Petersen},
  \citenamefont {Nelson}, \citenamefont {Kohli}, \citenamefont
  {Mori-S\'{a}nchez}, \citenamefont {Hassabis},\ and\ \citenamefont
  {Cohen}}]{Kirkpatrick2021}%
  \BibitemOpen
  \bibfield  {author} {\bibinfo {author} {\bibfnamefont {J.}~\bibnamefont
  {Kirkpatrick}}, \bibinfo {author} {\bibfnamefont {B.}~\bibnamefont
  {McMorrow}}, \bibinfo {author} {\bibfnamefont {D.~H.~P.}\ \bibnamefont
  {Turban}}, \bibinfo {author} {\bibfnamefont {A.~L.}\ \bibnamefont {Gaunt}},
  \bibinfo {author} {\bibfnamefont {J.~S.}\ \bibnamefont {Spencer}}, \bibinfo
  {author} {\bibfnamefont {A.~G. D.~G.}\ \bibnamefont {Matthews}}, \bibinfo
  {author} {\bibfnamefont {A.}~\bibnamefont {Obika}}, \bibinfo {author}
  {\bibfnamefont {L.}~\bibnamefont {Thiry}}, \bibinfo {author} {\bibfnamefont
  {M.}~\bibnamefont {Fortunato}}, \bibinfo {author} {\bibfnamefont
  {D.}~\bibnamefont {Pfau}}, \bibinfo {author} {\bibfnamefont {L.~R.}\
  \bibnamefont {Castellanos}}, \bibinfo {author} {\bibfnamefont
  {S.}~\bibnamefont {Petersen}}, \bibinfo {author} {\bibfnamefont {A.~W.~R.}\
  \bibnamefont {Nelson}}, \bibinfo {author} {\bibfnamefont {P.}~\bibnamefont
  {Kohli}}, \bibinfo {author} {\bibfnamefont {P.}~\bibnamefont
  {Mori-S\'{a}nchez}}, \bibinfo {author} {\bibfnamefont {D.}~\bibnamefont
  {Hassabis}},\ and\ \bibinfo {author} {\bibfnamefont {A.~J.}\ \bibnamefont
  {Cohen}},\ }\bibfield  {title} {\bibinfo {title} {Pushing the frontiers of
  density functionals by solving the fractional electron problem},\ }\href
  {https://doi.org/10.1126/science.abj6511} {\bibfield  {journal} {\bibinfo
  {journal} {Science}\ }\textbf {\bibinfo {volume} {374}},\ \bibinfo {pages}
  {1385} (\bibinfo {year} {2021})}\BibitemShut {NoStop}%
\bibitem [{\citenamefont {Bystrom}\ and\ \citenamefont
  {Kozinsky}(2022)}]{Bystrom2022}%
  \BibitemOpen
  \bibfield  {author} {\bibinfo {author} {\bibfnamefont {K.}~\bibnamefont
  {Bystrom}}\ and\ \bibinfo {author} {\bibfnamefont {B.}~\bibnamefont
  {Kozinsky}},\ }\bibfield  {title} {\bibinfo {title} {{CIDER}: An expressive,
  nonlocal feature set for machine learning density functionals with exact
  constraints},\ }\href {https://doi.org/10.1021/acs.jctc.1c00904} {\bibfield
  {journal} {\bibinfo  {journal} {Journal of Chemical Theory and Computation}\
  }\textbf {\bibinfo {volume} {18}},\ \bibinfo {pages} {2180} (\bibinfo {year}
  {2022})}\BibitemShut {NoStop}%
\bibitem [{\citenamefont {Bystrom}\ \emph {et~al.}(2024)\citenamefont
  {Bystrom}, \citenamefont {Falletta},\ and\ \citenamefont
  {Kozinsky}}]{Bystrom2024gaps}%
  \BibitemOpen
  \bibfield  {author} {\bibinfo {author} {\bibfnamefont {K.}~\bibnamefont
  {Bystrom}}, \bibinfo {author} {\bibfnamefont {S.}~\bibnamefont {Falletta}},\
  and\ \bibinfo {author} {\bibfnamefont {B.}~\bibnamefont {Kozinsky}},\
  }\bibfield  {title} {\bibinfo {title} {Training machine-learned density
  functionals on band gaps},\ }\href {https://doi.org/10.1021/acs.jctc.4c00999}
  {\bibfield  {journal} {\bibinfo  {journal} {Journal of Chemical Theory and
  Computation}\ }\textbf {\bibinfo {volume} {20}},\ \bibinfo {pages} {7516}
  (\bibinfo {year} {2024})}\BibitemShut {NoStop}%
\bibitem [{\citenamefont {Luise}\ \emph {et~al.}(2025)\citenamefont {Luise},
  \citenamefont {Huang}, \citenamefont {Vogels}, \citenamefont {Kooi},
  \citenamefont {Ehlert}, \citenamefont {Lanius}, \citenamefont {Giesbertz},
  \citenamefont {Karton}, \citenamefont {Gunceler}, \citenamefont {Stanley},
  \citenamefont {Bruinsma}, \citenamefont {Huang}, \citenamefont {Wei},
  \citenamefont {Torres}, \citenamefont {Katbashev}, \citenamefont {Zavaleta},
  \citenamefont {Máté}, \citenamefont {Kaba}, \citenamefont {Sordillo},
  \citenamefont {Chen}, \citenamefont {Williams-Young}, \citenamefont {Bishop},
  \citenamefont {Hermann}, \citenamefont {van~den Berg},\ and\ \citenamefont
  {Gori-Giorgi}}]{luise2025accurate}%
  \BibitemOpen
  \bibfield  {author} {\bibinfo {author} {\bibfnamefont {G.}~\bibnamefont
  {Luise}}, \bibinfo {author} {\bibfnamefont {C.-W.}\ \bibnamefont {Huang}},
  \bibinfo {author} {\bibfnamefont {T.}~\bibnamefont {Vogels}}, \bibinfo
  {author} {\bibfnamefont {D.~P.}\ \bibnamefont {Kooi}}, \bibinfo {author}
  {\bibfnamefont {S.}~\bibnamefont {Ehlert}}, \bibinfo {author} {\bibfnamefont
  {S.}~\bibnamefont {Lanius}}, \bibinfo {author} {\bibfnamefont {K.~J.~H.}\
  \bibnamefont {Giesbertz}}, \bibinfo {author} {\bibfnamefont {A.}~\bibnamefont
  {Karton}}, \bibinfo {author} {\bibfnamefont {D.}~\bibnamefont {Gunceler}},
  \bibinfo {author} {\bibfnamefont {M.}~\bibnamefont {Stanley}}, \bibinfo
  {author} {\bibfnamefont {W.~P.}\ \bibnamefont {Bruinsma}}, \bibinfo {author}
  {\bibfnamefont {L.}~\bibnamefont {Huang}}, \bibinfo {author} {\bibfnamefont
  {X.}~\bibnamefont {Wei}}, \bibinfo {author} {\bibfnamefont {J.~G.}\
  \bibnamefont {Torres}}, \bibinfo {author} {\bibfnamefont {A.}~\bibnamefont
  {Katbashev}}, \bibinfo {author} {\bibfnamefont {R.~C.}\ \bibnamefont
  {Zavaleta}}, \bibinfo {author} {\bibfnamefont {B.}~\bibnamefont {Máté}},
  \bibinfo {author} {\bibfnamefont {S.-O.}\ \bibnamefont {Kaba}}, \bibinfo
  {author} {\bibfnamefont {R.}~\bibnamefont {Sordillo}}, \bibinfo {author}
  {\bibfnamefont {Y.}~\bibnamefont {Chen}}, \bibinfo {author} {\bibfnamefont
  {D.~B.}\ \bibnamefont {Williams-Young}}, \bibinfo {author} {\bibfnamefont
  {C.~M.}\ \bibnamefont {Bishop}}, \bibinfo {author} {\bibfnamefont
  {J.}~\bibnamefont {Hermann}}, \bibinfo {author} {\bibfnamefont
  {R.}~\bibnamefont {van~den Berg}},\ and\ \bibinfo {author} {\bibfnamefont
  {P.}~\bibnamefont {Gori-Giorgi}},\ }\href {https://arxiv.org/abs/2506.14665}
  {\bibinfo {title} {Accurate and scalable exchange-correlation with deep
  learning}} (\bibinfo {year} {2025}),\ \Eprint
  {https://arxiv.org/abs/2506.14665} {arXiv:2506.14665 [physics.chem-ph]}
  \BibitemShut {NoStop}%
\bibitem [{\citenamefont {Dick}\ and\ \citenamefont
  {Fernandez-Serra}(2020)}]{Dick2020neuralxc}%
  \BibitemOpen
  \bibfield  {author} {\bibinfo {author} {\bibfnamefont {S.}~\bibnamefont
  {Dick}}\ and\ \bibinfo {author} {\bibfnamefont {M.}~\bibnamefont
  {Fernandez-Serra}},\ }\bibfield  {title} {\bibinfo {title} {Machine learning
  accurate exchange and correlation functionals of the electronic density},\
  }\href {https://doi.org/10.1038/s41467-020-17265-7} {\bibfield  {journal}
  {\bibinfo  {journal} {Nature Communications}\ }\textbf {\bibinfo {volume}
  {11}},\ \bibinfo {pages} {3509} (\bibinfo {year} {2020})}\BibitemShut
  {NoStop}%
\bibitem [{\citenamefont {Cuierrier}\ \emph {et~al.}(2021)\citenamefont
  {Cuierrier}, \citenamefont {Roy},\ and\ \citenamefont
  {Ernzerhof}}]{Cuierrier2021}%
  \BibitemOpen
  \bibfield  {author} {\bibinfo {author} {\bibfnamefont {E.}~\bibnamefont
  {Cuierrier}}, \bibinfo {author} {\bibfnamefont {P.-O.}\ \bibnamefont {Roy}},\
  and\ \bibinfo {author} {\bibfnamefont {M.}~\bibnamefont {Ernzerhof}},\
  }\bibfield  {title} {\bibinfo {title} {Constructing and representing
  exchange–correlation holes through artificial neural networks},\ }\href
  {https://doi.org/10.1063/5.0062940} {\bibfield  {journal} {\bibinfo
  {journal} {The Journal of Chemical Physics}\ }\textbf {\bibinfo {volume}
  {155}},\ \bibinfo {pages} {174121} (\bibinfo {year} {2021})}\BibitemShut
  {NoStop}%
\bibitem [{\citenamefont {Polak}\ \emph {et~al.}(2025)\citenamefont {Polak},
  \citenamefont {Zhao},\ and\ \citenamefont {Vuckovic}}]{Polak2025realspaceML}%
  \BibitemOpen
  \bibfield  {author} {\bibinfo {author} {\bibfnamefont {E.}~\bibnamefont
  {Polak}}, \bibinfo {author} {\bibfnamefont {H.}~\bibnamefont {Zhao}},\ and\
  \bibinfo {author} {\bibfnamefont {S.}~\bibnamefont {Vuckovic}},\ }\href
  {https://doi.org/10.26434/chemrxiv-2024-zk6hp-v3} {\bibinfo {title}
  {Real-space machine learning of correlation density functionals}} (\bibinfo
  {year} {2025})\BibitemShut {NoStop}%
\bibitem [{\citenamefont {Wellendorff}\ \emph {et~al.}(2012)\citenamefont
  {Wellendorff}, \citenamefont {Lundgaard}, \citenamefont {M\o{}gelh\o{}j},
  \citenamefont {Petzold}, \citenamefont {Landis}, \citenamefont {N\o{}rskov},
  \citenamefont {Bligaard},\ and\ \citenamefont
  {Jacobsen}}]{Wellendorff2012beefvdw}%
  \BibitemOpen
  \bibfield  {author} {\bibinfo {author} {\bibfnamefont {J.}~\bibnamefont
  {Wellendorff}}, \bibinfo {author} {\bibfnamefont {K.~T.}\ \bibnamefont
  {Lundgaard}}, \bibinfo {author} {\bibfnamefont {A.}~\bibnamefont
  {M\o{}gelh\o{}j}}, \bibinfo {author} {\bibfnamefont {V.}~\bibnamefont
  {Petzold}}, \bibinfo {author} {\bibfnamefont {D.~D.}\ \bibnamefont {Landis}},
  \bibinfo {author} {\bibfnamefont {J.~K.}\ \bibnamefont {N\o{}rskov}},
  \bibinfo {author} {\bibfnamefont {T.}~\bibnamefont {Bligaard}},\ and\
  \bibinfo {author} {\bibfnamefont {K.~W.}\ \bibnamefont {Jacobsen}},\
  }\bibfield  {title} {\bibinfo {title} {Density functionals for surface
  science: Exchange-correlation model development with bayesian error
  estimation},\ }\href {https://doi.org/10.1103/PhysRevB.85.235149} {\bibfield
  {journal} {\bibinfo  {journal} {Phys. Rev. B}\ }\textbf {\bibinfo {volume}
  {85}},\ \bibinfo {pages} {235149} (\bibinfo {year} {2012})}\BibitemShut
  {NoStop}%
\bibitem [{\citenamefont {Wellendorff}\ \emph {et~al.}(2014)\citenamefont
  {Wellendorff}, \citenamefont {Lundgaard}, \citenamefont {Jacobsen},\ and\
  \citenamefont {Bligaard}}]{Wellendorff2014}%
  \BibitemOpen
  \bibfield  {author} {\bibinfo {author} {\bibfnamefont {J.}~\bibnamefont
  {Wellendorff}}, \bibinfo {author} {\bibfnamefont {K.~T.}\ \bibnamefont
  {Lundgaard}}, \bibinfo {author} {\bibfnamefont {K.~W.}\ \bibnamefont
  {Jacobsen}},\ and\ \bibinfo {author} {\bibfnamefont {T.}~\bibnamefont
  {Bligaard}},\ }\bibfield  {title} {\bibinfo {title} {{mBEEF}: An accurate
  semi-local {Bayesian} error estimation density functional},\ }\href
  {https://doi.org/10.1063/1.4870397} {\bibfield  {journal} {\bibinfo
  {journal} {The Journal of Chemical Physics}\ }\textbf {\bibinfo {volume}
  {140}},\ \bibinfo {pages} {144107} (\bibinfo {year} {2014})}\BibitemShut
  {NoStop}%
\bibitem [{\citenamefont {Aldegunde}\ \emph {et~al.}(2016)\citenamefont
  {Aldegunde}, \citenamefont {Kermode},\ and\ \citenamefont
  {Zabaras}}]{Aldegunde2016}%
  \BibitemOpen
  \bibfield  {author} {\bibinfo {author} {\bibfnamefont {M.}~\bibnamefont
  {Aldegunde}}, \bibinfo {author} {\bibfnamefont {J.~R.}\ \bibnamefont
  {Kermode}},\ and\ \bibinfo {author} {\bibfnamefont {N.}~\bibnamefont
  {Zabaras}},\ }\bibfield  {title} {\bibinfo {title} {Development of an
  exchange\textendash{}correlation functional with uncertainty quantification
  capabilities for density functional theory},\ }\href
  {https://doi.org/10.1016/j.jcp.2016.01.034} {\bibfield  {journal} {\bibinfo
  {journal} {Journal of Computational Physics}\ }\textbf {\bibinfo {volume}
  {311}},\ \bibinfo {pages} {173} (\bibinfo {year} {2016})}\BibitemShut
  {NoStop}%
\bibitem [{\citenamefont {Kovács}\ \emph {et~al.}(2022)\citenamefont
  {Kovács}, \citenamefont {Tran}, \citenamefont {Blaha},\ and\ \citenamefont
  {Madsen}}]{Kovacs2022}%
  \BibitemOpen
  \bibfield  {author} {\bibinfo {author} {\bibfnamefont {P.}~\bibnamefont
  {Kovács}}, \bibinfo {author} {\bibfnamefont {F.}~\bibnamefont {Tran}},
  \bibinfo {author} {\bibfnamefont {P.}~\bibnamefont {Blaha}},\ and\ \bibinfo
  {author} {\bibfnamefont {G.~K.~H.}\ \bibnamefont {Madsen}},\ }\bibfield
  {title} {\bibinfo {title} {What is the optimal mgga exchange functional for
  solids?},\ }\href {https://doi.org/10.1063/5.0098787} {\bibfield  {journal}
  {\bibinfo  {journal} {The Journal of Chemical Physics}\ }\textbf {\bibinfo
  {volume} {157}},\ \bibinfo {pages} {094110} (\bibinfo {year}
  {2022})}\BibitemShut {NoStop}%
\bibitem [{\citenamefont {Borlido}\ \emph {et~al.}(2020)\citenamefont
  {Borlido}, \citenamefont {Schmidt}, \citenamefont {Huran}, \citenamefont
  {Tran}, \citenamefont {Marques},\ and\ \citenamefont {Botti}}]{Borlido2020}%
  \BibitemOpen
  \bibfield  {author} {\bibinfo {author} {\bibfnamefont {P.}~\bibnamefont
  {Borlido}}, \bibinfo {author} {\bibfnamefont {J.}~\bibnamefont {Schmidt}},
  \bibinfo {author} {\bibfnamefont {A.~W.}\ \bibnamefont {Huran}}, \bibinfo
  {author} {\bibfnamefont {F.}~\bibnamefont {Tran}}, \bibinfo {author}
  {\bibfnamefont {M.~A.~L.}\ \bibnamefont {Marques}},\ and\ \bibinfo {author}
  {\bibfnamefont {S.}~\bibnamefont {Botti}},\ }\bibfield  {title} {\bibinfo
  {title} {Exchange-correlation functionals for band gaps of solids: benchmark,
  reparametrization and machine learning},\ }\href
  {https://doi.org/10.1038/s41524-020-00360-0} {\bibfield  {journal} {\bibinfo
  {journal} {npj Computational Materials}\ }\textbf {\bibinfo {volume} {6}},\
  \bibinfo {pages} {96} (\bibinfo {year} {2020})}\BibitemShut {NoStop}%
\bibitem [{\citenamefont {Perdew}\ \emph {et~al.}(2009)\citenamefont {Perdew},
  \citenamefont {Ruzsinszky}, \citenamefont {Csonka}, \citenamefont
  {Constantin},\ and\ \citenamefont {Sun}}]{perdew2009revtpss}%
  \BibitemOpen
  \bibfield  {author} {\bibinfo {author} {\bibfnamefont {J.~P.}\ \bibnamefont
  {Perdew}}, \bibinfo {author} {\bibfnamefont {A.}~\bibnamefont {Ruzsinszky}},
  \bibinfo {author} {\bibfnamefont {G.~I.}\ \bibnamefont {Csonka}}, \bibinfo
  {author} {\bibfnamefont {L.~A.}\ \bibnamefont {Constantin}},\ and\ \bibinfo
  {author} {\bibfnamefont {J.}~\bibnamefont {Sun}},\ }\bibfield  {title}
  {\bibinfo {title} {Workhorse semilocal density functional for condensed
  matter physics and quantum chemistry},\ }\href
  {https://doi.org/10.1103/PhysRevLett.103.026403} {\bibfield  {journal}
  {\bibinfo  {journal} {Phys. Rev. Lett.}\ }\textbf {\bibinfo {volume} {103}},\
  \bibinfo {pages} {026403} (\bibinfo {year} {2009})}\BibitemShut {NoStop}%
\bibitem [{\citenamefont {Lebeda}\ \emph {et~al.}(2024)\citenamefont {Lebeda},
  \citenamefont {Aschebrock},\ and\ \citenamefont {K\"ummel}}]{Lebeda2024lak}%
  \BibitemOpen
  \bibfield  {author} {\bibinfo {author} {\bibfnamefont {T.}~\bibnamefont
  {Lebeda}}, \bibinfo {author} {\bibfnamefont {T.}~\bibnamefont {Aschebrock}},\
  and\ \bibinfo {author} {\bibfnamefont {S.}~\bibnamefont {K\"ummel}},\
  }\bibfield  {title} {\bibinfo {title} {Balancing the contributions to the
  gradient expansion: Accurate binding and band gaps with a nonempirical
  meta-gga},\ }\href {https://doi.org/10.1103/PhysRevLett.133.136402}
  {\bibfield  {journal} {\bibinfo  {journal} {Phys. Rev. Lett.}\ }\textbf
  {\bibinfo {volume} {133}},\ \bibinfo {pages} {136402} (\bibinfo {year}
  {2024})}\BibitemShut {NoStop}%
\bibitem [{\citenamefont {Bosoni}\ \emph {et~al.}(2023)\citenamefont {Bosoni},
  \citenamefont {Beal}, \citenamefont {Bercx}, \citenamefont {Blaha},
  \citenamefont {Bl\"{u}gel}, \citenamefont {Br\"{o}der}, \citenamefont
  {Callsen}, \citenamefont {Cottenier}, \citenamefont {Degomme}, \citenamefont
  {Dikan}, \citenamefont {Eimre}, \citenamefont {Flage-Larsen}, \citenamefont
  {Fornari}, \citenamefont {Garcia}, \citenamefont {Genovese}, \citenamefont
  {Giantomassi}, \citenamefont {Huber}, \citenamefont {Janssen}, \citenamefont
  {Kastlunger}, \citenamefont {Krack}, \citenamefont {Kresse}, \citenamefont
  {K\"{u}hne}, \citenamefont {Lejaeghere}, \citenamefont {Madsen},
  \citenamefont {Marsman}, \citenamefont {Marzari}, \citenamefont {Michalicek},
  \citenamefont {Mirhosseini}, \citenamefont {M\"{u}ller}, \citenamefont
  {Petretto}, \citenamefont {Pickard}, \citenamefont {Ponc\'{e}}, \citenamefont
  {Rignanese}, \citenamefont {Rubel}, \citenamefont {Ruh}, \citenamefont
  {Sluydts}, \citenamefont {Vanpoucke}, \citenamefont {Vijay}, \citenamefont
  {Wolloch}, \citenamefont {Wortmann}, \citenamefont {Yakutovich},
  \citenamefont {Yu}, \citenamefont {Zadoks}, \citenamefont {Zhu},\ and\
  \citenamefont {Pizzi}}]{Bosoni2023}%
  \BibitemOpen
  \bibfield  {author} {\bibinfo {author} {\bibfnamefont {E.}~\bibnamefont
  {Bosoni}}, \bibinfo {author} {\bibfnamefont {L.}~\bibnamefont {Beal}},
  \bibinfo {author} {\bibfnamefont {M.}~\bibnamefont {Bercx}}, \bibinfo
  {author} {\bibfnamefont {P.}~\bibnamefont {Blaha}}, \bibinfo {author}
  {\bibfnamefont {S.}~\bibnamefont {Bl\"{u}gel}}, \bibinfo {author}
  {\bibfnamefont {J.}~\bibnamefont {Br\"{o}der}}, \bibinfo {author}
  {\bibfnamefont {M.}~\bibnamefont {Callsen}}, \bibinfo {author} {\bibfnamefont
  {S.}~\bibnamefont {Cottenier}}, \bibinfo {author} {\bibfnamefont
  {A.}~\bibnamefont {Degomme}}, \bibinfo {author} {\bibfnamefont
  {V.}~\bibnamefont {Dikan}}, \bibinfo {author} {\bibfnamefont
  {K.}~\bibnamefont {Eimre}}, \bibinfo {author} {\bibfnamefont
  {E.}~\bibnamefont {Flage-Larsen}}, \bibinfo {author} {\bibfnamefont
  {M.}~\bibnamefont {Fornari}}, \bibinfo {author} {\bibfnamefont
  {A.}~\bibnamefont {Garcia}}, \bibinfo {author} {\bibfnamefont
  {L.}~\bibnamefont {Genovese}}, \bibinfo {author} {\bibfnamefont
  {M.}~\bibnamefont {Giantomassi}}, \bibinfo {author} {\bibfnamefont {S.~P.}\
  \bibnamefont {Huber}}, \bibinfo {author} {\bibfnamefont {H.}~\bibnamefont
  {Janssen}}, \bibinfo {author} {\bibfnamefont {G.}~\bibnamefont {Kastlunger}},
  \bibinfo {author} {\bibfnamefont {M.}~\bibnamefont {Krack}}, \bibinfo
  {author} {\bibfnamefont {G.}~\bibnamefont {Kresse}}, \bibinfo {author}
  {\bibfnamefont {T.~D.}\ \bibnamefont {K\"{u}hne}}, \bibinfo {author}
  {\bibfnamefont {K.}~\bibnamefont {Lejaeghere}}, \bibinfo {author}
  {\bibfnamefont {G.~K.~H.}\ \bibnamefont {Madsen}}, \bibinfo {author}
  {\bibfnamefont {M.}~\bibnamefont {Marsman}}, \bibinfo {author} {\bibfnamefont
  {N.}~\bibnamefont {Marzari}}, \bibinfo {author} {\bibfnamefont
  {G.}~\bibnamefont {Michalicek}}, \bibinfo {author} {\bibfnamefont
  {H.}~\bibnamefont {Mirhosseini}}, \bibinfo {author} {\bibfnamefont
  {T.~M.~A.}\ \bibnamefont {M\"{u}ller}}, \bibinfo {author} {\bibfnamefont
  {G.}~\bibnamefont {Petretto}}, \bibinfo {author} {\bibfnamefont {C.~J.}\
  \bibnamefont {Pickard}}, \bibinfo {author} {\bibfnamefont {S.}~\bibnamefont
  {Ponc\'{e}}}, \bibinfo {author} {\bibfnamefont {G.-M.}\ \bibnamefont
  {Rignanese}}, \bibinfo {author} {\bibfnamefont {O.}~\bibnamefont {Rubel}},
  \bibinfo {author} {\bibfnamefont {T.}~\bibnamefont {Ruh}}, \bibinfo {author}
  {\bibfnamefont {M.}~\bibnamefont {Sluydts}}, \bibinfo {author} {\bibfnamefont
  {D.~E.~P.}\ \bibnamefont {Vanpoucke}}, \bibinfo {author} {\bibfnamefont
  {S.}~\bibnamefont {Vijay}}, \bibinfo {author} {\bibfnamefont
  {M.}~\bibnamefont {Wolloch}}, \bibinfo {author} {\bibfnamefont
  {D.}~\bibnamefont {Wortmann}}, \bibinfo {author} {\bibfnamefont {A.~V.}\
  \bibnamefont {Yakutovich}}, \bibinfo {author} {\bibfnamefont
  {J.}~\bibnamefont {Yu}}, \bibinfo {author} {\bibfnamefont {A.}~\bibnamefont
  {Zadoks}}, \bibinfo {author} {\bibfnamefont {B.}~\bibnamefont {Zhu}},\ and\
  \bibinfo {author} {\bibfnamefont {G.}~\bibnamefont {Pizzi}},\ }\bibfield
  {title} {\bibinfo {title} {How to verify the precision of
  density-functional-theory implementations via reproducible and universal
  workflows},\ }\href {https://doi.org/10.1038/s42254-023-00655-3} {\bibfield
  {journal} {\bibinfo  {journal} {Nature Reviews Physics}\ }\textbf {\bibinfo
  {volume} {6}},\ \bibinfo {pages} {45} (\bibinfo {year} {2023})}\BibitemShut
  {NoStop}%
\bibitem [{\citenamefont
  {Ma{\'{z}}dziarz}(2024)}]{Mazdziarz2024incompatiblepseudo}%
  \BibitemOpen
  \bibfield  {author} {\bibinfo {author} {\bibfnamefont {M.}~\bibnamefont
  {Ma{\'{z}}dziarz}},\ }\bibfield  {title} {\bibinfo {title} {Uncertainty of
  dft calculated mechanical and structural properties of solids due to
  incompatibility of pseudopotentials and exchange--correlation functionals},\
  }\href {https://doi.org/10.1021/acs.jctc.4c01036} {\bibfield  {journal}
  {\bibinfo  {journal} {Journal of Chemical Theory and Computation}\ }\textbf
  {\bibinfo {volume} {20}},\ \bibinfo {pages} {9734} (\bibinfo {year}
  {2024})}\BibitemShut {NoStop}%
\bibitem [{\citenamefont {Lejaeghere}\ \emph {et~al.}(2016)\citenamefont
  {Lejaeghere}, \citenamefont {Bihlmayer}, \citenamefont {Bjorkman},
  \citenamefont {Blaha}, \citenamefont {Blugel}, \citenamefont {Blum},
  \citenamefont {Caliste}, \citenamefont {Castelli}, \citenamefont {Clark},
  \citenamefont {Dal~Corso}, \citenamefont {de~Gironcoli}, \citenamefont
  {Deutsch}, \citenamefont {Dewhurst}, \citenamefont {Di~Marco}, \citenamefont
  {Draxl}, \citenamefont {Du~ak}, \citenamefont {Eriksson}, \citenamefont
  {Flores-Livas}, \citenamefont {Garrity}, \citenamefont {Genovese},
  \citenamefont {Giannozzi}, \citenamefont {Giantomassi}, \citenamefont
  {Goedecker}, \citenamefont {Gonze}, \citenamefont {Granas}, \citenamefont
  {Gross}, \citenamefont {Gulans}, \citenamefont {Gygi}, \citenamefont
  {Hamann}, \citenamefont {Hasnip}, \citenamefont {Holzwarth}, \citenamefont
  {Iu~an}, \citenamefont {Jochym}, \citenamefont {Jollet}, \citenamefont
  {Jones}, \citenamefont {Kresse}, \citenamefont {Koepernik}, \citenamefont
  {Kucukbenli}, \citenamefont {Kvashnin}, \citenamefont {Locht}, \citenamefont
  {Lubeck}, \citenamefont {Marsman}, \citenamefont {Marzari}, \citenamefont
  {Nitzsche}, \citenamefont {Nordstrom}, \citenamefont {Ozaki}, \citenamefont
  {Paulatto}, \citenamefont {Pickard}, \citenamefont {Poelmans}, \citenamefont
  {Probert}, \citenamefont {Refson}, \citenamefont {Richter}, \citenamefont
  {Rignanese}, \citenamefont {Saha}, \citenamefont {Scheffler}, \citenamefont
  {Schlipf}, \citenamefont {Schwarz}, \citenamefont {Sharma}, \citenamefont
  {Tavazza}, \citenamefont {Thunstrom}, \citenamefont {Tkatchenko},
  \citenamefont {Torrent}, \citenamefont {Vanderbilt}, \citenamefont {van
  Setten}, \citenamefont {Van~Speybroeck}, \citenamefont {Wills}, \citenamefont
  {Yates}, \citenamefont {Zhang},\ and\ \citenamefont
  {Cottenier}}]{Lejaeghere2016}%
  \BibitemOpen
  \bibfield  {author} {\bibinfo {author} {\bibfnamefont {K.}~\bibnamefont
  {Lejaeghere}}, \bibinfo {author} {\bibfnamefont {G.}~\bibnamefont
  {Bihlmayer}}, \bibinfo {author} {\bibfnamefont {T.}~\bibnamefont {Bjorkman}},
  \bibinfo {author} {\bibfnamefont {P.}~\bibnamefont {Blaha}}, \bibinfo
  {author} {\bibfnamefont {S.}~\bibnamefont {Blugel}}, \bibinfo {author}
  {\bibfnamefont {V.}~\bibnamefont {Blum}}, \bibinfo {author} {\bibfnamefont
  {D.}~\bibnamefont {Caliste}}, \bibinfo {author} {\bibfnamefont {I.~E.}\
  \bibnamefont {Castelli}}, \bibinfo {author} {\bibfnamefont {S.~J.}\
  \bibnamefont {Clark}}, \bibinfo {author} {\bibfnamefont {A.}~\bibnamefont
  {Dal~Corso}}, \bibinfo {author} {\bibfnamefont {S.}~\bibnamefont
  {de~Gironcoli}}, \bibinfo {author} {\bibfnamefont {T.}~\bibnamefont
  {Deutsch}}, \bibinfo {author} {\bibfnamefont {J.~K.}\ \bibnamefont
  {Dewhurst}}, \bibinfo {author} {\bibfnamefont {I.}~\bibnamefont {Di~Marco}},
  \bibinfo {author} {\bibfnamefont {C.}~\bibnamefont {Draxl}}, \bibinfo
  {author} {\bibfnamefont {M.}~\bibnamefont {Du~ak}}, \bibinfo {author}
  {\bibfnamefont {O.}~\bibnamefont {Eriksson}}, \bibinfo {author}
  {\bibfnamefont {J.~A.}\ \bibnamefont {Flores-Livas}}, \bibinfo {author}
  {\bibfnamefont {K.~F.}\ \bibnamefont {Garrity}}, \bibinfo {author}
  {\bibfnamefont {L.}~\bibnamefont {Genovese}}, \bibinfo {author}
  {\bibfnamefont {P.}~\bibnamefont {Giannozzi}}, \bibinfo {author}
  {\bibfnamefont {M.}~\bibnamefont {Giantomassi}}, \bibinfo {author}
  {\bibfnamefont {S.}~\bibnamefont {Goedecker}}, \bibinfo {author}
  {\bibfnamefont {X.}~\bibnamefont {Gonze}}, \bibinfo {author} {\bibfnamefont
  {O.}~\bibnamefont {Granas}}, \bibinfo {author} {\bibfnamefont {E.~K.~U.}\
  \bibnamefont {Gross}}, \bibinfo {author} {\bibfnamefont {A.}~\bibnamefont
  {Gulans}}, \bibinfo {author} {\bibfnamefont {F.}~\bibnamefont {Gygi}},
  \bibinfo {author} {\bibfnamefont {D.~R.}\ \bibnamefont {Hamann}}, \bibinfo
  {author} {\bibfnamefont {P.~J.}\ \bibnamefont {Hasnip}}, \bibinfo {author}
  {\bibfnamefont {N.~A.~W.}\ \bibnamefont {Holzwarth}}, \bibinfo {author}
  {\bibfnamefont {D.}~\bibnamefont {Iu~an}}, \bibinfo {author} {\bibfnamefont
  {D.~B.}\ \bibnamefont {Jochym}}, \bibinfo {author} {\bibfnamefont
  {F.}~\bibnamefont {Jollet}}, \bibinfo {author} {\bibfnamefont
  {D.}~\bibnamefont {Jones}}, \bibinfo {author} {\bibfnamefont
  {G.}~\bibnamefont {Kresse}}, \bibinfo {author} {\bibfnamefont
  {K.}~\bibnamefont {Koepernik}}, \bibinfo {author} {\bibfnamefont
  {E.}~\bibnamefont {Kucukbenli}}, \bibinfo {author} {\bibfnamefont {Y.~O.}\
  \bibnamefont {Kvashnin}}, \bibinfo {author} {\bibfnamefont {I.~L.~M.}\
  \bibnamefont {Locht}}, \bibinfo {author} {\bibfnamefont {S.}~\bibnamefont
  {Lubeck}}, \bibinfo {author} {\bibfnamefont {M.}~\bibnamefont {Marsman}},
  \bibinfo {author} {\bibfnamefont {N.}~\bibnamefont {Marzari}}, \bibinfo
  {author} {\bibfnamefont {U.}~\bibnamefont {Nitzsche}}, \bibinfo {author}
  {\bibfnamefont {L.}~\bibnamefont {Nordstrom}}, \bibinfo {author}
  {\bibfnamefont {T.}~\bibnamefont {Ozaki}}, \bibinfo {author} {\bibfnamefont
  {L.}~\bibnamefont {Paulatto}}, \bibinfo {author} {\bibfnamefont {C.~J.}\
  \bibnamefont {Pickard}}, \bibinfo {author} {\bibfnamefont {W.}~\bibnamefont
  {Poelmans}}, \bibinfo {author} {\bibfnamefont {M.~I.~J.}\ \bibnamefont
  {Probert}}, \bibinfo {author} {\bibfnamefont {K.}~\bibnamefont {Refson}},
  \bibinfo {author} {\bibfnamefont {M.}~\bibnamefont {Richter}}, \bibinfo
  {author} {\bibfnamefont {G.-M.}\ \bibnamefont {Rignanese}}, \bibinfo {author}
  {\bibfnamefont {S.}~\bibnamefont {Saha}}, \bibinfo {author} {\bibfnamefont
  {M.}~\bibnamefont {Scheffler}}, \bibinfo {author} {\bibfnamefont
  {M.}~\bibnamefont {Schlipf}}, \bibinfo {author} {\bibfnamefont
  {K.}~\bibnamefont {Schwarz}}, \bibinfo {author} {\bibfnamefont
  {S.}~\bibnamefont {Sharma}}, \bibinfo {author} {\bibfnamefont
  {F.}~\bibnamefont {Tavazza}}, \bibinfo {author} {\bibfnamefont
  {P.}~\bibnamefont {Thunstrom}}, \bibinfo {author} {\bibfnamefont
  {A.}~\bibnamefont {Tkatchenko}}, \bibinfo {author} {\bibfnamefont
  {M.}~\bibnamefont {Torrent}}, \bibinfo {author} {\bibfnamefont
  {D.}~\bibnamefont {Vanderbilt}}, \bibinfo {author} {\bibfnamefont {M.~J.}\
  \bibnamefont {van Setten}}, \bibinfo {author} {\bibfnamefont
  {V.}~\bibnamefont {Van~Speybroeck}}, \bibinfo {author} {\bibfnamefont
  {J.~M.}\ \bibnamefont {Wills}}, \bibinfo {author} {\bibfnamefont {J.~R.}\
  \bibnamefont {Yates}}, \bibinfo {author} {\bibfnamefont {G.-X.}\ \bibnamefont
  {Zhang}},\ and\ \bibinfo {author} {\bibfnamefont {S.}~\bibnamefont
  {Cottenier}},\ }\bibfield  {title} {\bibinfo {title} {Reproducibility in
  density functional theory calculations of solids},\ }\href
  {https://doi.org/10.1126/science.aad3000} {\bibfield  {journal} {\bibinfo
  {journal} {Science}\ }\textbf {\bibinfo {volume} {351}},\ \bibinfo {pages}
  {aad3000} (\bibinfo {year} {2016})}\BibitemShut {NoStop}%
\bibitem [{\citenamefont {Prandini}\ \emph {et~al.}(2018)\citenamefont
  {Prandini}, \citenamefont {Marrazzo}, \citenamefont {Castelli}, \citenamefont
  {Mounet},\ and\ \citenamefont {Marzari}}]{Prandini2018}%
  \BibitemOpen
  \bibfield  {author} {\bibinfo {author} {\bibfnamefont {G.}~\bibnamefont
  {Prandini}}, \bibinfo {author} {\bibfnamefont {A.}~\bibnamefont {Marrazzo}},
  \bibinfo {author} {\bibfnamefont {I.~E.}\ \bibnamefont {Castelli}}, \bibinfo
  {author} {\bibfnamefont {N.}~\bibnamefont {Mounet}},\ and\ \bibinfo {author}
  {\bibfnamefont {N.}~\bibnamefont {Marzari}},\ }\bibfield  {title} {\bibinfo
  {title} {Precision and efficiency in solid-state pseudopotential
  calculations},\ }\href {https://doi.org/10.1038/s41524-018-0127-2} {\bibfield
   {journal} {\bibinfo  {journal} {npj Computational Materials}\ }\textbf
  {\bibinfo {volume} {4}},\ \bibinfo {pages} {72} (\bibinfo {year}
  {2018})}\BibitemShut {NoStop}%
\bibitem [{\citenamefont {Hamann}(2013)}]{hamann_optimized_2013}%
  \BibitemOpen
  \bibfield  {author} {\bibinfo {author} {\bibfnamefont {D.~R.}\ \bibnamefont
  {Hamann}},\ }\bibfield  {title} {\bibinfo {title} {Optimized norm-conserving
  {Vanderbilt} pseudopotentials},\ }\href
  {https://doi.org/10.1103/PhysRevB.88.085117} {\bibfield  {journal} {\bibinfo
  {journal} {Phys. Rev. B}\ }\textbf {\bibinfo {volume} {88}},\ \bibinfo
  {pages} {085117} (\bibinfo {year} {2013})}\BibitemShut {NoStop}%
\bibitem [{\citenamefont {Schlipf}\ and\ \citenamefont {Gygi}(2015)}]{SG15}%
  \BibitemOpen
  \bibfield  {author} {\bibinfo {author} {\bibfnamefont {M.}~\bibnamefont
  {Schlipf}}\ and\ \bibinfo {author} {\bibfnamefont {F.}~\bibnamefont {Gygi}},\
  }\bibfield  {title} {\bibinfo {title} {Optimization algorithm for the
  generation of {ONCV} pseudopotentials},\ }\href
  {https://doi.org/10.1016/j.cpc.2015.05.011} {\bibfield  {journal} {\bibinfo
  {journal} {Computer Physics Communications}\ }\textbf {\bibinfo {volume}
  {196}},\ \bibinfo {pages} {36} (\bibinfo {year} {2015})}\BibitemShut
  {NoStop}%
\bibitem [{\citenamefont {van Setten}\ \emph {et~al.}(2018)\citenamefont {van
  Setten}, \citenamefont {Giantomassi}, \citenamefont {Bousquet}, \citenamefont
  {Verstraete}, \citenamefont {Hamann}, \citenamefont {Gonze},\ and\
  \citenamefont {Rignanese}}]{VanSetten2018}%
  \BibitemOpen
  \bibfield  {author} {\bibinfo {author} {\bibfnamefont {M.}~\bibnamefont {van
  Setten}}, \bibinfo {author} {\bibfnamefont {M.}~\bibnamefont {Giantomassi}},
  \bibinfo {author} {\bibfnamefont {E.}~\bibnamefont {Bousquet}}, \bibinfo
  {author} {\bibfnamefont {M.}~\bibnamefont {Verstraete}}, \bibinfo {author}
  {\bibfnamefont {D.}~\bibnamefont {Hamann}}, \bibinfo {author} {\bibfnamefont
  {X.}~\bibnamefont {Gonze}},\ and\ \bibinfo {author} {\bibfnamefont {G.-M.}\
  \bibnamefont {Rignanese}},\ }\bibfield  {title} {\bibinfo {title} {The
  {PseudoDojo}: Training and grading a 85 element optimized norm-conserving
  pseudopotential table},\ }\href {https://doi.org/10.1016/j.cpc.2018.01.012}
  {\bibfield  {journal} {\bibinfo  {journal} {Computer Physics Communications}\
  }\textbf {\bibinfo {volume} {226}},\ \bibinfo {pages} {39} (\bibinfo {year}
  {2018})}\BibitemShut {NoStop}%
\bibitem [{\citenamefont {Shojaei}\ \emph {et~al.}(2023)\citenamefont
  {Shojaei}, \citenamefont {Pask}, \citenamefont {Medford},\ and\ \citenamefont
  {Suryanarayana}}]{SHOJAEI2023multiobjectiveONCV}%
  \BibitemOpen
  \bibfield  {author} {\bibinfo {author} {\bibfnamefont {M.~F.}\ \bibnamefont
  {Shojaei}}, \bibinfo {author} {\bibfnamefont {J.~E.}\ \bibnamefont {Pask}},
  \bibinfo {author} {\bibfnamefont {A.~J.}\ \bibnamefont {Medford}},\ and\
  \bibinfo {author} {\bibfnamefont {P.}~\bibnamefont {Suryanarayana}},\
  }\bibfield  {title} {\bibinfo {title} {Soft and transferable pseudopotentials
  from multi-objective optimization},\ }\href
  {https://doi.org/https://doi.org/10.1016/j.cpc.2022.108594} {\bibfield
  {journal} {\bibinfo  {journal} {Computer Physics Communications}\ }\textbf
  {\bibinfo {volume} {283}},\ \bibinfo {pages} {108594} (\bibinfo {year}
  {2023})}\BibitemShut {NoStop}%
\bibitem [{\citenamefont {Hansen}\ \emph {et~al.}(2025)\citenamefont {Hansen},
  \citenamefont {Mortensen}, \citenamefont {Bligaard},\ and\ \citenamefont
  {Jacobsen}}]{hansen_uncertainty-aware_2025}%
  \BibitemOpen
  \bibfield  {author} {\bibinfo {author} {\bibfnamefont {T.}~\bibnamefont
  {Hansen}}, \bibinfo {author} {\bibfnamefont {J.~J.}\ \bibnamefont
  {Mortensen}}, \bibinfo {author} {\bibfnamefont {T.}~\bibnamefont
  {Bligaard}},\ and\ \bibinfo {author} {\bibfnamefont {K.~W.}\ \bibnamefont
  {Jacobsen}},\ }\bibfield  {title} {\bibinfo {title} {Uncertainty-aware
  electronic density-functional distributions},\ }\href
  {https://doi.org/10.1103/yhly-wxhv} {\bibfield  {journal} {\bibinfo
  {journal} {Phys. Rev. B}\ }\textbf {\bibinfo {volume} {112}},\ \bibinfo
  {pages} {075412} (\bibinfo {year} {2025})}\BibitemShut {NoStop}%
\bibitem [{\citenamefont {Canc\`{e}s}\ \emph {et~al.}(2022)\citenamefont
  {Canc\`{e}s}, \citenamefont {Dusson}, \citenamefont {Kemlin},\ and\
  \citenamefont {Levitt}}]{Cances2022}%
  \BibitemOpen
  \bibfield  {author} {\bibinfo {author} {\bibfnamefont {E.}~\bibnamefont
  {Canc\`{e}s}}, \bibinfo {author} {\bibfnamefont {G.}~\bibnamefont {Dusson}},
  \bibinfo {author} {\bibfnamefont {G.}~\bibnamefont {Kemlin}},\ and\ \bibinfo
  {author} {\bibfnamefont {A.}~\bibnamefont {Levitt}},\ }\bibfield  {title}
  {\bibinfo {title} {Practical error bounds for properties in plane-wave
  electronic structure calculations},\ }\href
  {https://doi.org/10.1137/21m1456224} {\bibfield  {journal} {\bibinfo
  {journal} {SIAM Journal on Scientific Computing}\ }\textbf {\bibinfo {volume}
  {44}},\ \bibinfo {pages} {B1312} (\bibinfo {year} {2022})}\BibitemShut
  {NoStop}%
\bibitem [{\citenamefont {Mazitov}\ \emph {et~al.}(2025)\citenamefont
  {Mazitov}, \citenamefont {Bigi}, \citenamefont {Kellner}, \citenamefont
  {Pegolo}, \citenamefont {Tisi}, \citenamefont {Fraux}, \citenamefont
  {Pozdnyakov}, \citenamefont {Loche},\ and\ \citenamefont
  {Ceriotti}}]{mazitov2025petmaduniversalinteratomicpotential}%
  \BibitemOpen
  \bibfield  {author} {\bibinfo {author} {\bibfnamefont {A.}~\bibnamefont
  {Mazitov}}, \bibinfo {author} {\bibfnamefont {F.}~\bibnamefont {Bigi}},
  \bibinfo {author} {\bibfnamefont {M.}~\bibnamefont {Kellner}}, \bibinfo
  {author} {\bibfnamefont {P.}~\bibnamefont {Pegolo}}, \bibinfo {author}
  {\bibfnamefont {D.}~\bibnamefont {Tisi}}, \bibinfo {author} {\bibfnamefont
  {G.}~\bibnamefont {Fraux}}, \bibinfo {author} {\bibfnamefont
  {S.}~\bibnamefont {Pozdnyakov}}, \bibinfo {author} {\bibfnamefont
  {P.}~\bibnamefont {Loche}},\ and\ \bibinfo {author} {\bibfnamefont
  {M.}~\bibnamefont {Ceriotti}},\ }\href {https://arxiv.org/abs/2503.14118}
  {\bibinfo {title} {{PET}-{MAD}, a universal interatomic potential for
  advanced materials modeling}} (\bibinfo {year} {2025}),\ \Eprint
  {https://arxiv.org/abs/2503.14118} {arXiv:2503.14118 [cond-mat.mtrl-sci]}
  \BibitemShut {NoStop}%
\bibitem [{\citenamefont {Blondel}\ and\ \citenamefont
  {Roulet}(2024)}]{blondel_elements_2024}%
  \BibitemOpen
  \bibfield  {author} {\bibinfo {author} {\bibfnamefont {M.}~\bibnamefont
  {Blondel}}\ and\ \bibinfo {author} {\bibfnamefont {V.}~\bibnamefont
  {Roulet}},\ }\href {http://arxiv.org/abs/2403.14606} {\bibinfo {title} {The
  {Elements} of {Differentiable} {Programming}}} (\bibinfo {year} {2024}),\
  \bibinfo {note} {arXiv:2403.14606 [cs]}\BibitemShut {NoStop}%
\bibitem [{\citenamefont {Togo}\ \emph {et~al.}(2024)\citenamefont {Togo},
  \citenamefont {Shinohara},\ and\ \citenamefont {Tanaka}}]{togo_spglib_2024}%
  \BibitemOpen
  \bibfield  {author} {\bibinfo {author} {\bibfnamefont {A.}~\bibnamefont
  {Togo}}, \bibinfo {author} {\bibfnamefont {K.}~\bibnamefont {Shinohara}},\
  and\ \bibinfo {author} {\bibfnamefont {I.}~\bibnamefont {Tanaka}},\
  }\bibfield  {title} {\bibinfo {title} {Spglib: a software library for crystal
  symmetry search},\ }\href {https://doi.org/10.1080/27660400.2024.2384822}
  {\bibfield  {journal} {\bibinfo  {journal} {Science and Technology of
  Advanced Materials: Methods}\ }\textbf {\bibinfo {volume} {4}},\ \bibinfo
  {pages} {2384822} (\bibinfo {year} {2024})}\BibitemShut {NoStop}%
\bibitem [{\citenamefont {Knyazev}(2001)}]{knyazev2001lobpcg}%
  \BibitemOpen
  \bibfield  {author} {\bibinfo {author} {\bibfnamefont {A.~V.}\ \bibnamefont
  {Knyazev}},\ }\bibfield  {title} {\bibinfo {title} {Toward the optimal
  preconditioned eigensolver: Locally optimal block preconditioned conjugate
  gradient method},\ }\href {https://doi.org/10.1137/S1064827500366124}
  {\bibfield  {journal} {\bibinfo  {journal} {SIAM Journal on Scientific
  Computing}\ }\textbf {\bibinfo {volume} {23}},\ \bibinfo {pages} {517}
  (\bibinfo {year} {2001})},\ \Eprint
  {https://arxiv.org/abs/https://doi.org/10.1137/S1064827500366124}
  {https://doi.org/10.1137/S1064827500366124} \BibitemShut {NoStop}%
\bibitem [{\citenamefont {Walker}\ and\ \citenamefont
  {Ni}(2011)}]{walker2011anderson_gmres}%
  \BibitemOpen
  \bibfield  {author} {\bibinfo {author} {\bibfnamefont {H.~F.}\ \bibnamefont
  {Walker}}\ and\ \bibinfo {author} {\bibfnamefont {P.}~\bibnamefont {Ni}},\
  }\bibfield  {title} {\bibinfo {title} {Anderson acceleration for fixed-point
  iterations},\ }\href {https://doi.org/10.1137/10078356X} {\bibfield
  {journal} {\bibinfo  {journal} {SIAM Journal on Numerical Analysis}\ }\textbf
  {\bibinfo {volume} {49}},\ \bibinfo {pages} {1715} (\bibinfo {year}
  {2011})},\ \Eprint {https://arxiv.org/abs/https://doi.org/10.1137/10078356X}
  {https://doi.org/10.1137/10078356X} \BibitemShut {NoStop}%
\bibitem [{\citenamefont {Anderson}\ \emph {et~al.}(1999)\citenamefont
  {Anderson}, \citenamefont {Bai}, \citenamefont {Bischof}, \citenamefont
  {Blackford}, \citenamefont {Demmel}, \citenamefont {Dongarra}, \citenamefont
  {Du~Croz}, \citenamefont {Greenbaum}, \citenamefont {Hammarling},
  \citenamefont {McKenney},\ and\ \citenamefont {Sorensen}}]{laug}%
  \BibitemOpen
  \bibfield  {author} {\bibinfo {author} {\bibfnamefont {E.}~\bibnamefont
  {Anderson}}, \bibinfo {author} {\bibfnamefont {Z.}~\bibnamefont {Bai}},
  \bibinfo {author} {\bibfnamefont {C.}~\bibnamefont {Bischof}}, \bibinfo
  {author} {\bibfnamefont {S.}~\bibnamefont {Blackford}}, \bibinfo {author}
  {\bibfnamefont {J.}~\bibnamefont {Demmel}}, \bibinfo {author} {\bibfnamefont
  {J.}~\bibnamefont {Dongarra}}, \bibinfo {author} {\bibfnamefont
  {J.}~\bibnamefont {Du~Croz}}, \bibinfo {author} {\bibfnamefont
  {A.}~\bibnamefont {Greenbaum}}, \bibinfo {author} {\bibfnamefont
  {S.}~\bibnamefont {Hammarling}}, \bibinfo {author} {\bibfnamefont
  {A.}~\bibnamefont {McKenney}},\ and\ \bibinfo {author} {\bibfnamefont
  {D.}~\bibnamefont {Sorensen}},\ }\href@noop {} {\emph {\bibinfo {title}
  {{LAPACK} Users' Guide}}},\ \bibinfo {edition} {3rd}\ ed.\ (\bibinfo
  {publisher} {Society for Industrial and Applied Mathematics},\ \bibinfo
  {address} {Philadelphia, PA},\ \bibinfo {year} {1999})\BibitemShut {NoStop}%
\bibitem [{\citenamefont {Larsen}\ \emph {et~al.}(2017)\citenamefont {Larsen},
  \citenamefont {Mortensen}, \citenamefont {Blomqvist}, \citenamefont
  {Castelli}, \citenamefont {Christensen}, \citenamefont {Dułak},
  \citenamefont {Friis}, \citenamefont {Groves}, \citenamefont {Hammer},
  \citenamefont {Hargus}, \citenamefont {Hermes}, \citenamefont {Jennings},
  \citenamefont {Jensen}, \citenamefont {Kermode}, \citenamefont {Kitchin},
  \citenamefont {Kolsbjerg}, \citenamefont {Kubal}, \citenamefont {Kaasbjerg},
  \citenamefont {Lysgaard}, \citenamefont {Maronsson}, \citenamefont {Maxson},
  \citenamefont {Olsen}, \citenamefont {Pastewka}, \citenamefont {Peterson},
  \citenamefont {Rostgaard}, \citenamefont {Schiøtz}, \citenamefont {Schütt},
  \citenamefont {Strange}, \citenamefont {Thygesen}, \citenamefont {Vegge},
  \citenamefont {Vilhelmsen}, \citenamefont {Walter}, \citenamefont {Zeng},\
  and\ \citenamefont {Jacobsen}}]{ase-paper}%
  \BibitemOpen
  \bibfield  {author} {\bibinfo {author} {\bibfnamefont {A.~H.}\ \bibnamefont
  {Larsen}}, \bibinfo {author} {\bibfnamefont {J.~J.}\ \bibnamefont
  {Mortensen}}, \bibinfo {author} {\bibfnamefont {J.}~\bibnamefont
  {Blomqvist}}, \bibinfo {author} {\bibfnamefont {I.~E.}\ \bibnamefont
  {Castelli}}, \bibinfo {author} {\bibfnamefont {R.}~\bibnamefont
  {Christensen}}, \bibinfo {author} {\bibfnamefont {M.}~\bibnamefont {Dułak}},
  \bibinfo {author} {\bibfnamefont {J.}~\bibnamefont {Friis}}, \bibinfo
  {author} {\bibfnamefont {M.~N.}\ \bibnamefont {Groves}}, \bibinfo {author}
  {\bibfnamefont {B.}~\bibnamefont {Hammer}}, \bibinfo {author} {\bibfnamefont
  {C.}~\bibnamefont {Hargus}}, \bibinfo {author} {\bibfnamefont {E.~D.}\
  \bibnamefont {Hermes}}, \bibinfo {author} {\bibfnamefont {P.~C.}\
  \bibnamefont {Jennings}}, \bibinfo {author} {\bibfnamefont {P.~B.}\
  \bibnamefont {Jensen}}, \bibinfo {author} {\bibfnamefont {J.}~\bibnamefont
  {Kermode}}, \bibinfo {author} {\bibfnamefont {J.~R.}\ \bibnamefont
  {Kitchin}}, \bibinfo {author} {\bibfnamefont {E.~L.}\ \bibnamefont
  {Kolsbjerg}}, \bibinfo {author} {\bibfnamefont {J.}~\bibnamefont {Kubal}},
  \bibinfo {author} {\bibfnamefont {K.}~\bibnamefont {Kaasbjerg}}, \bibinfo
  {author} {\bibfnamefont {S.}~\bibnamefont {Lysgaard}}, \bibinfo {author}
  {\bibfnamefont {J.~B.}\ \bibnamefont {Maronsson}}, \bibinfo {author}
  {\bibfnamefont {T.}~\bibnamefont {Maxson}}, \bibinfo {author} {\bibfnamefont
  {T.}~\bibnamefont {Olsen}}, \bibinfo {author} {\bibfnamefont
  {L.}~\bibnamefont {Pastewka}}, \bibinfo {author} {\bibfnamefont
  {A.}~\bibnamefont {Peterson}}, \bibinfo {author} {\bibfnamefont
  {C.}~\bibnamefont {Rostgaard}}, \bibinfo {author} {\bibfnamefont
  {J.}~\bibnamefont {Schiøtz}}, \bibinfo {author} {\bibfnamefont
  {O.}~\bibnamefont {Schütt}}, \bibinfo {author} {\bibfnamefont
  {M.}~\bibnamefont {Strange}}, \bibinfo {author} {\bibfnamefont {K.~S.}\
  \bibnamefont {Thygesen}}, \bibinfo {author} {\bibfnamefont {T.}~\bibnamefont
  {Vegge}}, \bibinfo {author} {\bibfnamefont {L.}~\bibnamefont {Vilhelmsen}},
  \bibinfo {author} {\bibfnamefont {M.}~\bibnamefont {Walter}}, \bibinfo
  {author} {\bibfnamefont {Z.}~\bibnamefont {Zeng}},\ and\ \bibinfo {author}
  {\bibfnamefont {K.~W.}\ \bibnamefont {Jacobsen}},\ }\bibfield  {title}
  {\bibinfo {title} {The atomic simulation environment—a {Python} library for
  working with atoms},\ }\href
  {http://stacks.iop.org/0953-8984/29/i=27/a=273002} {\bibfield  {journal}
  {\bibinfo  {journal} {Journal of Physics: Condensed Matter}\ }\textbf
  {\bibinfo {volume} {29}},\ \bibinfo {pages} {273002} (\bibinfo {year}
  {2017})}\BibitemShut {NoStop}%
\bibitem [{\citenamefont {Mogensen}\ and\ \citenamefont
  {Riseth}(2018)}]{mogensen2018optim}%
  \BibitemOpen
  \bibfield  {author} {\bibinfo {author} {\bibfnamefont {P.~K.}\ \bibnamefont
  {Mogensen}}\ and\ \bibinfo {author} {\bibfnamefont {A.~N.}\ \bibnamefont
  {Riseth}},\ }\bibfield  {title} {\bibinfo {title} {Optim: A mathematical
  optimization package for {Julia}},\ }\href
  {https://doi.org/10.21105/joss.00615} {\bibfield  {journal} {\bibinfo
  {journal} {Journal of Open Source Software}\ }\textbf {\bibinfo {volume}
  {3}},\ \bibinfo {pages} {615} (\bibinfo {year} {2018})}\BibitemShut {NoStop}%
\bibitem [{\citenamefont {Canc\`{e}s}\ \emph
  {et~al.}(2023{\natexlab{b}})\citenamefont {Canc\`{e}s}, \citenamefont
  {Hassan},\ and\ \citenamefont {Vidal}}]{Cances2023modif}%
  \BibitemOpen
  \bibfield  {author} {\bibinfo {author} {\bibfnamefont {E.}~\bibnamefont
  {Canc\`{e}s}}, \bibinfo {author} {\bibfnamefont {M.}~\bibnamefont {Hassan}},\
  and\ \bibinfo {author} {\bibfnamefont {L.}~\bibnamefont {Vidal}},\ }\bibfield
   {title} {\bibinfo {title} {Modified-operator method for the calculation of
  band diagrams of crystalline materials},\ }\href
  {https://doi.org/10.1090/mcom/3897} {\bibfield  {journal} {\bibinfo
  {journal} {Mathematics of Computation}\ }\textbf {\bibinfo {volume} {93}},\
  \bibinfo {pages} {1203} (\bibinfo {year} {2023}{\natexlab{b}})}\BibitemShut
  {NoStop}%
\bibitem [{\citenamefont {Goedecker}\ \emph {et~al.}(1996)\citenamefont
  {Goedecker}, \citenamefont {Teter},\ and\ \citenamefont
  {Hutter}}]{Goedecker1996}%
  \BibitemOpen
  \bibfield  {author} {\bibinfo {author} {\bibfnamefont {S.}~\bibnamefont
  {Goedecker}}, \bibinfo {author} {\bibfnamefont {M.}~\bibnamefont {Teter}},\
  and\ \bibinfo {author} {\bibfnamefont {J.}~\bibnamefont {Hutter}},\
  }\bibfield  {title} {\bibinfo {title} {Separable dual-space gaussian
  pseudopotentials},\ }\href {https://doi.org/10.1103/PhysRevB.54.1703}
  {\bibfield  {journal} {\bibinfo  {journal} {Phys. Rev. B}\ }\textbf {\bibinfo
  {volume} {54}},\ \bibinfo {pages} {1703} (\bibinfo {year}
  {1996})}\BibitemShut {NoStop}%
\bibitem [{\citenamefont {Krack}(2005)}]{krack_pseudopotentials_2005}%
  \BibitemOpen
  \bibfield  {author} {\bibinfo {author} {\bibfnamefont {M.}~\bibnamefont
  {Krack}},\ }\bibfield  {title} {\bibinfo {title} {Pseudopotentials for {H} to
  {Kr} optimized for gradient-corrected exchange-correlation functionals},\
  }\href {https://doi.org/10.1007/s00214-005-0655-y} {\bibfield  {journal}
  {\bibinfo  {journal} {Theoretical Chemistry Accounts}\ }\textbf {\bibinfo
  {volume} {114}},\ \bibinfo {pages} {145} (\bibinfo {year}
  {2005})}\BibitemShut {NoStop}%
\bibitem [{\citenamefont {Garrity}\ \emph {et~al.}(2014)\citenamefont
  {Garrity}, \citenamefont {Bennett}, \citenamefont {Rabe},\ and\ \citenamefont
  {Vanderbilt}}]{garrity_pseudopotentials_2014}%
  \BibitemOpen
  \bibfield  {author} {\bibinfo {author} {\bibfnamefont {K.~F.}\ \bibnamefont
  {Garrity}}, \bibinfo {author} {\bibfnamefont {J.~W.}\ \bibnamefont
  {Bennett}}, \bibinfo {author} {\bibfnamefont {K.~M.}\ \bibnamefont {Rabe}},\
  and\ \bibinfo {author} {\bibfnamefont {D.}~\bibnamefont {Vanderbilt}},\
  }\bibfield  {title} {\bibinfo {title} {Pseudopotentials for high-throughput
  {DFT} calculations},\ }\href
  {https://doi.org/10.1016/j.commatsci.2013.08.053} {\bibfield  {journal}
  {\bibinfo  {journal} {Computational Materials Science}\ }\textbf {\bibinfo
  {volume} {81}},\ \bibinfo {pages} {446} (\bibinfo {year} {2014})}\BibitemShut
  {NoStop}%
\bibitem [{\citenamefont {Schmitz}\ \emph {et~al.}(2025)\citenamefont
  {Schmitz}, \citenamefont {Ploumhans},\ and\ \citenamefont
  {Herbst}}]{schmitz_2025_17084313}%
  \BibitemOpen
  \bibfield  {author} {\bibinfo {author} {\bibfnamefont {N.~F.}\ \bibnamefont
  {Schmitz}}, \bibinfo {author} {\bibfnamefont {B.}~\bibnamefont {Ploumhans}},\
  and\ \bibinfo {author} {\bibfnamefont {M.~F.}\ \bibnamefont {Herbst}},\
  }\href {https://doi.org/10.5281/zenodo.17084313} {\bibinfo {title}
  {niklasschmitz/ad-dfpt}} (\bibinfo {year} {2025}),\ \Eprint
  {https://arxiv.org/abs/17084313} {zenodo:17084313} \BibitemShut {NoStop}%
\end{thebibliography}%


\begin{thebibliography}{3}%
\makeatletter
\providecommand \@ifxundefined [1]{%
 \@ifx{#1\undefined}
}%
\providecommand \@ifnum [1]{%
 \ifnum #1\expandafter \@firstoftwo
 \else \expandafter \@secondoftwo
 \fi
}%
\providecommand \@ifx [1]{%
 \ifx #1\expandafter \@firstoftwo
 \else \expandafter \@secondoftwo
 \fi
}%
\providecommand \natexlab [1]{#1}%
\providecommand \enquote  [1]{``#1''}%
\providecommand \bibnamefont  [1]{#1}%
\providecommand \bibfnamefont [1]{#1}%
\providecommand \citenamefont [1]{#1}%
\providecommand \href@noop [0]{\@secondoftwo}%
\providecommand \href [0]{\begingroup \@sanitize@url \@href}%
\providecommand \@href[1]{\@@startlink{#1}\@@href}%
\providecommand \@@href[1]{\endgroup#1\@@endlink}%
\providecommand \@sanitize@url [0]{\catcode `\\12\catcode `\$12\catcode
  `\&12\catcode `\#12\catcode `\^12\catcode `\_12\catcode `\%12\relax}%
\providecommand \@@startlink[1]{}%
\providecommand \@@endlink[0]{}%
\providecommand \url  [0]{\begingroup\@sanitize@url \@url }%
\providecommand \@url [1]{\endgroup\@href {#1}{\urlprefix }}%
\providecommand \urlprefix  [0]{URL }%
\providecommand \Eprint [0]{\href }%
\providecommand \doibase [0]{https://doi.org/}%
\providecommand \selectlanguage [0]{\@gobble}%
\providecommand \bibinfo  [0]{\@secondoftwo}%
\providecommand \bibfield  [0]{\@secondoftwo}%
\providecommand \translation [1]{[#1]}%
\providecommand \BibitemOpen [0]{}%
\providecommand \bibitemStop [0]{}%
\providecommand \bibitemNoStop [0]{.\EOS\space}%
\providecommand \EOS [0]{\spacefactor3000\relax}%
\providecommand \BibitemShut  [1]{\csname bibitem#1\endcsname}%
\let\auto@bib@innerbib\@empty
\bibitem [{\citenamefont {de~Jong}\ \emph {et~al.}(2015)\citenamefont
  {de~Jong}, \citenamefont {Chen}, \citenamefont {Angsten}, \citenamefont
  {Jain}, \citenamefont {Notestine}, \citenamefont {Gamst}, \citenamefont
  {Sluiter}, \citenamefont {Krishna~Ande}, \citenamefont {van~der Zwaag},
  \citenamefont {Plata}, \citenamefont {Toher}, \citenamefont {Curtarolo},
  \citenamefont {Ceder}, \citenamefont {Persson},\ and\ \citenamefont
  {Asta}}]{de_jong_charting_2015}%
  \BibitemOpen
  \bibfield  {author} {\bibinfo {author} {\bibfnamefont {M.}~\bibnamefont
  {de~Jong}}, \bibinfo {author} {\bibfnamefont {W.}~\bibnamefont {Chen}},
  \bibinfo {author} {\bibfnamefont {T.}~\bibnamefont {Angsten}}, \bibinfo
  {author} {\bibfnamefont {A.}~\bibnamefont {Jain}}, \bibinfo {author}
  {\bibfnamefont {R.}~\bibnamefont {Notestine}}, \bibinfo {author}
  {\bibfnamefont {A.}~\bibnamefont {Gamst}}, \bibinfo {author} {\bibfnamefont
  {M.}~\bibnamefont {Sluiter}}, \bibinfo {author} {\bibfnamefont
  {C.}~\bibnamefont {Krishna~Ande}}, \bibinfo {author} {\bibfnamefont
  {S.}~\bibnamefont {van~der Zwaag}}, \bibinfo {author} {\bibfnamefont {J.~J.}\
  \bibnamefont {Plata}}, \bibinfo {author} {\bibfnamefont {C.}~\bibnamefont
  {Toher}}, \bibinfo {author} {\bibfnamefont {S.}~\bibnamefont {Curtarolo}},
  \bibinfo {author} {\bibfnamefont {G.}~\bibnamefont {Ceder}}, \bibinfo
  {author} {\bibfnamefont {K.~A.}\ \bibnamefont {Persson}},\ and\ \bibinfo
  {author} {\bibfnamefont {M.}~\bibnamefont {Asta}},\ }\bibfield  {title}
  {\bibinfo {title} {Charting the complete elastic properties of inorganic
  crystalline compounds},\ }\href {https://doi.org/10.1038/sdata.2015.9}
  {\bibfield  {journal} {\bibinfo  {journal} {Scientific Data}\ }\textbf
  {\bibinfo {volume} {2}},\ \bibinfo {pages} {150009} (\bibinfo {year}
  {2015})}\BibitemShut {NoStop}%
\bibitem [{\citenamefont {Lin}\ \emph {et~al.}(2024)\citenamefont {Lin},
  \citenamefont {Poncé}, \citenamefont {Macheda}, \citenamefont {Mauri},\ and\
  \citenamefont {Marzari}}]{lin2024elasticconstantsbendingrigidities}%
  \BibitemOpen
  \bibfield  {author} {\bibinfo {author} {\bibfnamefont {C.}~\bibnamefont
  {Lin}}, \bibinfo {author} {\bibfnamefont {S.}~\bibnamefont {Poncé}},
  \bibinfo {author} {\bibfnamefont {F.}~\bibnamefont {Macheda}}, \bibinfo
  {author} {\bibfnamefont {F.}~\bibnamefont {Mauri}},\ and\ \bibinfo {author}
  {\bibfnamefont {N.}~\bibnamefont {Marzari}},\ }\href
  {https://arxiv.org/abs/2412.18482} {\bibinfo {title} {Elastic constants and
  bending rigidities from long-wavelength perturbation expansions}} (\bibinfo
  {year} {2024}),\ \Eprint {https://arxiv.org/abs/2412.18482} {arXiv:2412.18482
  [cond-mat.mtrl-sci]} \BibitemShut {NoStop}%
\bibitem [{\citenamefont {Togo}\ \emph {et~al.}(2024)\citenamefont {Togo},
  \citenamefont {Shinohara},\ and\ \citenamefont {Tanaka}}]{togo_spglib_2024}%
  \BibitemOpen
  \bibfield  {author} {\bibinfo {author} {\bibfnamefont {A.}~\bibnamefont
  {Togo}}, \bibinfo {author} {\bibfnamefont {K.}~\bibnamefont {Shinohara}},\
  and\ \bibinfo {author} {\bibfnamefont {I.}~\bibnamefont {Tanaka}},\
  }\bibfield  {title} {\bibinfo {title} {Spglib: a software library for crystal
  symmetry search},\ }\href {https://doi.org/10.1080/27660400.2024.2384822}
  {\bibfield  {journal} {\bibinfo  {journal} {Science and Technology of
  Advanced Materials: Methods}\ }\textbf {\bibinfo {volume} {4}},\ \bibinfo
  {pages} {2384822} (\bibinfo {year} {2024})}\BibitemShut {NoStop}%
\end{thebibliography}%

\section*{Acknowledgements}
\noindent
This work was supported by
the Swiss National Science Foundation (SNSF, Grant No. 221186)
as well as the NCCR MARVEL, a National Centre of Competence in Research,
funded by the SNSF (Grant No. 205602).
The funder played no role in study design, data collection,
analysis and interpretation of data, or the writing of this manuscript.
Fruitful discussions with Andrea Azzali, Gaspard Kemlin, Antoine Levitt,
Uwe Naumann, Étienne Polack, and Markus Towara
on the technical aspects of our AD-DFPT implementation and with Giovanni Pizzi
regarding the pseudopotential training example are gratefully acknowledged.

\section*{Author contributions}
\noindent
N.F.S. implemented AD-DFPT in DFTK and contributed the examples of elasticity, inverse design, XC learning, pseudopotential optimization, and XC uncertainty propagation.
B.P. contributed the plane-wave error estimation example.
M.F.H. supervised the project, and provided theoretical and implementation support.
All authors analyzed the results and contributed to the writing of the manuscript.

\section*{Competing interests}
\noindent
The authors declare no competing interests.

\end{document}


\title{
    Supplementary information for\\
    ``Algorithmic differentiation for plane-wave DFT:\\
    materials design, error control and learning model parameters''
}

\author{Niklas Frederik Schmitz}
\email{niklas.schmitz@epfl.ch}
\affiliation{Mathematics for Materials Modelling (MatMat)\char`,{} Institute of Mathematics \& Institute of Materials,
École Polytechnique Fédérale de Lausanne, 1015 Lausanne, Switzerland}
\affiliation{National Centre for Computational Design and Discovery of Novel Materials (MARVEL),
\'Ecole Polytechnique F\'ed\'erale de Lausanne, 1015 Lausanne, Switzerland}

\author{Bruno Ploumhans}
\affiliation{Mathematics for Materials Modelling (MatMat)\char`,{} Institute of Mathematics \& Institute of Materials,
École Polytechnique Fédérale de Lausanne, 1015 Lausanne, Switzerland}
\affiliation{National Centre for Computational Design and Discovery of Novel Materials (MARVEL),
\'Ecole Polytechnique F\'ed\'erale de Lausanne, 1015 Lausanne, Switzerland}

\author{Michael F. Herbst}
\email{michael.herbst@epfl.ch}
\affiliation{Mathematics for Materials Modelling (MatMat)\char`,{} Institute of Mathematics \& Institute of Materials,
École Polytechnique Fédérale de Lausanne, 1015 Lausanne, Switzerland}
\affiliation{National Centre for Computational Design and Discovery of Novel Materials (MARVEL),
\'Ecole Polytechnique F\'ed\'erale de Lausanne, 1015 Lausanne, Switzerland}
\maketitle

\appendix
\setcounter{section}{0}
\renewcommand{\thesection}{S\arabic{section}}
\setcounter{figure}{0}
\renewcommand{\thefigure}{S\arabic{figure}} 
\setcounter{table}{0}
\renewcommand{\thetable}{S\arabic{table}}
\setcounter{equation}{0}

\renewcommand{\figurename}{Supplementary Figure}
\renewcommand{\tablename}{Supplementary Table}

\newcommand{\figelastic}{Figure~3\xspace} 

\onecolumngrid  

\section{Elastic constants}

\noindent
Complementing \figelastic{} of the main text,
Supplementary \cref{tab:elastic} contains the computed clamped-ion elastic constants for the
tightest used SCF tolerance $10^{-12}$.
The table demonstrates an excellent agreement of 7--8 digits between our AD-DFPT
approach and finite differences.
In~\figelastic we therefore use the AD-DFPT results of Table~\ref{tab:elastic}
as the reference values for our error computation.
However, we remark that due to the tight agreement all relative errors
larger than $10^{-7}$ are to leading order unchanged if the finite difference values were used as the reference.
In particular none of the finite difference curves shown in~\figelastic would look visually different
if the accurate finite-difference result from Table \ref{tab:elastic} was employed as the reference.

To obtain this data we employed the same 
computational parameters as described in the Methods section.
These settings (e.g.~the $k$-mesh) have been chosen to keep the computational cost
small when comparing to finite differences, and do not provide
a fully converged result.

\begin{table*}[ht]
    \centering
    \begin{tabular}{l l l l}
        \toprule
            & \multicolumn{3}{l}{\textbf{diamond}} \\
            & $C_{11}$ & $C_{12}$ & $C_{44}$ \\
        \midrule
            AD-DFPT & \textbf{1056.8864}147 & \textbf{125.1813}098 & \textbf{564.62397}27 \\
            FD      & \textbf{1056.8864}246 & \textbf{125.1813}203 & \textbf{564.62397}43 \\
            MP~\cite{de_jong_charting_2015} & 1053 & 126 & 561 \\
        \midrule
            & \multicolumn{3}{l}{\textbf{silicon}} \\
            & $C_{11}$ & $C_{12}$ & $C_{44}$ \\
        \midrule
            AD-DFPT & \textbf{153.05037}55 & \textbf{56.44207}2822 & \textbf{99.72657}670 \\
            FD      & \textbf{153.05037}61 & \textbf{56.44207}3688 & \textbf{99.72657}734 \\
            MP      & 153 & 57 & 74 \\
        \midrule
            & \multicolumn{3}{l}{\textbf{caesium chloride}} \\
            & $C_{11}$ & $C_{12}$ & $C_{44}$ \\
        \midrule
            AD-DFPT & \textbf{32.60613}704 & \textbf{5.50990}2647 & \textbf{4.975469}06 \\
            FD      & \textbf{32.60613}226 & \textbf{5.50990}3687 & \textbf{4.975469}17 \\
            MP      & 33 & 6 & 5 \\
        \bottomrule
    \end{tabular}
    \caption{\textbf{Elasticity.} Computed PBE clamped-ion elastic constants $C_{ij}$ (GPa) from our AD-DFPT framework versus finite difference (FD) computation of the stress (step size $h=10^{-5}$). Elastic constants from the Materials Project (MP) database~\cite{de_jong_charting_2015} are shown for additional context. Note that unlike our computations, the MP quantities employ PAW potentials as well as ionic relaxations, explaining the large discrepancy in $C_{44}$ for silicon~\cite{lin2024elasticconstantsbendingrigidities}.}
    \label{tab:elastic}
\end{table*}

\section{Derivative instability: Fermi-Dirac example}

\noindent
As highlighted in the Discussion, the standard rules of the AD system can lead to numerically unstable derivative computations.
To restore stability, a custom rule can be defined,
but the preferable solution is to switch to another mathematically equivalent expression.
In this section, we illustrate this point on the Fermi-Dirac function:
\begin{equation}
    \label{eqn:ffd}
    f_\text{FD}(x) = \frac{1}{1 + e^x}.
\end{equation}
An AD system will differentiate $f_\text{FD}$ to
\begin{equation}
    f_\text{FD}'(x) = \frac{e^x}{(1+e^x)^2}.
\end{equation}
For large $x$ (and overflowing exponential) this leads to the floating-point
operation $\text{Inf}/\text{Inf} = \text{NaN}$,
instead of the expected answer $f_\text{FD}'(\infty) = 0$.

To circumvent this problem in DFTK we employ the equivalent expression
\begin{equation}
    \label{eqn:ffdpos}
    f_\text{FD}(x) = \frac{e^{-x}}{e^{-x} + 1} \qquad \text{for $x > 0$}
\end{equation}
whenever $x$ is positive (while we keep the original expression \eqref{eqn:ffd} for $x\leq0$).
The AD-computed derivative of \eqref{eqn:ffdpos} becomes
\begin{equation}
    f_\text{FD}'(x) = \frac{-e^{-x}}{(e^{-x}+1)^2},
\end{equation}
which still features
an underflow of $e^{-x}$ for large $x$. However, this time the final result
in finite-precision arithmetic will remain the correct answer $0$.

\section{Symmetry-breaking crystal perturbations}

\noindent
In the Discussion, we mentioned that symmetry analysis must account for crystal perturbations. Consider a conventional diamond silicon unit cell, and let us compute the response of the electronic density wrt. the displacement of a single atom along the $x$ direction. This perturbation breaks some symmetries such as \SI{90}{\degree} rotation around the silicon atom situated in the $xy$ plane.

\begin{figure}[h!]
    \centering
    \includegraphics[width=\linewidth]{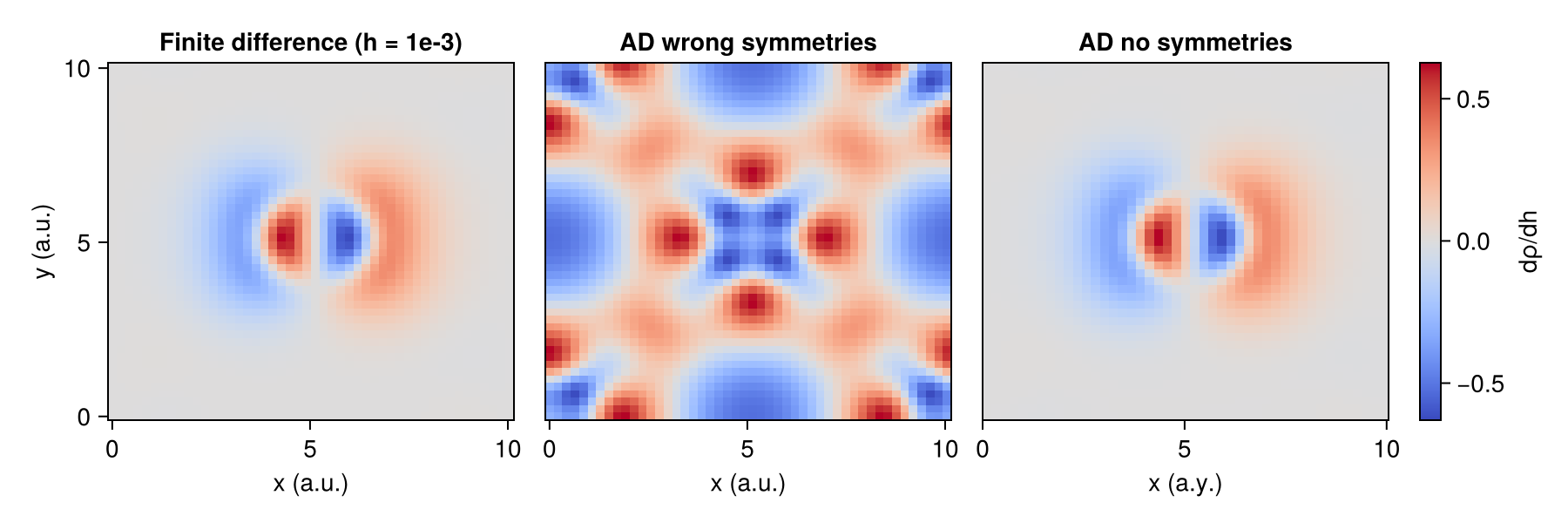}
    \caption{\textbf{Symmetry-breaking perturbation}. Derivative of the electronic density wrt.~the displacement of an atom along the $x$ direction. This perturbation breaks some symmetries of the diamond silicon crystal. Finite differences as well as symmetry-disabled AD-DFPT agree qualitatively, whereas AD-DFPT with naive symmetry analysis leads to an over-symmetrized density response. The density derivatives are shown in the $xy$ plane that contains the moved silicon atom.}
    \label{fig:symmetrybreaking}
\end{figure}

Supplementary \cref{fig:symmetrybreaking} compares three approaches to compute this derivative.
The first (leftmost) approach uses finite differences.
For each of the two SCF computations, the symmetry analysis is
performed separately on the respective input structures, and will thus disregard any broken perturbation.
The second approach is AD-DFPT with automatically detected symmetries. In our
current implementation, automatic symmetry analysis is performed by
Spglib~\cite{togo_spglib_2024} using only the \textit{unperturbed} lattice
parameters and atom positions; these symmetries are then used for both the SCF
and AD-DFPT computations. While simple, this naive approach leads to
over-symmetrization of the density response, leading to a result in qualitative disagreement with finite differences.
The third (rightmost) approach is AD-DFPT with all symmetries disabled. This
avoids erroneous over-symmetrization and provides the expected agreement with
finite differences, but comes with additional computational cost.

For maximal efficiency, AD-DFPT computations should be performed with the symmetry group of the perturbed crystal. Determining this symmetry group can be done in two practical alternative ways.
First, the symmetry group computed from a finite perturbation of the lattice and atoms positions with a small step size can be passed to our setup. We used this simple approach for the efficient implementation of elastic constants, which indeed require symmetry-breaking lattice strains.
Second, one may avoid this extra step and directly perform a symmetry analysis of the perturbation components inside of a custom differentiation rule. An implementation of this second approach would make the first approach obsolete and is currently work in progress.

\newpage
\bibliography{literature}